\documentclass[manuscript]{aastex}

\def\lax{{$\mathrel{\hbox{\rlap{\hbox{\lower4pt\hbox{$\sim$}}}\hbox{$<$}}}$}}
\def\gax{{$\mathrel{\hbox{\rlap{\hbox{\lower4pt\hbox{$\sim$}}}\hbox{$>$}}}$}}

\usepackage{color}

\usepackage{lscape}
\usepackage{physics}
\usepackage{graphicx}
\usepackage{ulem}

\slugcomment{}

\begin{document}
\title{Tidal disruption flares from stars on marginally bound and unbound orbits}
 \author{
Gwanwoo \textsc{Park}\altaffilmark{1} and Kimitake \textsc{Hayasaki}\altaffilmark{1}
}
\altaffiltext{1}{Department of Astronomy and Space Science, Chungbuk National University, Cheongju 361-763, Korea}
\email{gpark@cbnu.ac.kr}
%
\begin{abstract}
We study the mass fallback rate of tidally disrupted stars on marginally 
bound and unbound orbits around a supermassive black hole (SMBH) by performing 
three-dimensional smoothed particle hydrodynamic (SPH) simulations with three key parameters. 
The star is modeled by a polytrope with two different indexes ($n=1.5$ and $3$). The stellar orbital 
properties are characterized by five orbital eccentricities ranging from $e=0.98$ to $1.02$ and five 
different penetration factors ranging from $\beta=1$ to $3$, where $\beta$ represents the ratio of 
the tidal disruption to pericenter distance radii. We derive analytic formulae for the mass fallback 
rate as a function of the stellar density profile, orbital eccentricity, and penetration factor. Moreover, 
two critical eccentricities to classify tidal disruption events (TDEs) into five different types: 
eccentric ($e<e_{\rm crit,1}$), marginally eccentric ($e_{\rm crit,1}\lesssim{e}<1$), purely parabolic 
($e=1$), marginally hyperbolic ($1<e<e_{\rm crit,2}$), and hyperbolic ($e\gtrsim{e_{\rm crit,2}}$) TDEs, 
are reevaluated as $e_{\rm crit,1}=1-2q^{-1/3}\beta^{k-1}$ and $e_{\rm crit,2}=1+2q^{-1/3}\beta^{k-1}$, 
where $q$ is the ratio of the SMBH to stellar masses and $0<k\lesssim2$. We find the 
asymptotic slope of the mass fallback rate varies with the TDE type. The asymptotic 
slope approaches $-5/3$ for the parabolic TDEs, is steeper for the marginally eccentric TDEs, and 
is flatter for the marginally hyperbolic TDEs. For the marginally eccentric TDEs, the peak of mass 
fallback rates can be about one order of magnitude larger than the parabolic TDE case. For marginally 
hyperbolic TDEs, the mass fallback rates can be much lower than the Eddington accretion rate, 
which can lead to the formation of a radiatively inefficient accretion flow, while hyperbolic TDEs 
lead to failed TDEs. Marginally unbound TDEs could be an origin of a very low density gas disk 
around a dormant SMBH. 
\end{abstract}
%
\section{Introduction}
%

%
%

There is growing evidence that supermassive black holes (SMBHs) ubiquitously reside at the center of galaxies, based on observations of stellar proper motion, stellar velocity dispersion or accretion luminosity \citep{2013ARAA..51..511K}. 
Tidal disruption events (TDEs) provide a distinct opportunity to probe dormant 
SMBHs in inactive galaxies. Once a star approaches a SMBH and enters inside the 
tidal sphere, the star is tidally disrupted by the SMBH. The stellar debris then falls back 
to the SMBH at a super-Eddington rate, leading to a prominent flaring event 
with a luminosity exceeding the Eddington luminosity for weeks to months \citep{rees88,p89,ek89}.

%
%
Tidal disruption flares have been observed over the broad range of waveband from optical (\citealt{2011ApJ...741...73V}; \citealt{sg+12}; \citealt{2016MNRAS.455.2918H}) to ultraviolet \citep{2006ApJ...653L..25G,2008ApJ...676..944G,2014ApJ...780...44C} to soft X-ray (\citealt{kb99}; \citealt{2012A&A...541A.106S}; \citealt{2013MNRAS.435.1904M}; \citealt{2017ApJ...838..149A}) wavelengths. 
The TDE rates have been estimated to be $10^{-5}-10^{-4}$ per year per galaxy for soft-X-ray selected TDEs \citep{dbeb02,2008A&A...489..543E} and for optical-selected TDEs \citep{vf14,2016MNRAS.455.2918H,2018ApJ...852...72V,2018ApJS..238...15H}.
The observed rates are consistent with the theoretically expected rates (\citealt{jd04,sm16}; 
see also \citealt{2015JHEAp...7..148K} and \citealt{2020SSRv..216...35S} for a recent review). 
On the other hand, high-energy jetted TDEs have been detected through non-thermal emissions at radio \citep{baz+11,2016ApJ...819L..25A,2016Sci...351...62V} and/or hard X-ray \citep{dnb+11,2015MNRAS.452.4297B} wavelengths with much lower event rate than the thermal TDE case \citep{2014arXiv1411.0704F}. Spectroscopic studies have confirmed ${\rm H \,I}$ and ${\rm He\, II}$ (\citealt{ar14}) as well as metal lines (\citealt{2019arXiv190303120L}). Recently discovered blue-shifted (0.05c) broad absorption lines are 
likely to result from a high velocity outflow produced by the candidate TDE AT2018zr (\citealt{2019arXiv190305637H}).


%
%
It still remains under debate how the standard, theoretical mass fallback rate, 
which is proportional to $t^{-5/3}$ \citep{rees88,p89,ek89}, can be translated 
into the observed light curves. 
Dozens of X-ray TDEs have light curves shallower than $t^{-5/3}$ \citep{2017ApJ...838..149A}, while many optical/UV TDEs are well fit by $t^{-5/3}$ (e.g. \citealt{2017ApJ...842...29H}) but some deviate from $t^{-5/3}$ \citep{sg+12,2014ApJ...780...44C,ar14,ho14}. Specifically, the slope of the light curve depends on when measurements are taken relative to the peak. PS1-10JH shows a light curve more consistent with $t^{-5/3}$ at late times as shown by \citet{2015ApJ...815L...5G}. The decay rate becomes flatter at very late times \citep{2019ApJ...878...82V}. This 
flattening likely reflects the evolution of a viscously spreading disk rather than the continued evolution of the mass fallback rate \citep{1990ApJ...351...38C,2009ApJ...700.1047C,2014ApJ...784...87S}.

There have been some arguments that the mass fallback rate itself can deviate from the $t^{-5/3}$ decay rate. \cite{2009MNRAS.392..332L} demonstrated that the stellar internal structure makes the mass fallback rate deviate from the standard fallback rate in an early time. 
When the star is simply modeled by a polytrope, the stellar density profile is characterized by a polytropic index. In this case, the polytrope index is a key parameter to determine the mass fallback rate.

The penetration factor, which is the ratio of the tidal disruption to pericenter radii, 
is also an important parameter for TDEs. 
\cite{2013ApJ...767...25G} showed that resultant light curves 
can be steeper because of the centrally condensed core surviving after 
the partial disruption of the star. It would happen if the penetration factor 
is relatively low ($\beta\lesssim2$ for the case that the polytropic index equals to 3) 
\citep{2017A&A...600A.124M}. The self-gravity of the survived core can change the trajectories of the striped material and therefore the resultant mass fallback rate as well \citep{2012ApJ...757..134M}. The penetration factor also plays an important role in the process of debris circularization. The energy dissipation of a strong shock by a collision between the debris head and tail naturally leads to the formation of an accretion disk \citep{2013MNRAS.434..909H,2016MNRAS.455.2253B,2016MNRAS.461.3760H}, although the detailed dissipation mechanism is still under debate \citep{2013MNRAS.435.1809S,2015ApJ...804...85S,2015ApJ...806..164P}. Moreover, the penetration factor is a key parameter to determine the size and temperature of the circularized accretion disk \citep{2015ApJ...812L..39D}. Recent X-ray observation suggests that the TDE would happen with very high penetration factor \citep{2018arXiv181010713P}.

In the case of tidal disruption of a star on an elliptical orbit, 
\citet{2013MNRAS.434..909H} found that the resultant mass fallback rate is higher 
with smaller orbital eccentricity because the fallback timescale is much shorter and 
more material is bound to the SMBH compared to the parabolic orbit case. In contrast, 
the mass fallback rate can be much smaller than the Eddington rate or even zero for 
a hyperbolic orbit case. \citet{2018ApJ...855..129H} proposed that there are two critical 
eccentricities to classify TDEs into five different types by the stellar orbit: eccentric 
($e<e_{\rm crit,1}$), marginally eccentric ($e_{\rm crit,1}\lesssim{e}<1$), purely parabolic 
($e=1$), marginally hyperbolic ($1<e<e_{\rm crit,2}$), and hyperbolic ($e\gtrsim{e_{\rm crit,2}}$) 
TDEs, respectively, where $e_{\rm crit,1}=1-2q^{-1/3}\beta^{-1}$ and $e_{\rm crit,2}=1+2q^{-1/3}\beta^{-1}$, 
and $q$ is the ratio of the SMBH to stellar masses. Based on this classification, they also 
examined the frequency of each TDE by N-body experiments. 
They pointed out that stars on marginally elliptical and hyperbolic orbits 
can be a main TDE source in a spherical star cluster. Therefore, it is clear that the orbital 
eccentricity (and semi-major axis through the penetration factor) is also a key parameter 
to make the mass fallback rate deviate from the standard $t^{-5/3}$ decay rate.

%
%
However, there is still little known about how the three key parameters: polytropic index, penetration factor, 
and orbital eccentricity, and their combinations affect the mass fallback rate. In this paper, we therefore revisit the mass fallback rate onto an SMBH or IMBH by taking account of the three key parameters. In Section~\ref{sec:2}, we give a new condition to classify the TDEs by the stellar orbital type based on the assumption that the spread in debris energy is proportional to the $k$-th power of the penetration factor, where $k$ is presumed to range for $0<k<2$ (see \citealt{2013MNRAS.435.1809S}). In addition, we derive analytically the formula of the mass fallback rates, which includes the effect of the three key parameters plus $k$ on them. In Section~\ref{sec:results}, we describe our numerical simulation approach and compare the semi-analytical solutions of the mass fallback rates with the simulation results. Finally, Section~\ref{sec:con} is devoted to the conclusion of our scenario. 

%
\section{Revisit of mass fallback rates}
\label{sec:2}
%

In this section, we revisit the mass fallback rate by taking account of a stellar density profile \citep{2009MNRAS.392..332L} 
and the orbital eccentricity \citep{2018ApJ...855..129H}, including the dependence of the penetration factor, $\beta=r_{\rm t}/r _{\rm p}$, where $r_{\rm p}$ is the pericenter distance, on a spread in debris specific energy.
As a star approaches to a SMBH or an IMBH, it is torn apart by the tidal force of the black hole, 
which dominates the self-gravity of the star at the tidal disruption radius:
\begin{equation}
 r_{\rm t}=\left(\frac{M_{\rm bh}}{m_*}\right)^{1/3}r_{*}\approx
24\left(\frac{M_{\rm bh}}{10^6\,{M}_\odot}\right)^{-2/3}
\left(\frac{m_*}{{M}_\odot}\right)^{-1/3}
\left(\frac{r_*}{{R}_\odot}\right)
r_{\rm S}.
\label{eq:rt}
\end{equation} 
Here we denote the black hole mass as $M_{\rm bh}$, stellar mass and radius as $m_*$ and $r_*$, 
and the Schwarzschild radius as $r_{\rm S}=2{\rm G}M_{\rm bh}/{\rm c}^2$, where $G$ and 
$c$ are Newton's gravitational constant and the speed of light, 
respectively. 

Following \cite{2013MNRAS.435.1809S}, the tidal force produces a spread in specific
energy of the stellar debris:
\begin{equation}
\Delta \mathcal{E} = \beta^{k}\Delta\epsilon,
\label{eq:delebeta}
\end{equation}
where $k$ is the power-law index of the penetration factor in the leading order term 
of the tidal potential energy (hereafter, tidal spread energy index) and 
\begin{equation}
\Delta\epsilon= \frac{GM_{\rm bh}r_*}{r_{\rm t}^2}
\label{eq:spreade}
\end{equation}
is the standard spread energy \citep{rees88,ek89}.
If $\beta=1$ or $k=0$, Equation (\ref{eq:delebeta}) reduces to the standard equation. 
The possible range of $k$ has been taken as $0\le{k}\le2$.

The mass fallback rate can be written by the chain rule as
\begin{eqnarray}
\frac{dM}{dt}=\frac{dM}{d\epsilon}\frac{d\epsilon}{dt},
\label{eq:dmdt}
\end{eqnarray}
where $dM/d\epsilon_{\rm }$ is the differential mass distribution of the stellar 
debris with specific energy $\epsilon$. Because the thermal energy of the stellar debris 
is negligible compared with the debris binding energy, $\epsilon\approx\epsilon_{\rm d}$:
\begin{equation}
\epsilon_{\rm d}\equiv-\frac{GM_{\rm bh}}{2a_{\rm d}},
\label{eq:ed}
\end{equation}
where $a_{\rm d}$ is the semi-major axis of the stellar debris. 
Applying the Kepler's third law to equation (\ref{eq:ed}), we obtain that \begin{equation}
\frac{d\epsilon_{\rm d}}{dt}=\frac{1}{3}(2\pi{GM}_{\rm bh})^{2/3}t^{-5/3}.
\label{eq:dedt}
\end{equation}

%
\subsection{Effect of stellar density profiles}
%

\cite{2009MNRAS.392..332L} included the effect of the stellar density profile on 
the differential mass distribution of the stellar debris as
\begin{equation}
\frac{dM}{d\epsilon_{\rm d}}=\frac{dM}{d\Delta r}\frac{d\Delta r}{d\epsilon_{\rm d}},
\label{eq:lkp09dmde}
\end{equation}
where $\Delta r$ is the radial width of the star.
In our case, the relation between the radial width and 
the debris specific binding energy is given by
\begin{equation}
\frac{\Delta r}{r_*}=
\frac{|\epsilon_{\rm d}|}{\Delta \mathcal{E}}=\frac{\mathcal{A}_c}{a_{\rm d}},
\label{eq:lkp09drde}
\end{equation}
where $\mathcal{A}_{\rm c}$ is the critical semi-major axis:
\begin{eqnarray}
\mathcal{A}_{\rm c}=a_{\rm c}\beta^{-k},
\label{eq:ac}
\end{eqnarray}
with $a_{\rm c}\equiv(M_{\rm bh}/m_*)^{1/3}r_{\rm t}/2$. If $\beta=1$ or $k=0$ is adopted here, 
equation (\ref{eq:lkp09drde}) reduces to that of \cite{2009MNRAS.392..332L}. Moreover, the radial width depends on the 
orbital period of the stellar debris through the binding energy, i.e., $\Delta{r}\propto{a}_{\rm d}^{-1}\propto{t^{-2/3}}$. 

%
%
The internal density structure of the star is given by the radial integral of the stellar density
\begin{equation}
\frac{dM}{d\Delta r} = 
2\pi \int_{\Delta r}^{r_{*}} \rho\left(r\right) r {\rm d}r,
\label{eq:dmdr}
\end{equation}
where $\rho \left(r\right)$ is the spherically symmetric mass density of the star and 
the polytropes with no stellar rotation are considered. We can integrate equation~(\ref{eq:dmdr}) by solving
the Lane-Emden equation: 
$\left(
1/\xi^{2}\right)
d
\left(
\xi^{2}d\theta/d\xi\right)/d\xi=-\theta^{n}$, 
where $\theta\left(\xi\right)=\rho/\rho_{\rm c}$, 
$\xi=r/r_{\rm c}$, $\rho_{\rm c}$ is the normalization density, 
$r_{\rm c}=\sqrt{(n+1)K\rho_{\rm c}^{1/n-1}/4\pi{G}}$ is the 
normalization radius, $n$ is a polytropic index, $K$ is a polytropic constant, respectively \citep{chan67}. 
Substituting equations~(\ref{eq:lkp09drde}) and (\ref{eq:dmdr}) into equation~(\ref{eq:lkp09dmde}),
we obtain the differential mass distribution as
\begin{equation}
\frac{dM}{d\epsilon_{\rm d}}=\frac{3}{2}
\left(\frac{\rho_{\rm c}}{\bar{\rho}}\right)
\left(\frac{r_{\rm c}}{r_{*}}\right)^{2}
\left(\frac{m_{*}}{\Delta \mathcal{E}}\right)
\int_{\Delta{r}/r_{\rm c}}^{r_*/r_{\rm c}} \theta\left(\xi\right)\xi {\rm d}\xi,
\label{eq:semidmde}
\end{equation}
where $\bar{\rho}=m_*/(4\pi{r_*^3}/3)$ is the mean density of the star. 
Because $\theta(\xi)$ is obtained by solving the Lane-Emden equation numerically, 
$dM/d\epsilon_{\rm d}$ is semi-analytically determined (see also Figure~\ref{fig:dmde2}). 
Following the Kepler's third law, we can estimate the orbital period of the most tightly bound debris as 
\begin{equation} 
t_{\rm mtb}^{'}=2\pi\sqrt{\frac{\mathcal{A}_{\rm c}^3}{GM_{\rm bh}}}=t_{\rm mtb}\beta^{-3k/2},
\label{eq:tmtb}
\end{equation}
where $t_{\rm mtb}=2\pi\sqrt{a_{c}^{3}/GM_{\rm bh}}$ corresponds to the $\beta=1$ case. 
Substituting equations~(\ref{eq:dedt}) and (\ref{eq:semidmde}) into equation~(\ref{eq:dmdt}), 
we obtain the mass fallback rate:
\begin{eqnarray}
\frac{dM}{dt}
&=&
\left(\frac{\rho_{\rm c}}{\bar{\rho}}\right)
\left(\frac{r_{\rm c}}{r_{*}}\right)^{2}
\left(\frac{m_{*}}{t_{\rm mtb}^{'}}\right)
 \left(\frac{t}{t_{\rm mtb}^{'}}\right)^{-5/3}
  \int_{\Delta{r}/r_{\rm c}}^{r_*/r_{\rm c}} \theta\left(\xi\right)\xi {\rm d}\xi
  \nonumber \\
&=& 
\left(\frac{\rho_{\rm c}}{\bar{\rho}}\right)
\left(\frac{r_{\rm c}}{r_{*}}\right)^{2}
\left(\frac{1}{\beta^{k}}\right)
\left(\frac{m_{*}}{t_{\rm mtb}}\right)
\left(\frac{t}{t_{\rm mtb}}\right)^{-5/3}
\int_{\Delta r/r_{\rm c}}^{r_{*}/r_{\rm c}}\theta(\xi) \xi d \xi
\label{eq:dmdt1}
\end{eqnarray}
For $n=3$ and $\xi \ll1$, the normalized density can be expanded to be $\theta(\xi) \approx 1 - \xi^{2}/6 + \mathcal{O}(\xi^{4})$. Since we obtain from equation~(\ref{eq:lkp09drde}) and (\ref{eq:tmtb}) that $\Delta{r}/r_{\rm c}=(r_*/r_{\rm c})(t/t_{\rm mtb})^{-2/3}\beta^{-k}$, we can approximately estimate the mass fallback rate as $dM/dt\approx(1/2)(\rho_{c}/\bar{\rho})(1/\beta^{k})(m_{*}/t_{\rm mtb})(t/t_{\rm mtb})^{-5/3}[1-(t/t_{\rm mtb})^{-4/3}\beta^{-2k}][1-(1/12)(r_{*}/r_{\rm c})^{2}(1+(t/t_{\rm mtb})^{-4/3}\beta^{-2k})]$. We find that the mass fallback rate depends on not only the stellar density profile but also the penetration factor and the tidal spread energy index, which leads to the deviation from $t^{-5/3}$.

%
\subsection{Effect of stellar orbital properties}
\label{sec:orbitalprop}
%
In this section, we investigate stars that approach the SMBH on parabolic, 
eccentric, and hyperbolic orbits. The specific energy of the stellar debris is in the range of 
$-\Delta \mathcal{E} -GM/(2a) \le {\epsilon_{\rm d}} \le \Delta \mathcal{E} +GM/(2a)$, 
where $a$ is the orbital semi-major axis of the approaching star.

Following Hayasaki et al. (2018), the TDEs are classified by the critical eccentricities 
in terms of the orbital eccentricity of the star:
\begin{eqnarray}
\left\{ \begin{array}{ll}
0 \le e < e_{\rm crit,1} & {\rm eccentric \,\, TDEs} \\
e_{\rm crit,1} \le e < 1 & {\rm marginally \,\, eccentric \,\,TDEs} \\
e = 1 & {\rm parabolic \,\,TDEs} \\
1 < e \le e_{\rm crit,2} & {\rm marginally \,\, hyperbolic \,\,TDEs} \\
e_{\rm crit,2} < e & {\rm hyperbolic \,\,TDEs}, \\
\end{array} \right.
\label{eq:eclass}
\end{eqnarray}
where $e_{\rm crit, 1}$ and $e_{\rm crit,2}$ are modified as
\begin{eqnarray}
e_{\rm crit,1}&=&1-2q^{-1/3}\beta^{k-1}, \nonumber \\
e_{\rm crit,2}&=&1+2q^{-1/3}\beta^{k-1}
\label{eq:ec1}
\end{eqnarray}
with $q\equiv{M}_{\rm bh}/m_{*}$, respectively.
If $\beta=1$ or $k=0$, these two terms reduce to the previously defined critical eccentricities (see equations (5) and (6) of \citealt{2018ApJ...855..129H}). The modified specific binding energy of the most tightly bound stellar debris for eccentric or hyperbolic stellar orbits is given by
\begin{equation}
\epsilon_{\rm mtb}=-\Delta \mathcal{E} \pm \frac{GM_{\rm bh}}{2a}=-\left(
1\mp\frac{\mathcal{A}_{\rm c}}{a}
\right)\Delta \mathcal{E},
\label{eq:kimiemtb}
\end{equation}
where the negative and positive signs of the specific orbital energy of the star indicate 
the eccentric and hyperbolic orbit cases, respectively. 
The orbital period of the most tightly bound debris is also changed as
\begin{eqnarray}
\tau_{\rm mtb}=2\pi\sqrt{\frac{\mathcal{A}_{\rm c}^3}{GM_{\rm bh}}}
\left(1\mp\frac{\mathcal{A}_{\rm c}}{a}\right)^{-3/2}
=t_{\rm mtb}\left(\beta^{-3k/2}\right)\left(1\mp\frac{\mathcal{A}_{\rm c}}{a}\right)^{-3/2},
\label{eq:kimitmtb}
\end{eqnarray}
where we use equation (\ref{eq:tmtb}) and $\mathcal{A}_{\rm c}/a$ should 
be smaller than unity for the upper negative sign (hyperbolic TDE) case. 
The differential mass distribution is thus changed from equation (\ref{eq:semidmde}) to
\begin{eqnarray}
\frac{dM}{d\epsilon_{\rm d}}=
\frac{3}{2}
\left(
\frac{\rho_{\rm c}}{\bar{\rho}}\right)
\left(\frac{r_{\rm c}}{r_{*}}\right)^{2}
\left(\frac{m_{*}}{\Delta \mathcal{E}}\right)
\int_{\Delta{r}^{'}/r_{\rm c}}^{r_*/r_{\rm c}} \theta\left(\xi\right)\xi {\rm d}\xi,
\label{eq:kimidmde}
\end{eqnarray}
where $\Delta{r}^{'}$ is the newly defined radial width of the star and is given by 
\begin{equation}
\frac{\Delta{r}^{'}}{r_*}
=\frac{|\epsilon_{\rm d}^{'}|}{\Delta \mathcal{E}}=\frac{\mathcal{A}_{\rm c}}{a_{\rm d}}\left(1\mp\frac{a_{\rm d}}{a}\right)
\label{eq:mod-deltar}
\end{equation}
with the modification of the debris binding energy, i.e., 
$\epsilon_{\rm d}^{'}=-GM/(2a_{\rm d})\pm{GM}/(2a)$. 
Note that $\theta(\xi)=0$ if $\Delta{r}^{'}/r_*$ is greater than unity because there is no stellar gas there.

Substituting equations~(\ref{eq:dedt}) and (\ref{eq:kimidmde}) into equation~(\ref{eq:dmdt}) 
and applying equations~(\ref{eq:kimiemtb}) and (\ref{eq:kimitmtb}), we can obtain the modified mass fallback rate as
\begin{align}
\frac{dM}{dt}
&=
\left(\frac{\rho_{\rm c}}{\bar{\rho}}\right)
\left(\frac{r_{\rm c}}{r_{*}}\right)^{2}
\left(1\mp\frac{\mathcal{A}_{\rm c}}{a}\right)
\left(\frac{m_{*}}{\tau_{\rm mtb}}\right)
\left(\frac{t}{\tau_{\rm mtb}}\right)^{-5/3}
\int_{\Delta{r}^{'}/r_{\rm c}}^{r_*/r_{\rm c}} \theta\left(\xi\right)\xi {\rm d}\xi.
\nonumber 
\\
&=
\left(\frac{\rho_{\rm c}}{\bar{\rho}}\right)
\left(\frac{r_{\rm c}}{r_{*}}\right)^{2}
\left(\frac{1}{\beta^{k}}\right)
\left(\frac{m_{*}}{t_{\rm mtb}}\right)
\left(\frac{t}{t_{\rm mtb}}\right)^{-5/3}
\int_{\Delta{r}^{'}/r_{\rm c}}^{r_*/r_{\rm c}} \theta\left(\xi\right)\xi {\rm d}\xi.
\label{eq:dmdt2}
\end{align}
Applying $\theta(\xi) \simeq 1 - \xi^{2}/6 + O(\xi^{4})$ to equation~(\ref{eq:dmdt2}) 
for the $n=3$ and $\xi\ll1$ case, we approximately estimate the mass fallback rate as 
$dM/dt\approx(\rho_{c}/\bar{\rho})(1/\beta^{k})(m_{*}/t_{\rm mtb})(t/t_{\rm mtb})^{-5/3}
[1-(t/t_{\rm mtb})^{-4/3}\beta^{-2k}[1+(t/t_{\rm mtb})^{2/3}(a_{\rm c}/a)]^{2}]
[1-(1/12)(r_{*}/r_{\rm c})^{2}(1+(t/t_{\rm mtb})^{-4/3}\beta^{-2k}[1+(t/t_{\rm mtb})^{2/3}(a_{\rm c}/a)]^{2})]
$. 
This is applied only for eccentric to parabolic orbit cases and
is not valid for the hyperbolic orbit case, 
because the expansion formula of $\theta(\xi)$ corresponds to the density profile of the central part 
of the approaching star, which is unbound to the black hole if the star is on a hyperbolic orbit.
We find the mass fallback rate depends on the stellar density profile with the orbital eccentricity (or semi-major axis), 
the penetration factor, and the tidal spread energy index, leading to the deviation from $t^{-5/3}$. 
We test this hypothesis by numerical simulations and describe our results in Section~\ref{sec:results}.

%
\subsection{Numerical simulations}
\label{sec:3}
%

%
%
We evaluate how well the analytical solution matches the numerical simulations 
by using a three-dimensional (3D) Smoothed Particle Hydrodynamics (SPH) code. 
The SPH code is developed based on the original version of \citet{benz90a,benz90b} 
and substantially modified as described in \cite{bate95} and parallelized using both
 OpenMP and MPI.
 
%
%
Two-stage simulations are performed to model a tidal interaction between a star 
and a black hole. In the first-stage simulation, we model the star by a polytrope 
with $n=1.5$ and $n=3$ for a solar-type star. We run the simulations until the 
polytrope is virialized. In the second-stage simulation, the star is initially located 
at a distance of three times the tidal disruption radii and approaches the SMBH 
following Kepler's third law for five orbital eccentricities and five different penetration 
factors per each orbital eccentricity. In summary, we run a total 50 simulations in 
the second-stage. The stellar mass, stellar radius, and black hole mass are held 
constant throughout the simulations at  $m_*=1\,M_\odot$, $r_*=1\,R_\odot$, and 
$M_{\rm{bh}}=10^6\,M_\odot$, respectively. The total number of SPH particles used 
in each simulation is slightly more than $10^6$ and the run time is measured in units 
of $\Omega_{*}^{-1}=\sqrt{r_*^3/(Gm_*)}\simeq1.6\times10^3\,\rm{s}$.

%
%
Tables~\ref{tbl:1} and \ref{tbl:2} present a summary of the SPH simulation models 
and the corresponding spread energy indices obtained from the simulations. 
For both tables, the first to third columns show the polytropic index ($n$), 
the penetration factor ($\beta$), and the orbital eccentricity ($e$), respectively.
The fourth and fifth columns show two critical eccentricities 
$e_{\rm crit, 1}$ and $e_{\rm crit, 2}$, respectively (see equation~\ref{eq:ec1}). 
The final column presents the tidal spread energy index ($k$), which is estimated 
by fitting the simulation data (see the detail in Section~\ref{subsec:k}). 
We find from Tables~\ref{tbl:1} and \ref{tbl:2} that the $e=0.98$ and $e=0.99$, 
$e=1.0$, $e=1.01$, and $e=1.02$ cases correspond to marginally eccentric, 
parabolic, marginally hyperbolic, hyperbolic TDEs, respectively.


%
 \clearpage
 %

\begin{table}

  \caption{
Summary for parameters of our simulations. 
The first column shows the polytrope index ($n$). 
The second and third columns present the penetration 
factor ($\beta$) and the orbital eccentricity ($e$), respectively.
The fourth and fifth columns show the two critical eccentricities 
$e_{\rm crit, 1}$ and $e_{\rm crit, 2}$, respectively (see equation~\ref{eq:ec1}). 
The final column presents the specific energy index ($k$), which is obtained 
by fitting the simulation data (see equation~\ref{eq:k-eva}. 
}
\begin{minipage}{.45\linewidth}
 \centering
  \begin{tabular}{@{}cccccccccc@{}}
\cline{1-6}
\cline{1-6}
$n$ & $\beta$ & $e$ & $e_{\rm crit,1}$ & $e_{\rm crit,2}$  & $k$ \\
\cline{1-6}
\cline{1-6}
       &1 &         & 0.980 & 1.02   &  $-$  \\ 
\cline{2-2}\cline{4-9}
      &1.5 &       & 0.983 & 1.02    & 0.595   \\ 
\cline{2-2}\cline{4-9}
1.5 & 2 & 0.98 & 0.987 & 1.01     & 0.332  \\ 
\cline{2-2}\cline{4-9}
      & 2.5 &      & 0.989 & 1.01   & 0.293  \\ 
\cline{2-2}\cline{4-9}                          
       & 3 &        & 0.982 & 1.02    & 0.902  \\ 

\cline{2-6}
       &1 &       & 0.980  & 1.02 &   $-$  \\ 
\cline{2-2}\cline{4-9}
      &1.5 &      & 0.983  & 1.02 & 0.593 \\ 
\cline{2-2}\cline{4-9}
1.5 & 2 &    0.99       & 0.987  & 1.00 & 0.327 \\ 
\cline{2-2}\cline{4-9}
      & 2.5 &      & 0.989 & 1.01  & 0.287 \\ 
\cline{2-2}\cline{4-9}                      
        & 3 &       & 0.982 & 1.02 & 0.897 \\ 

\cline{2-6}
       &1 &      & 0.980  & 1.02  & $-$  \\ 
\cline{2-2}\cline{4-9}
       &1.5 &     & 0.983  & 1.02 & 0.646 \\ 
\cline{2-2}\cline{4-9}
1.5 & 2 &  1.0 & 0.987 & 1.01  & 0.324 \\ 
\cline{2-2}\cline{4-9}
      & 2.5 &     & 0.989 & 1.01 & 0.273 \\ 
\cline{2-2}\cline{4-9}                      
       & 3 &     & 0.990 &  1.01  & 0.323 \\
\cline{1-6}
\cline{1-6}       
 \end{tabular}
 \end{minipage}
 %
 %
 \begin{minipage}{.45\linewidth}
\centering

 \begin{tabular}{@{}cccccccccc@{}}
 \cline{1-6}
\cline{1-6}
$n$ & $\beta$ & $e$ & $e_{\rm crit,1}$ & $e_{\rm crit,2}$  & $k$ \\
\cline{1-6}
\cline{1-6}
\cline{1-6}
& 1 &         & 0.980 & 1.02 &  $-$ \\
\cline{2-2}\cline{4-9} 
& 1.5 &      & 0.983 & 1.02  & 0.584 \\
\cline{2-2}\cline{4-9}   
1.5 & 2 & 1.01 & 0.988 & 1.01  & 0.317  \\
\cline{2-2}\cline{4-9}
& 2.5 &         &0.989 & 1.01  & 0.275  \\ 
\cline{2-2}\cline{4-9}  
& 3 &            & 0.991 & 1.01 & 0.318 \\

\cline{2-6} 
      & 1 &       & 0.980 & 1.02   & $-$ \\
\cline{2-2}\cline{4-9} 
      & 1.5 &       & 0.983 & 1.02 &  0.580 \\
\cline{2-2}\cline{4-9}   
1.5 & 2 &   1.02   & 0.988 & 1.01 & 0.312 \\
\cline{2-2}\cline{4-9}
       & 2.5 &      & 0.989 & 1.01  & 0.270 \\ 
\cline{2-2}\cline{4-9}  
& 3 &            &0.991 & 1.01 &  0.313 \\
\cline{1-6}
\cline{1-6}
\end{tabular}
\label{tbl:1}
\end{minipage}
\end{table}

 %
 %
\begin{table*}
  \caption{
Summary for parameters of our simulations. 
The format of each column is the same as Table~\ref{tbl:1}, 
but for the $n=3$ case.
}
\begin{minipage}{.45\linewidth}
\centering
\begin{tabular}{@{}ccccccccccccc@{}}
 \cline{1-6} 
 \cline{1-6} 
$n$ & $\beta$ & $e$ & $e_{\rm crit,1}$ & $e_{\rm crit,2}$ & k \\
\cline{1-6} 
\cline{1-6} 
& 1 &     & 0.980 & 1.02 & $-$ & \\
\cline{2-2}\cline{4-9}
&1.5 &  & 0.968 & 1.03   & 2.17 &  \\ 
\cline{2-2}\cline{4-9}
3 & 2 &   0.98     & 0.970 & 1.03 &  1.58 & \\ 
\cline{2-2}\cline{4-9}
& 2.5 &     & 0.976 & 1.02 &  1.19 & \\ 
\cline{2-2}\cline{4-9}                    
& 3 &        & 0.980 & 1.02 & \phn{0.981} & \\ 

\cline{1-6}
& 1 &         & 0.980 & 1.02 &  $-$ & \\
\cline{2-2}\cline{4-9}
&1.5 &  & 0.968 & 1.03   &   2.14 &  \\ 
\cline{2-2}\cline{4-9}
3 & 2 &  0.99     & 0.970 & 1.03 & 1.57 & \\ 
\cline{2-2}\cline{4-9}
& 2.5 &       & 0.976 & 1.02  &  1.19 & \\ 
\cline{2-2}\cline{4-9}                    
& 3 &         & 0.980 & 1.02 &  \phn{0.978} & \\ 

\cline{1-6}
& 1 &      & 0.980  & 1.02  & $-$ &\\
\cline{2-2}\cline{4-9}
&1.5 &  & 0.968 & 1.03   &  2.12 &  \\ 
\cline{2-2}\cline{4-9}
3 & 2 &   1.0    & 0.970 & 1.03 &  1.57 & \\ 
\cline{2-2}\cline{4-9}
& 2.5 &      & 0.976 & 1.02  & 1.19 & \\ 
\cline{2-2}\cline{4-9}                    
& 3 &       & 0.981 & 1.02 & \phn{0.974} & \\ 
\cline{1-6} 
\cline{1-6} 

 \end{tabular}
 \end{minipage}
%
%
 \begin{minipage}{.45\linewidth}
 \centering
 \begin{tabular}{@{}cccccccccccc@{}}
 \cline{1-6}
\cline{1-6}
$n$ & $\beta$ & $e$ & $e_{\rm crit,1}$ & $e_{\rm crit,2}$  & $k$ \\
\cline{1-6}
\cline{1-6} 
& 1 &         & 0.980  & 1.02  & $-$ & \\
\cline{2-2}\cline{4-9}
&1.5 &  & 0.969 & 1.03   & 2.10 &  \\  
\cline{2-2}\cline{4-9}
3 & 2 &   1.01   & 0.971 & 1.03 & 1.56 & \\ 
\cline{2-2}\cline{4-9}
& 2.5 &       & 0.976 & 1.02  &  1.19 & \\ 
\cline{2-2}\cline{4-9}                    
& 3 &         & 0.981 & 1.02 & \phn{0.971} & \\ 

\cline{1-6} 
& 1 &        & 0.980 & 1.02  & $-$ & \\
\cline{2-2}\cline{4-9}
&1.5 &  & 0.969 & 1.03  & 2.07 &  \\ 
\cline{2-2}\cline{4-9}
3 & 2 &  1.02     & 0.971 & 1.03  & 1.55 & \\ 
\cline{2-2}\cline{4-9}
& 2.5 &      & 0.976 & 1.02  & 1.18 & \\ 
\cline{2-2}\cline{4-9}                    
& 3 &         & 0.981 & 1.02 &  \phn{0.967} & \\ 
\cline{1-6} 
\cline{1-6} 

\end{tabular}
\label{tbl:2}
\end{minipage}
\end{table*}

%
\clearpage
\section{Results}
\label{sec:results}
%

In this section, we describe the simulation results in order to compare 
our semi-analytical prediction with that of the SPH simulations. 

%
\subsection{Differential mass distribution of stellar debris}
%

We first compare the simulated differential mass distribution of the stellar debris over 
the specific energy measured at a run time of $t=4$ with the Gaussian-fitted distributions; 
we then compare our findings with the semi-analytical solution given by equation (\ref{eq:dmdt2}). 
The relevance of the fitting is also discussed.

\begin{figure}[!ht]

\includegraphics[width=8cm]{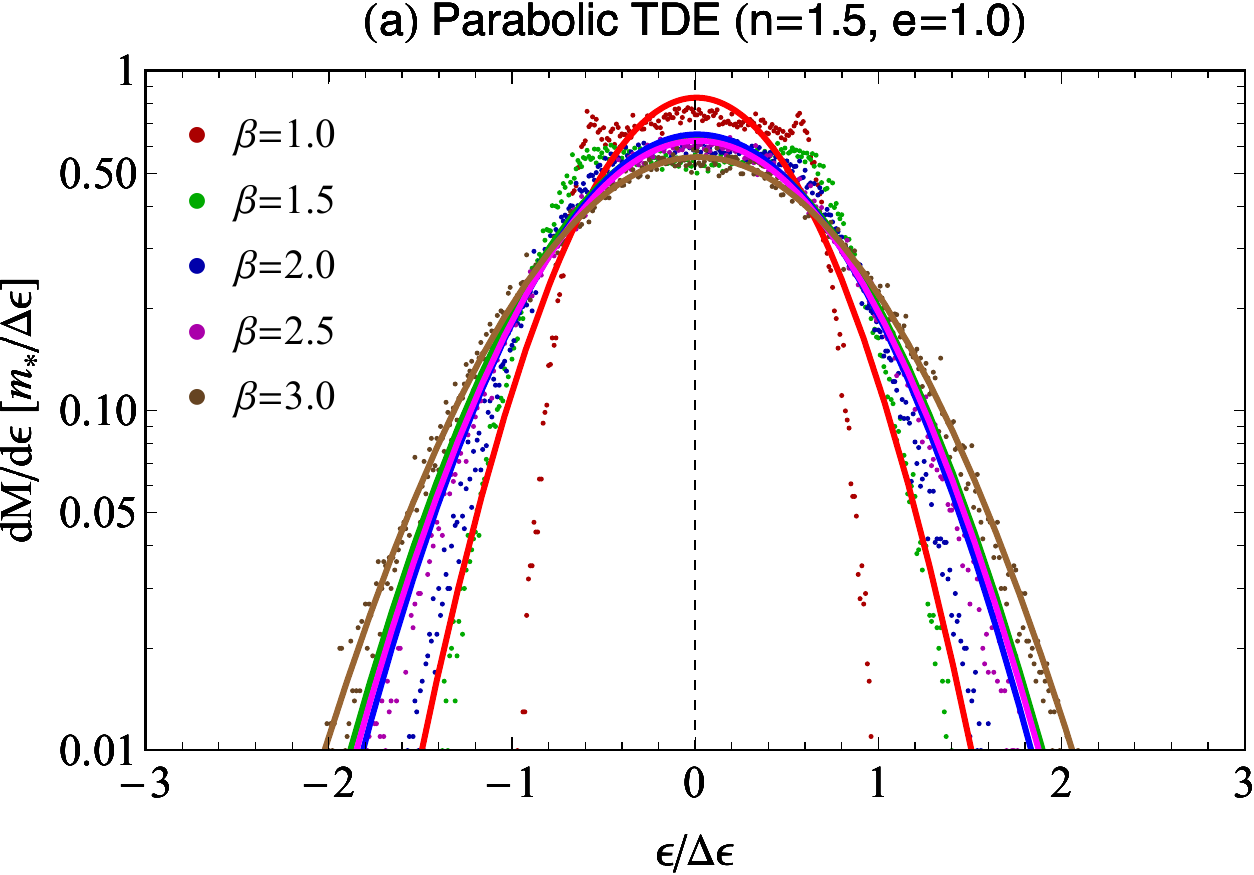}
\includegraphics[width=8cm]{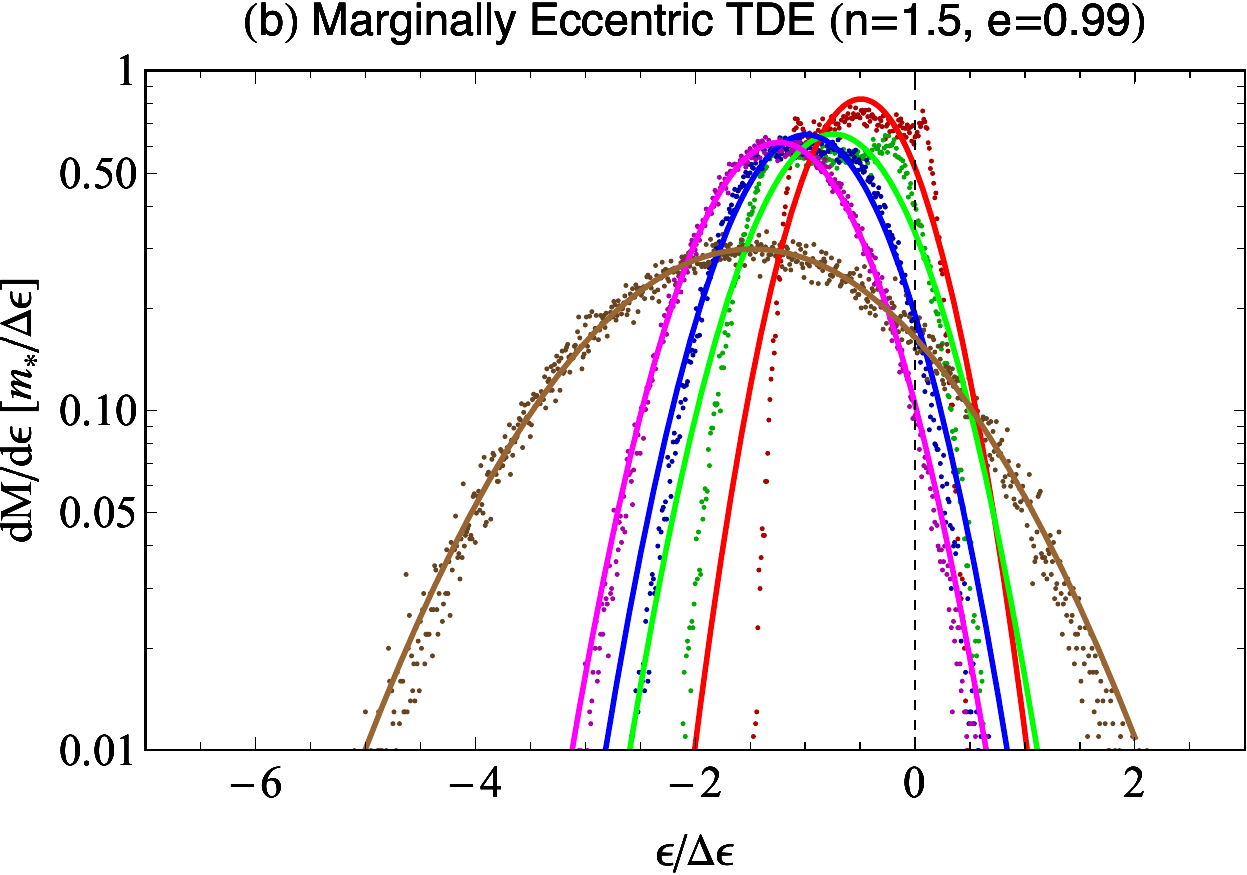}
\\
\includegraphics[width=8cm]{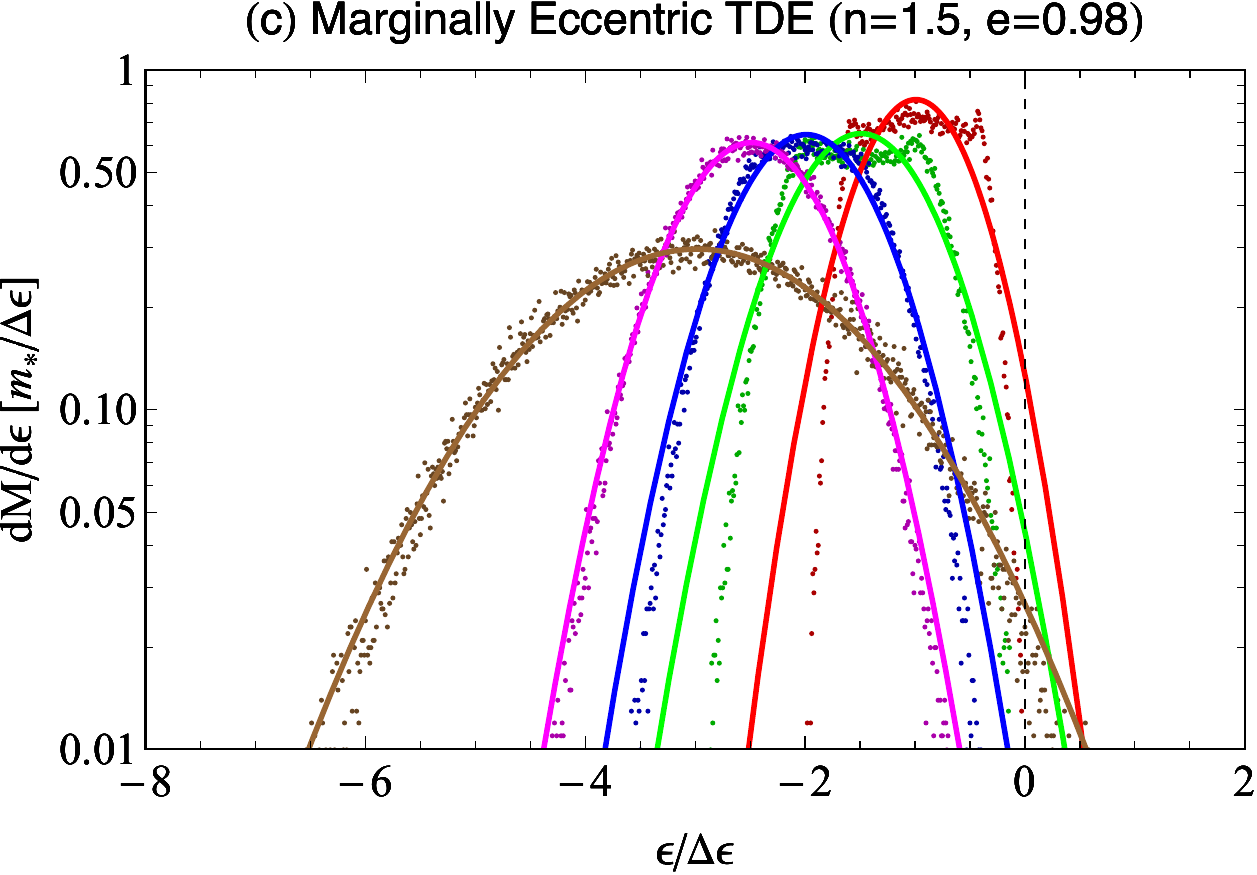}
\includegraphics[width=8cm]{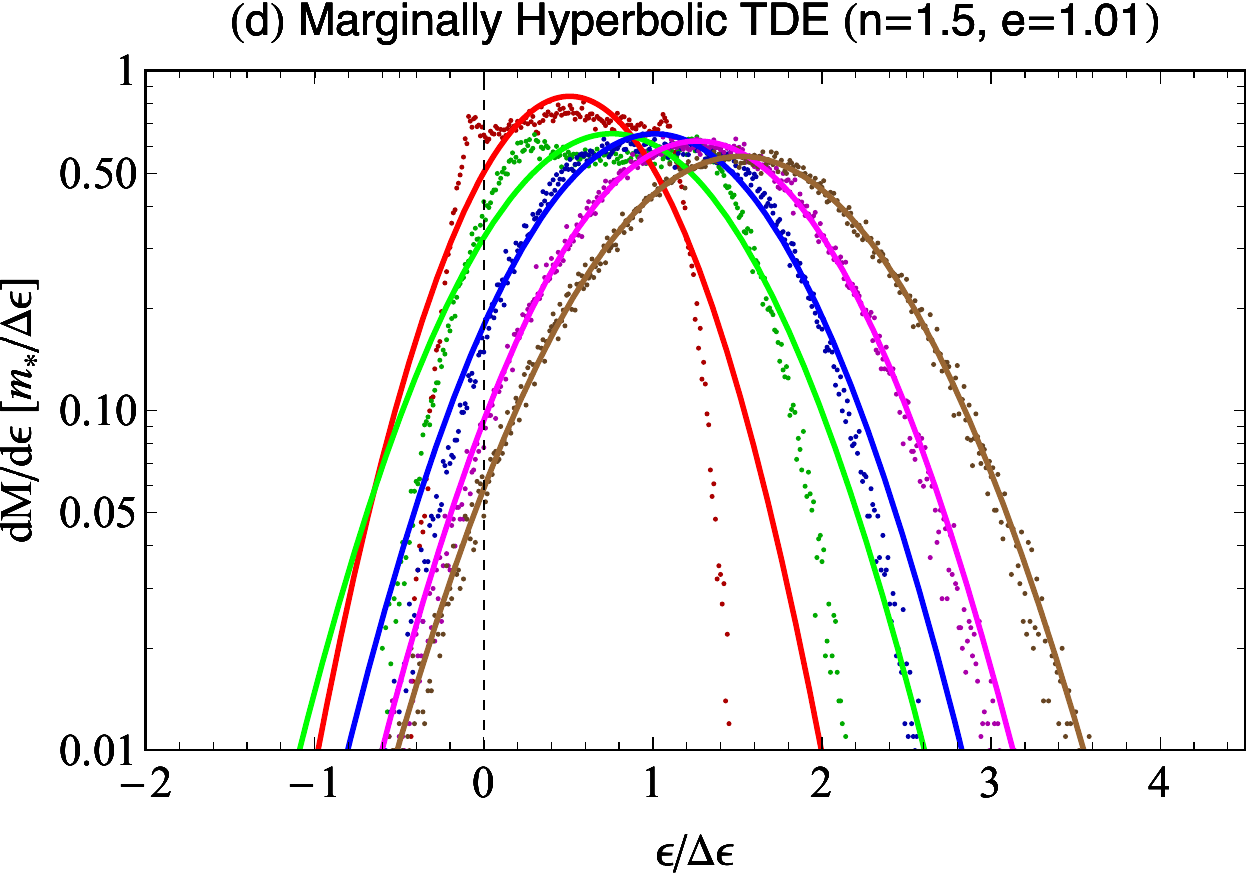}
\\
\includegraphics[width=8cm]{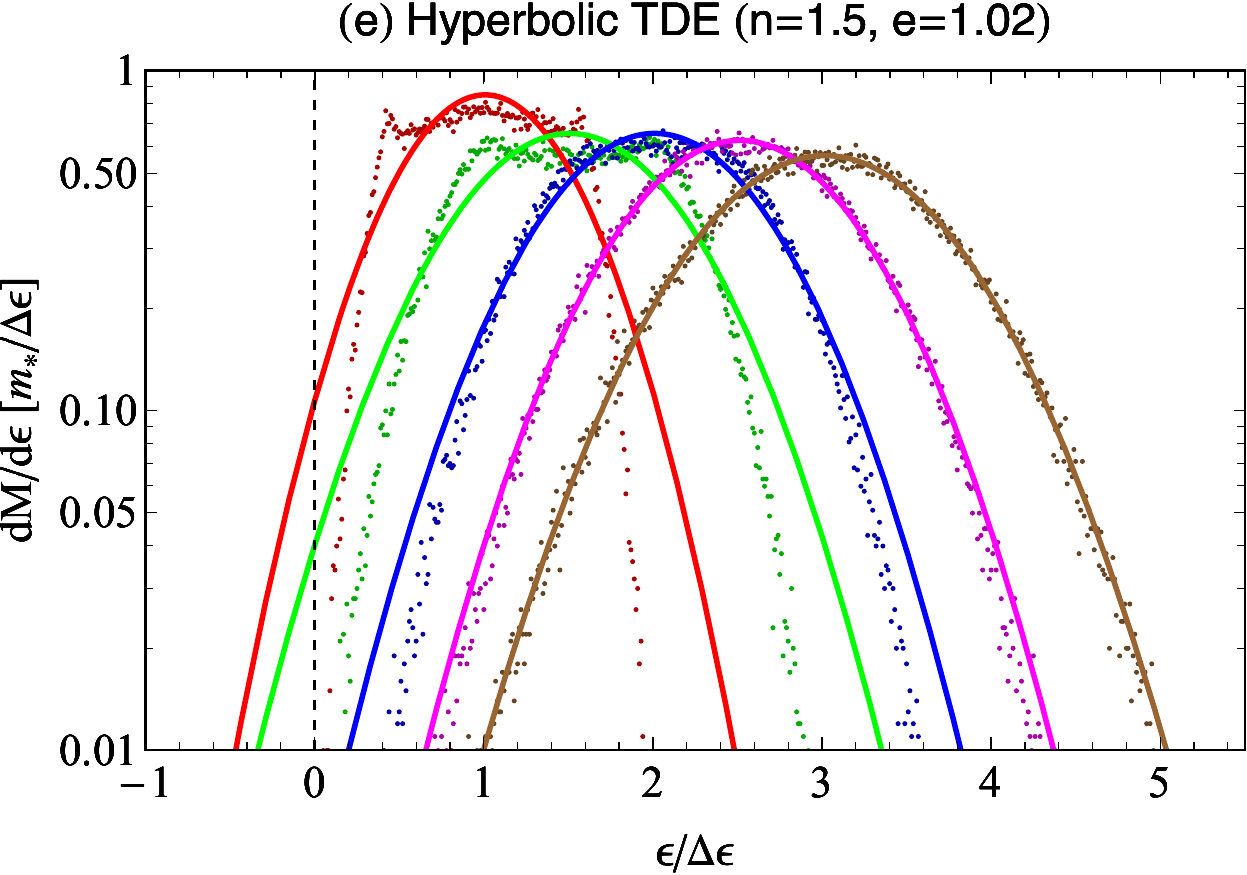}

\caption{
Simulated energy distribution of stellar debris for a $n=1.5$ polytrope (the differential debris mass is normalized 
by $m_{*}/\Delta \epsilon$). Each panel shows a different orbital eccentricity. 
The different colors correspond to different penetration factors ($\beta$). 
The solid lines are Gaussian fits to the simulation data.
}
\label{fig:dmde1}
\end{figure}

Figure~\ref{fig:dmde1} shows the energy distribution of the debris 
for $n=1.5$ with five different penetration factors. 
Panel (a) shows the differential mass distribution for the standard, parabolic case ($e=1.0$). 
Panels (b)-(e) show results for the marginally eccentric ($e=0.98$ and $0.99$) and 
marginally hyperbolic ($e=1.01$ and $e=1.02$) TDE cases, respectively. In all the panels, the 
red, green, blue, magenta, and brown color points are the differential mass distributions for 
$\beta=1$, $\beta=1.5$, $\beta=2$, $\beta=2.5$, and $\beta=3$, respectively.
The corresponding fitted curves are obtained using the {\it FindFit} model provided 
by {\it Mathematica} and are represented in the same color format.
It can find a non-linear fitting with the gaussian function, 
$f(x)=(1/\sqrt{2 \pi}\sigma)\exp[-(1/2)([x-\mu]/\sigma)^{2}]$, where $\mu$ is the position of the center of the peak 
and $\sigma$ is the standard deviation which is proportional to a half width at half maximum (HWHM).
We also simply evaluate the accuracy of the fitting by using the 
root mean square (RMS) and its square is given by 
${\rm RMS}^2\,=\sum_{i=1}^{N_{\rm d}}\,( [ y_i - f(x_i) ] / f(x_i) )^2 / N_{\rm d}$, 
where $N_{\rm d}$ is the number of data points, and $x_i$ and $y_i$ are the 
normalized specific energy at the $i$-th data point and the corresponding 
differential mass distribution, respectively.
We evaluate the RMS using only points within $1\,\sigma$ of the Gaussian fitted curve, $\Delta\mathcal{E}_{\rm sim}$.
We confirm that for parabolic TDEs, 
the differential mass is distributed around a specific energy of zero 
and also a half of the debris mass, independently of $\beta$, is bound by the SMBH. 
As we predicted in Section~\ref{sec:orbitalprop}, the position of the peak of the 
differential mass is shifted in the negative direction for eccentric TDEs and the resultant 
differential mass is distributed over there, while the position of the peak is positively shifted 
for the hyperbolic TDE cases. The deviation of the peak position corresponds to $a_{\rm c}/a$ 
to an accuracy of less than $2\%$. Most of the debris mass can fallback to the SMBH because of their negative 
binding energy in marginally eccentric TDEs, whereas most of the debris mass moves far away from the 
SMBH because of their positive binding energy in marginally hyperbolic TDE case.
The debris mass becomes more widely distributed for the debris specific energy as 
the penetration factor increases, particularly for the marginally eccentric and hyperbolic TDEs. 
In all cases, the peak of the differential mass distribution is smaller as the penetration factor is larger.

\begin{figure}[!ht]

\includegraphics[width=8cm]{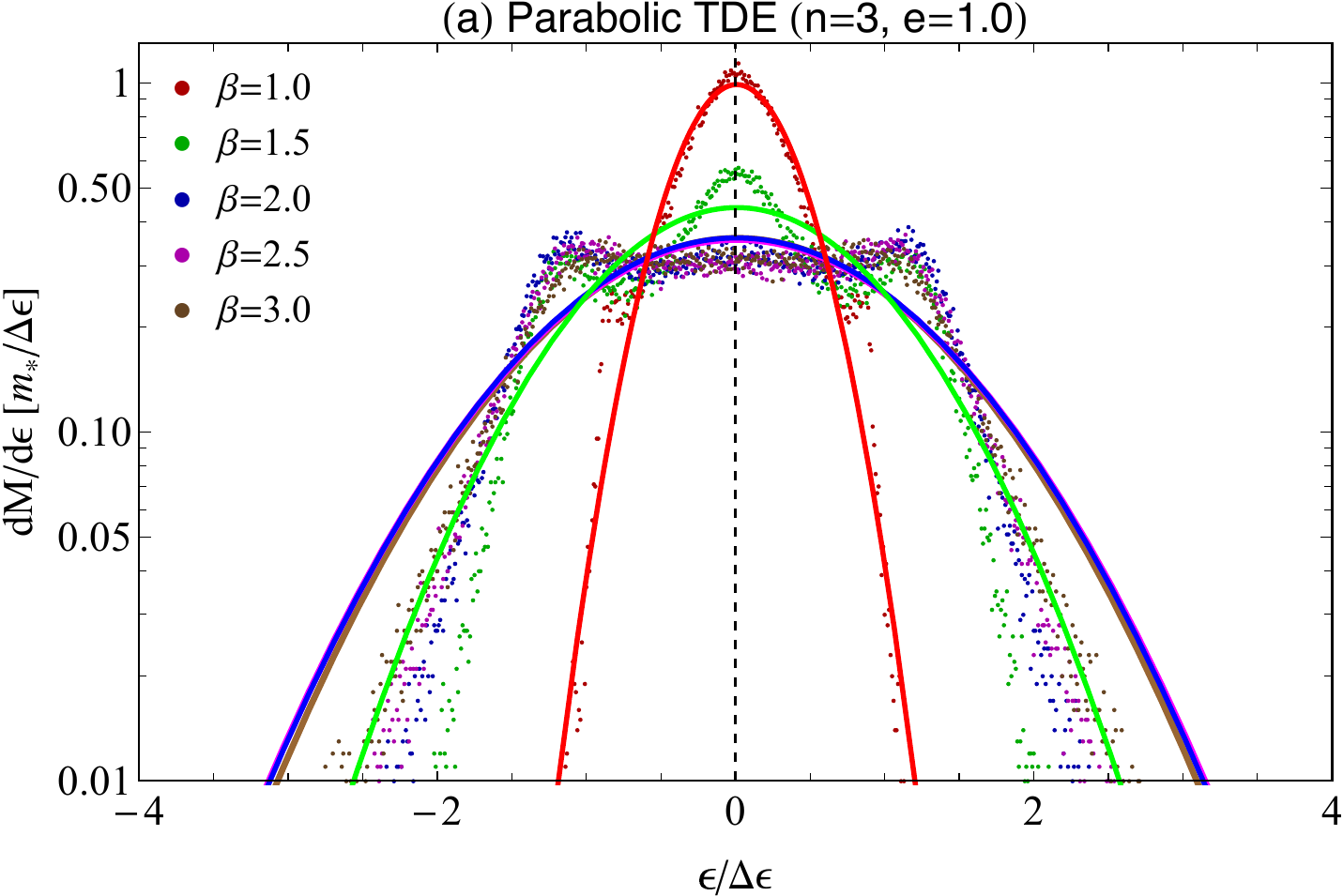}
\includegraphics[width=8cm]{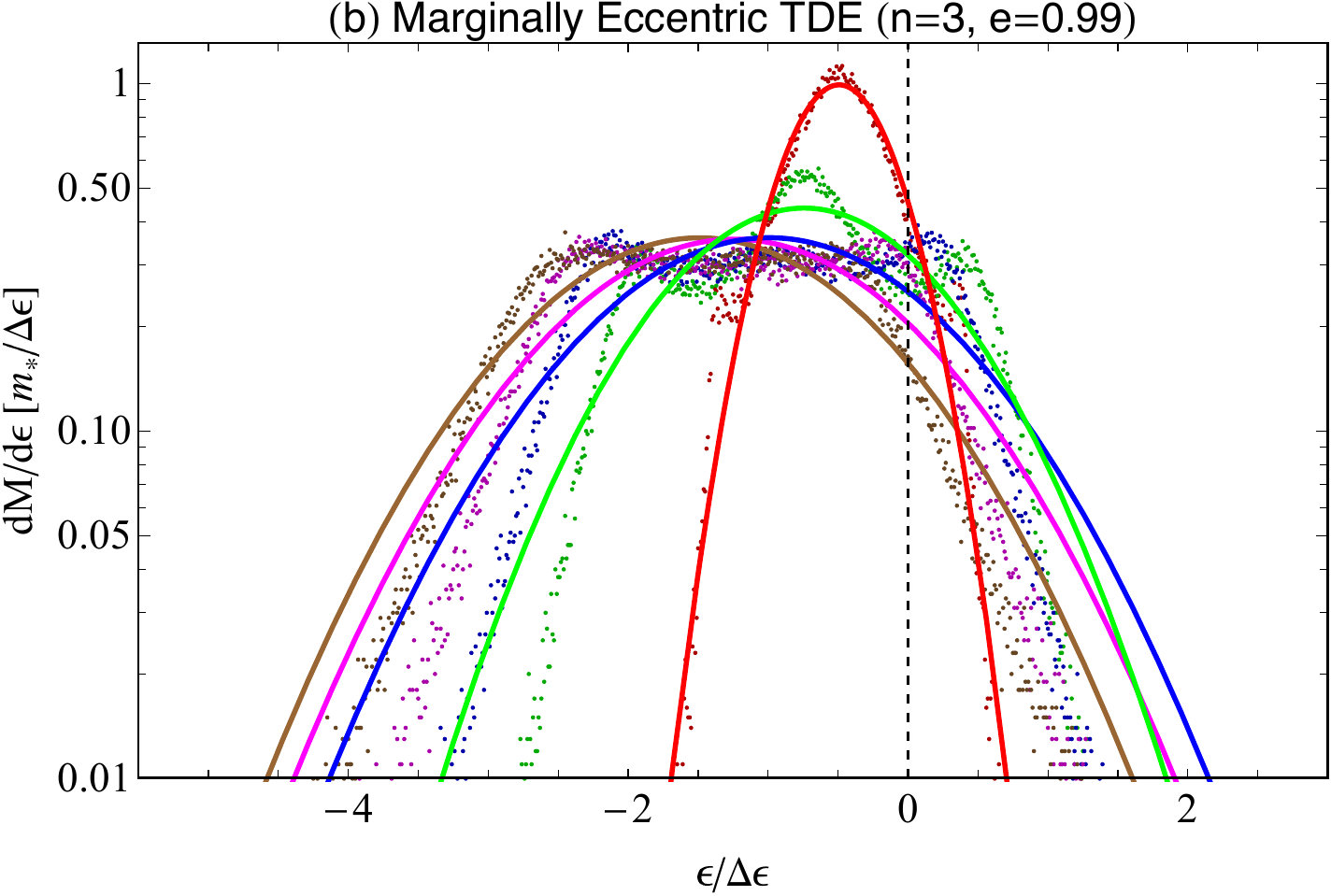}
\\
\includegraphics[width=8cm]{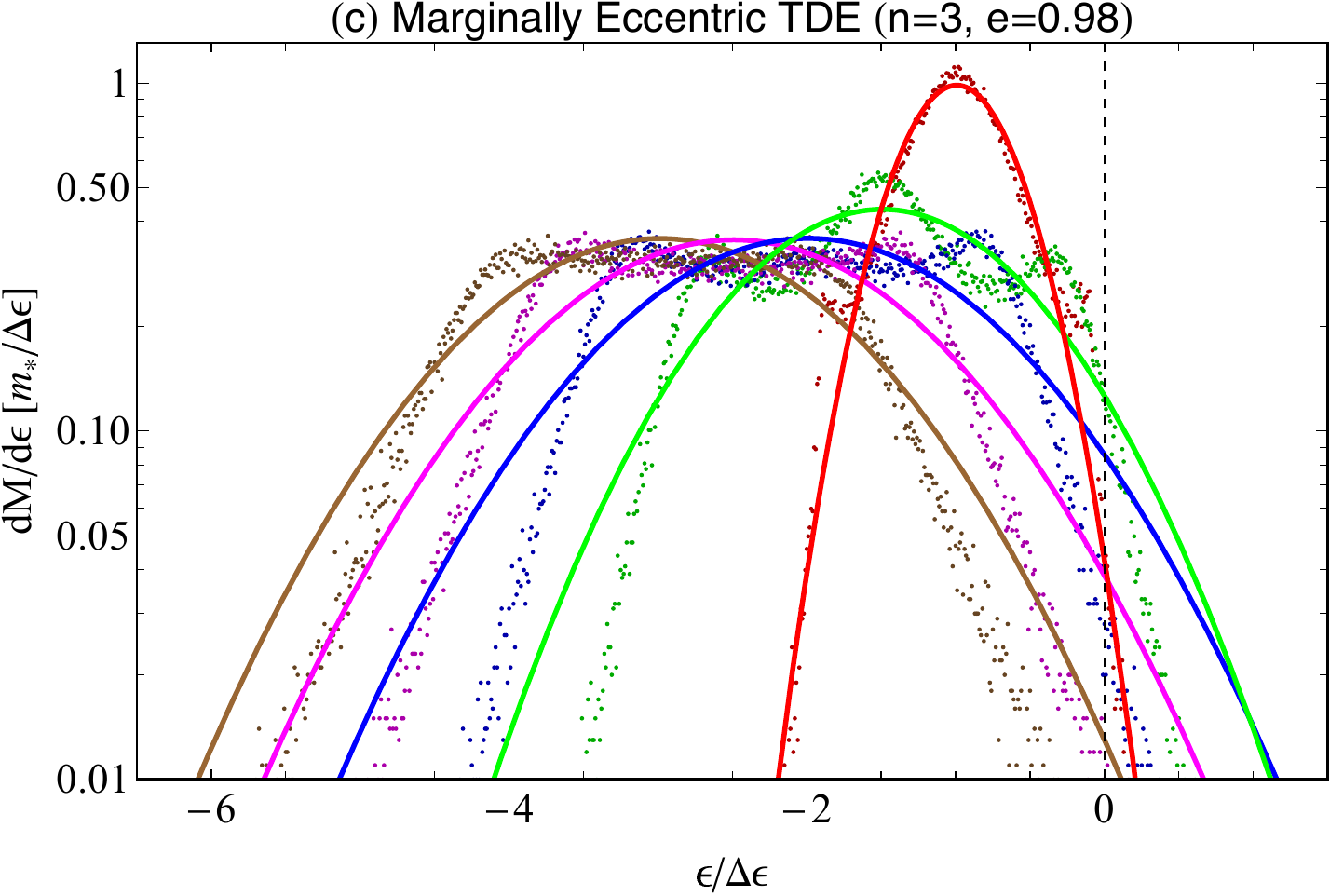}
\includegraphics[width=8cm]{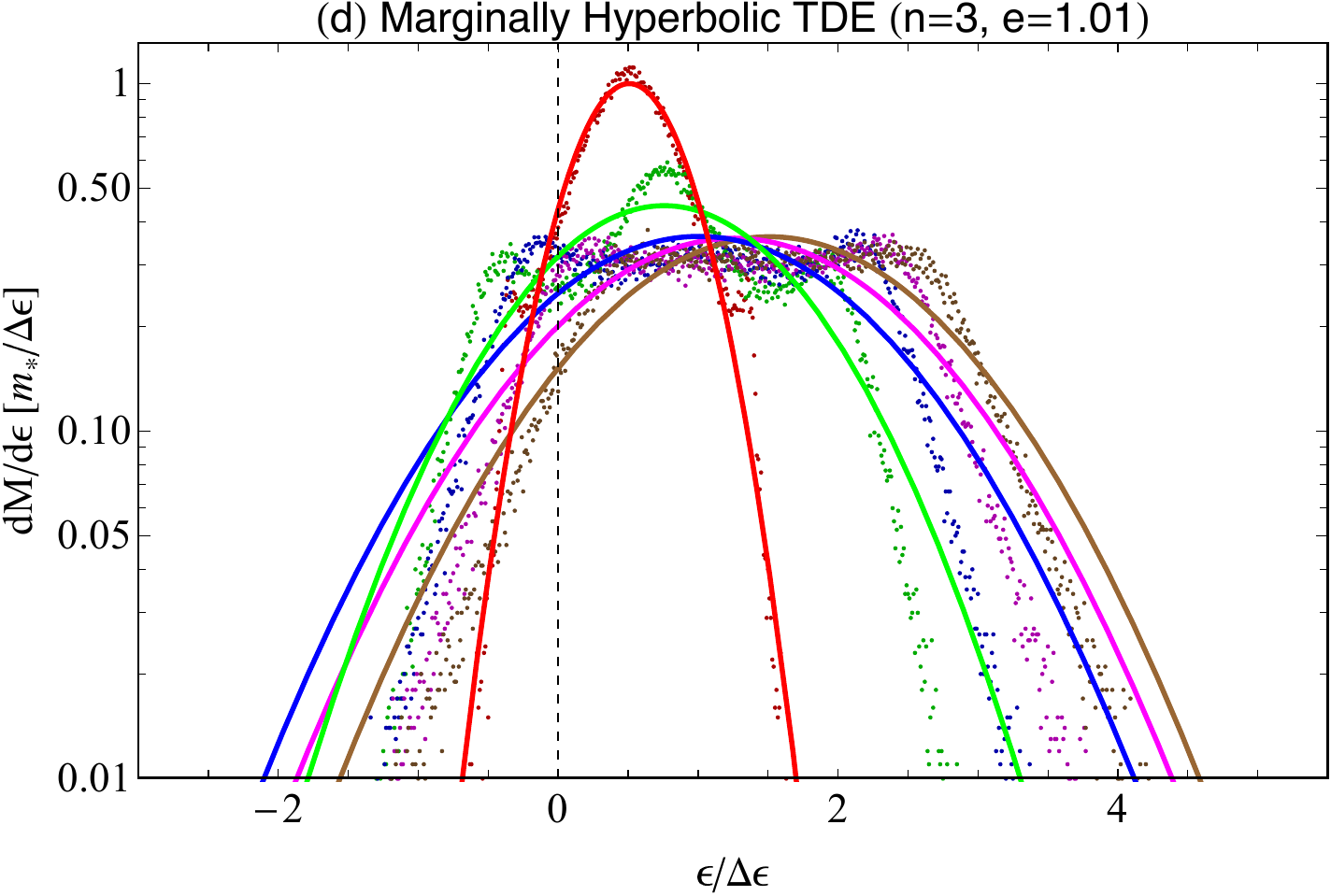}
\\
\includegraphics[width=8cm]{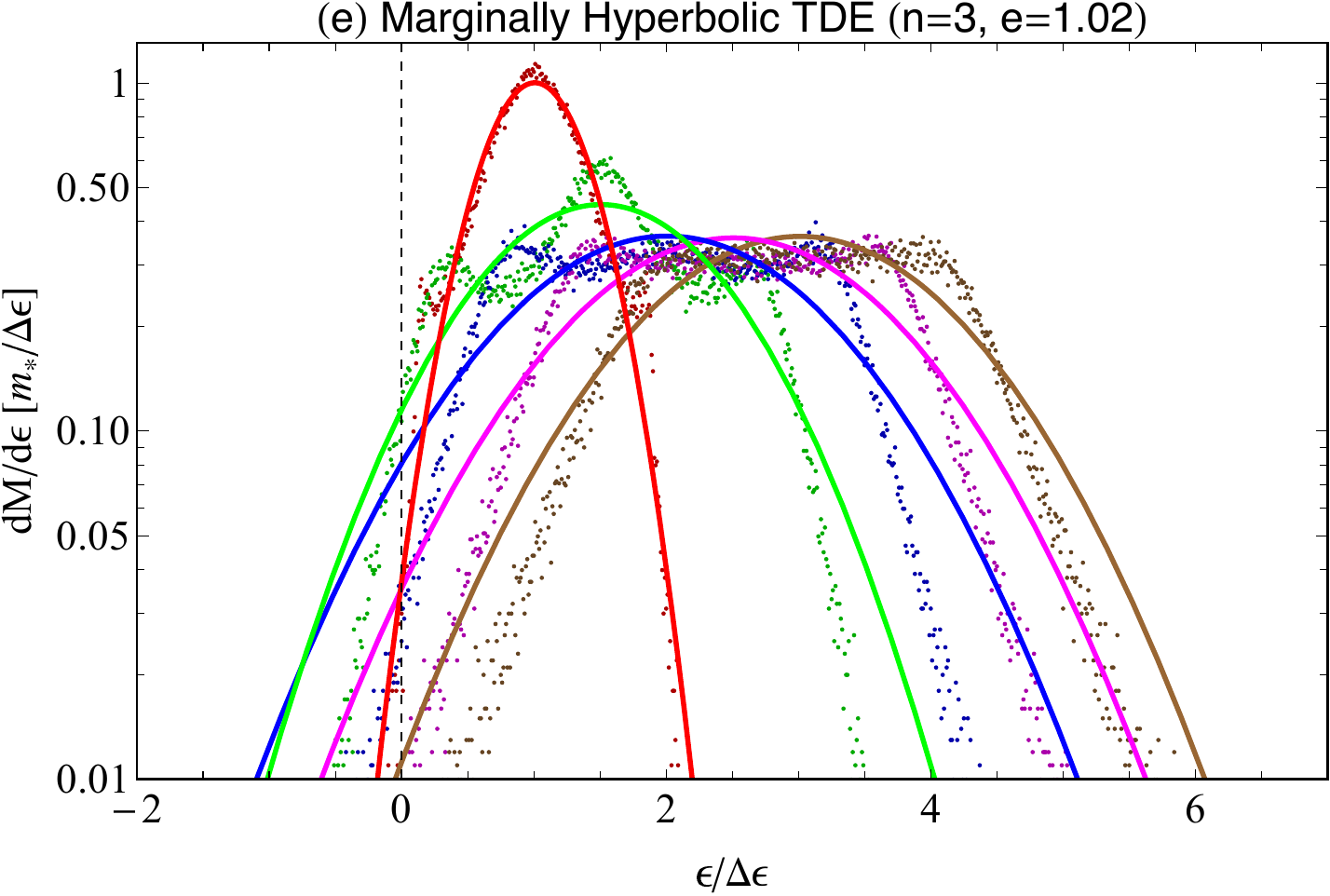}

\caption{
The same format as Figure~\ref{fig:dmde1}, but for the case of $n=3$
}
\label{fig:dmde2}
\end{figure}

Figure~\ref{fig:dmde2} is the same format as Figure~\ref{fig:dmde1} but for the $n=3$ case. 
For the parabolic TDEs, a half of stellar debris is bound, whereas other half is unbound to 
the SMBH. The differential mass distributions for $\beta=2$, $\beta=2.5$, and $\beta=3$ 
are similar as those of $n=1.5$, while the distribution of $\beta=1$ is steeper than that of 
$n=1.5$ because the central density of $n=3$ polytrope is one order magnitude 
higher than $n=1.5$ case, leading to the partial disruption of the star. This tendency appears 
to be independent of the orbital eccentricity.

\begin{figure}[!ht]
\center
\includegraphics[width=12cm]{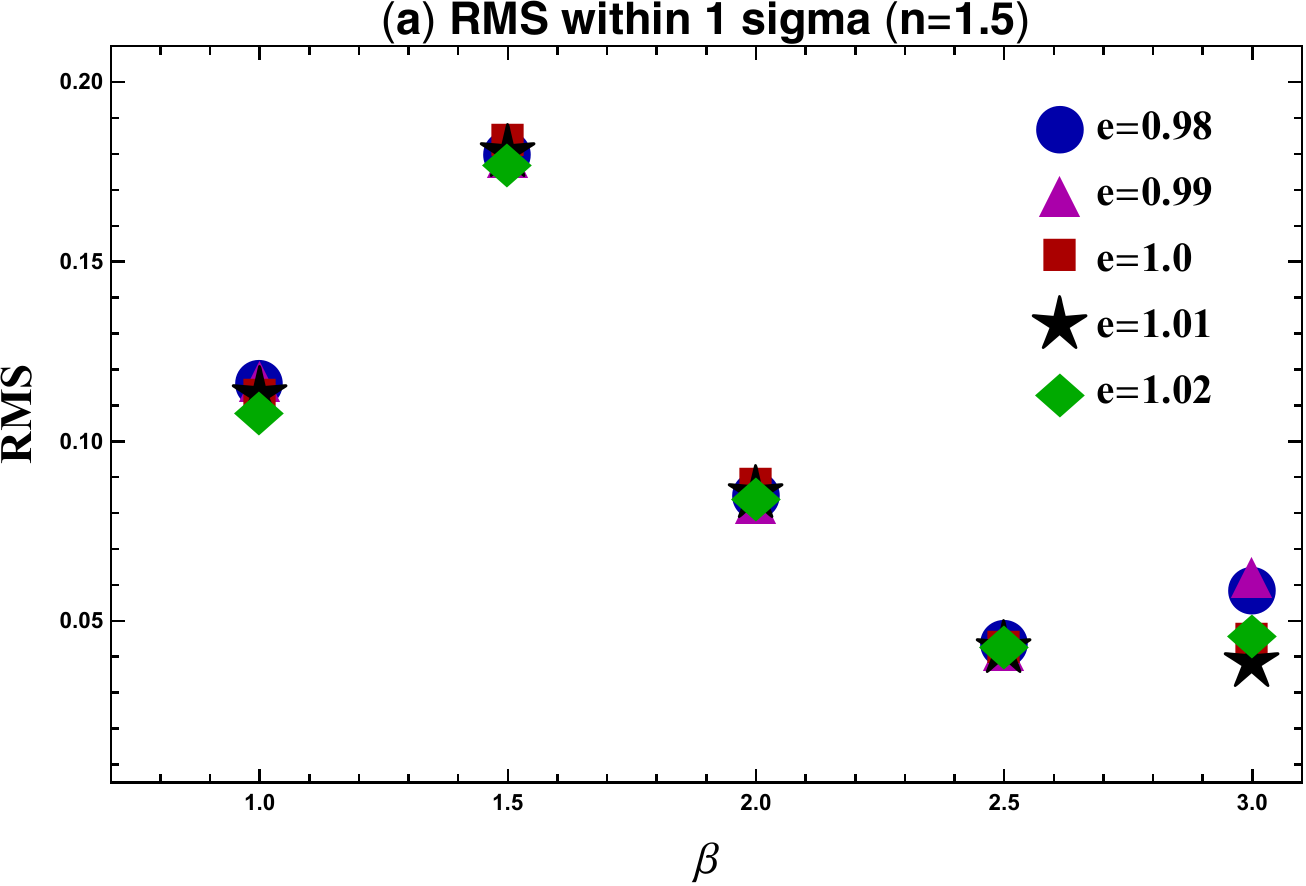}\\
\includegraphics[width=12cm]{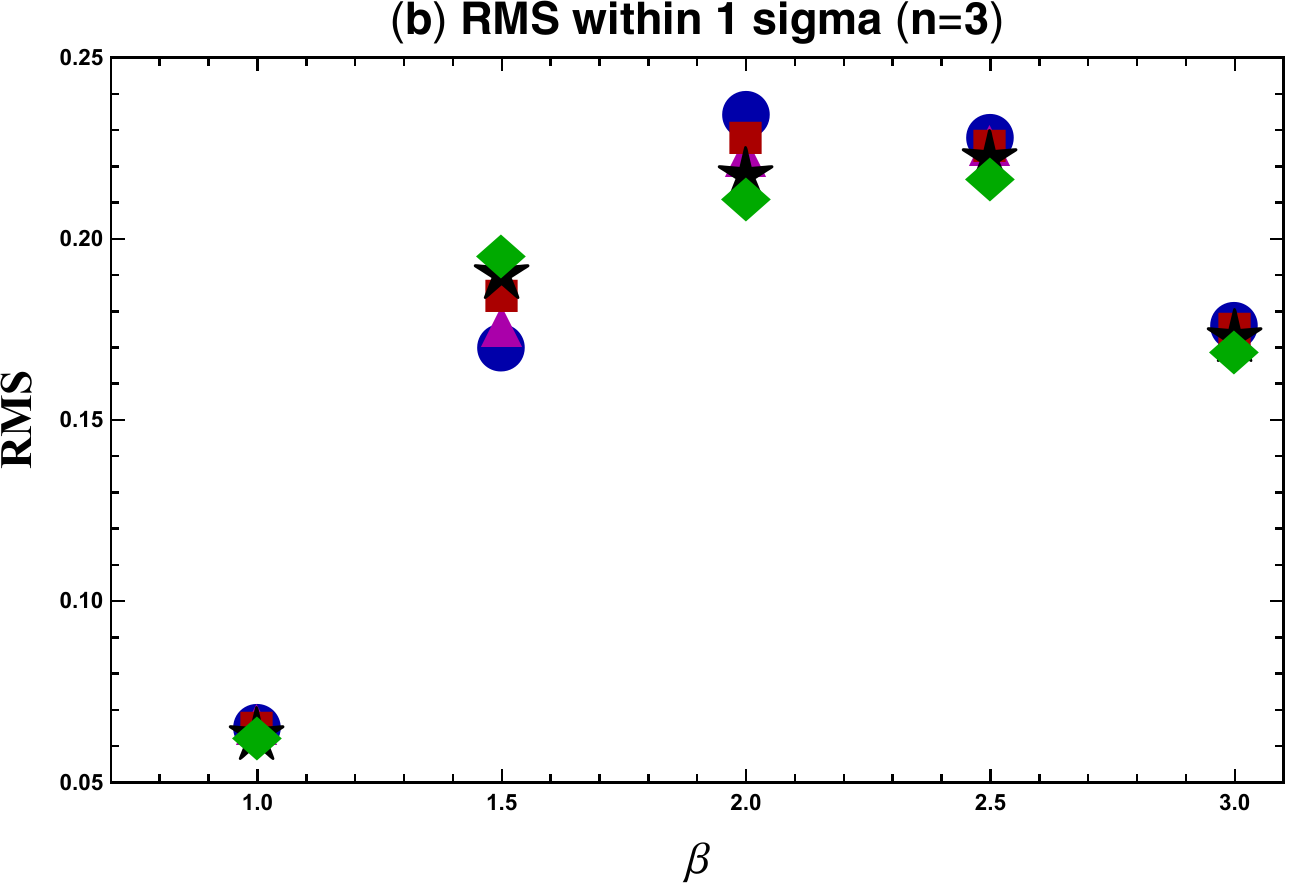}
\caption{
The root-mean square (RMS) between simulated data points and the Gaussian-fitted curves. 
Panels (a) and (b) panel show a $n=1.5$ and $n=3$ polytrope cases, respectively. 
The definition of RMS is shown in the second paragraph of Section 3.1.
}
\label{fig:rms1}
\end{figure}

\begin{figure}[!ht]

\includegraphics[width=8cm]{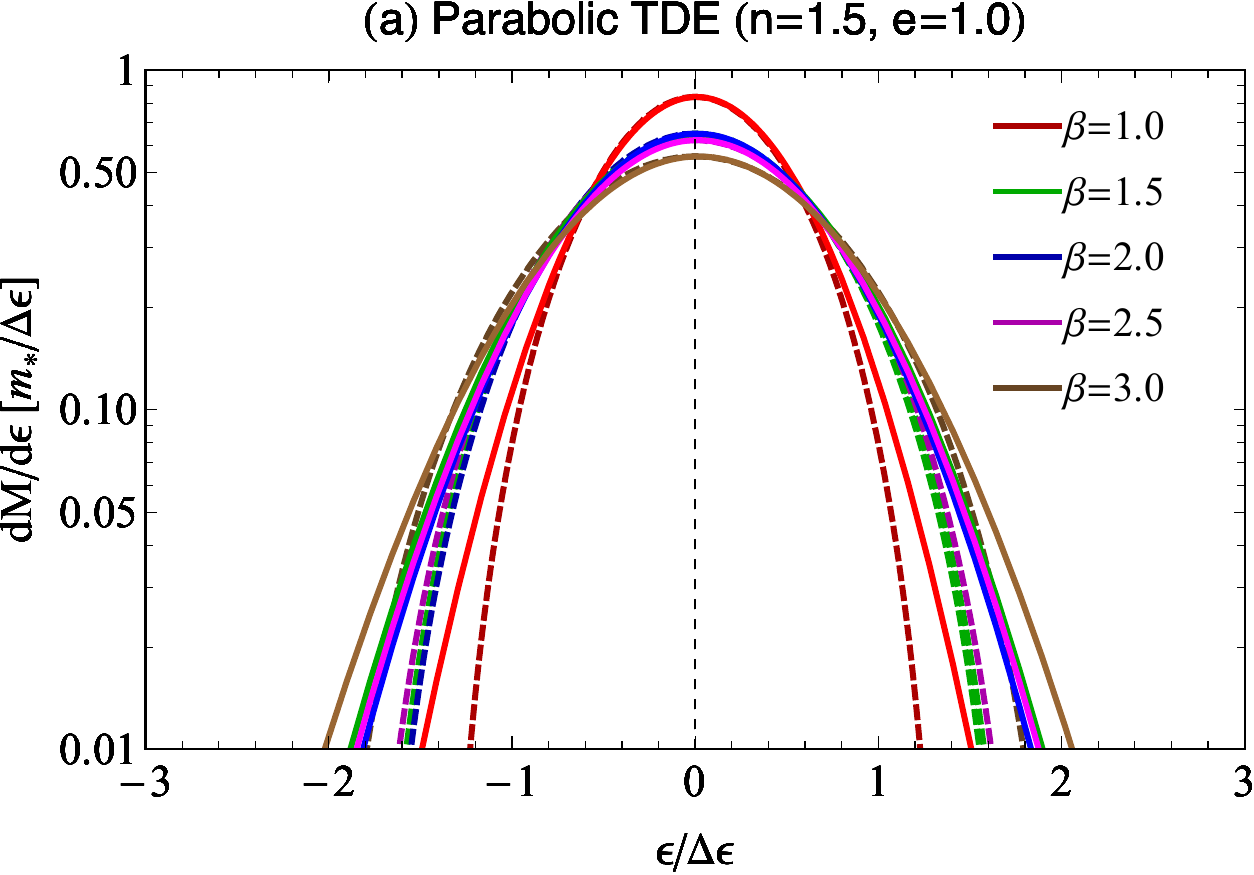}
\includegraphics[width=8cm]{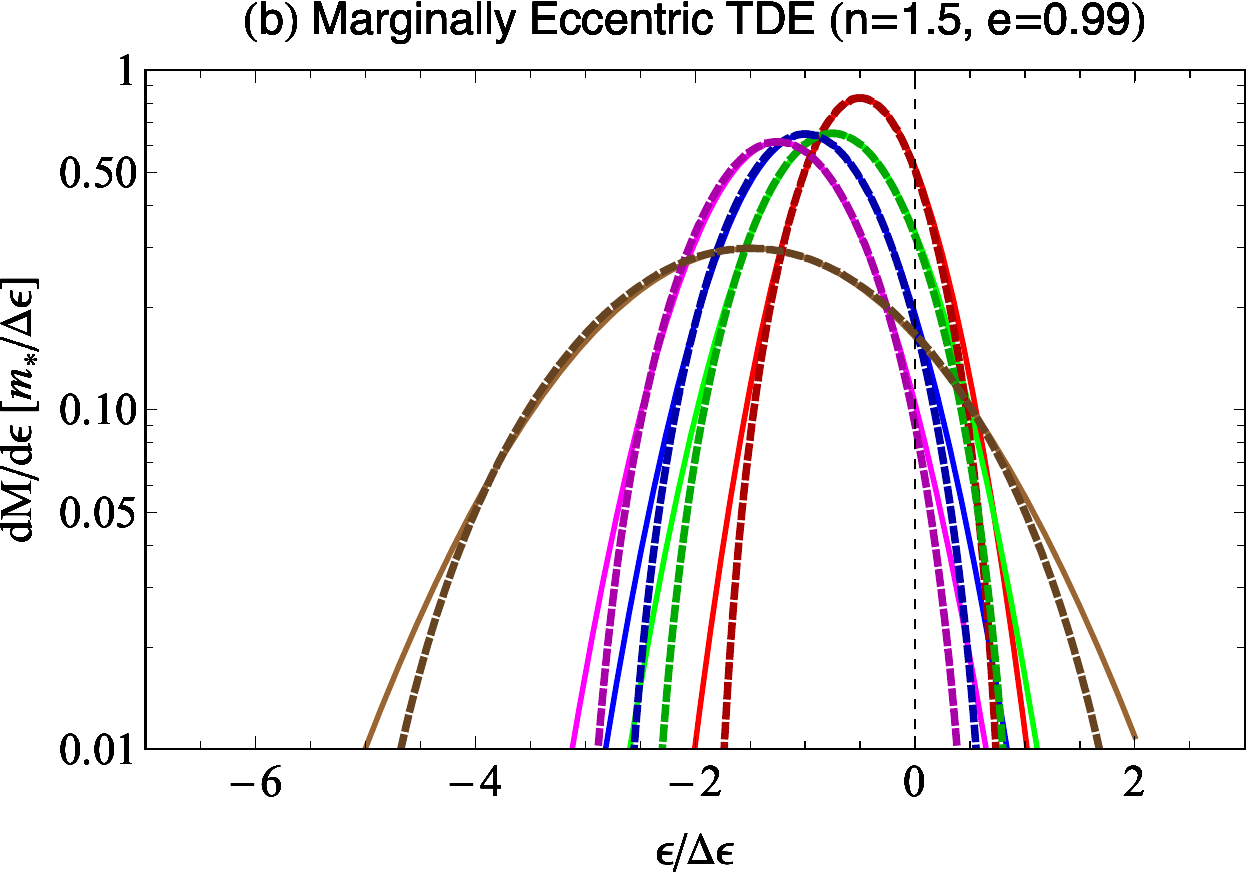}
\\
\includegraphics[width=8cm]{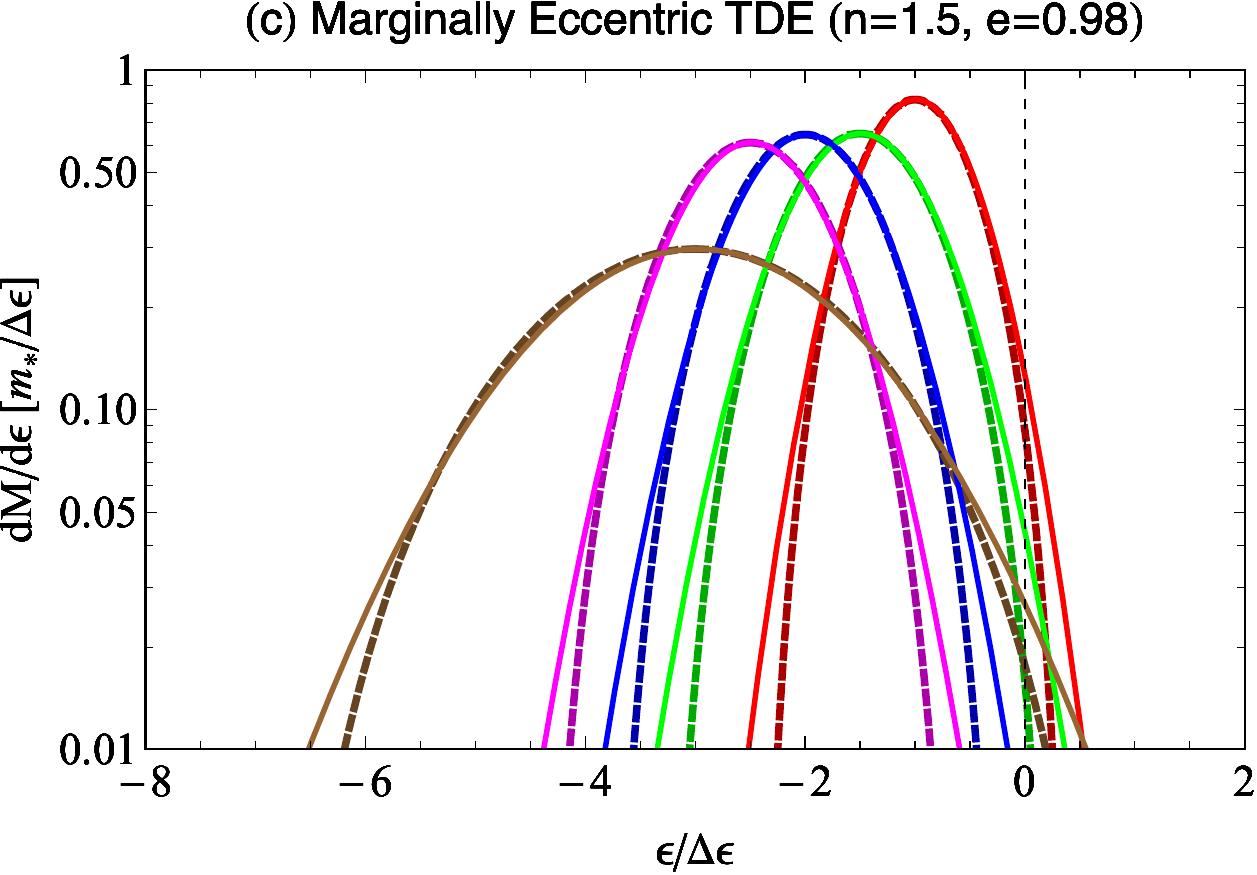}
\includegraphics[width=8cm]{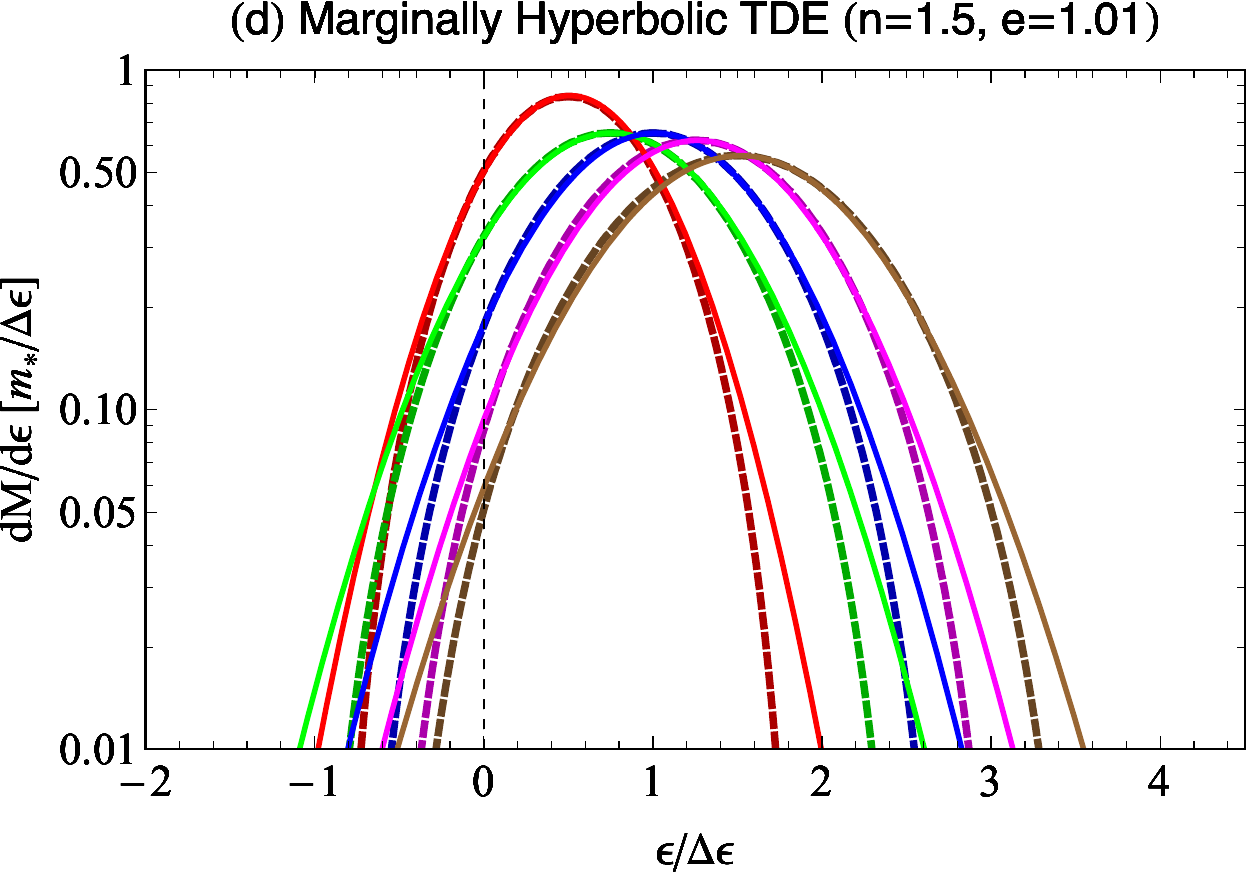}
\\
\includegraphics[width=8cm]{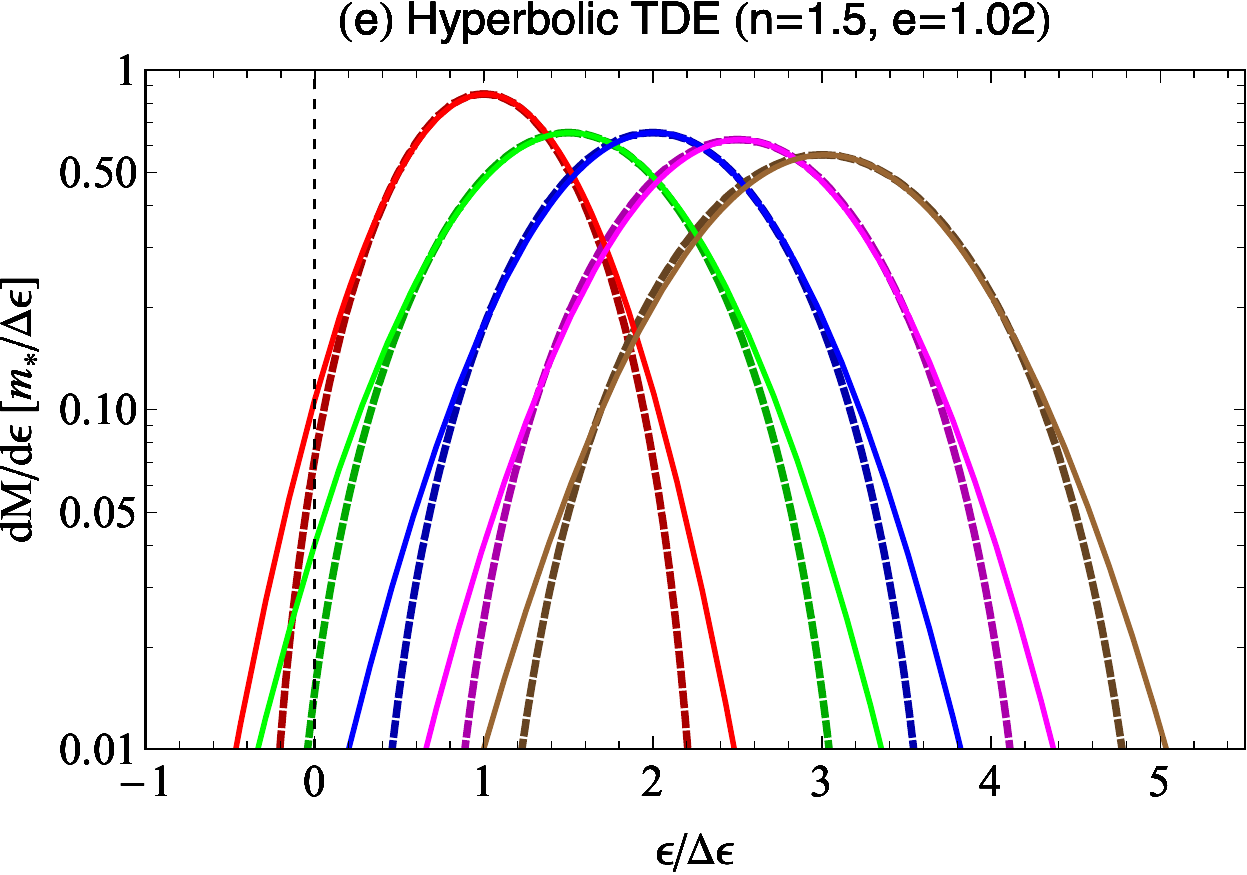}

\caption{
Comparison between the Gaussian fits to the simulation data 
and the corresponding semi-analytical distributions for an 
$n=1.5$ polytrope. Each panel shows a different orbital eccentricity.
The solid and dashed lines show the Gaussian-fits and the 
semi-analytical distributions, respectively.}
The color format of each line is 
the same as that of Figure~\ref{fig:dmde1}.
\label{fig:dmde3}
\end{figure}

\begin{figure}[!ht]

\includegraphics[width=8cm]{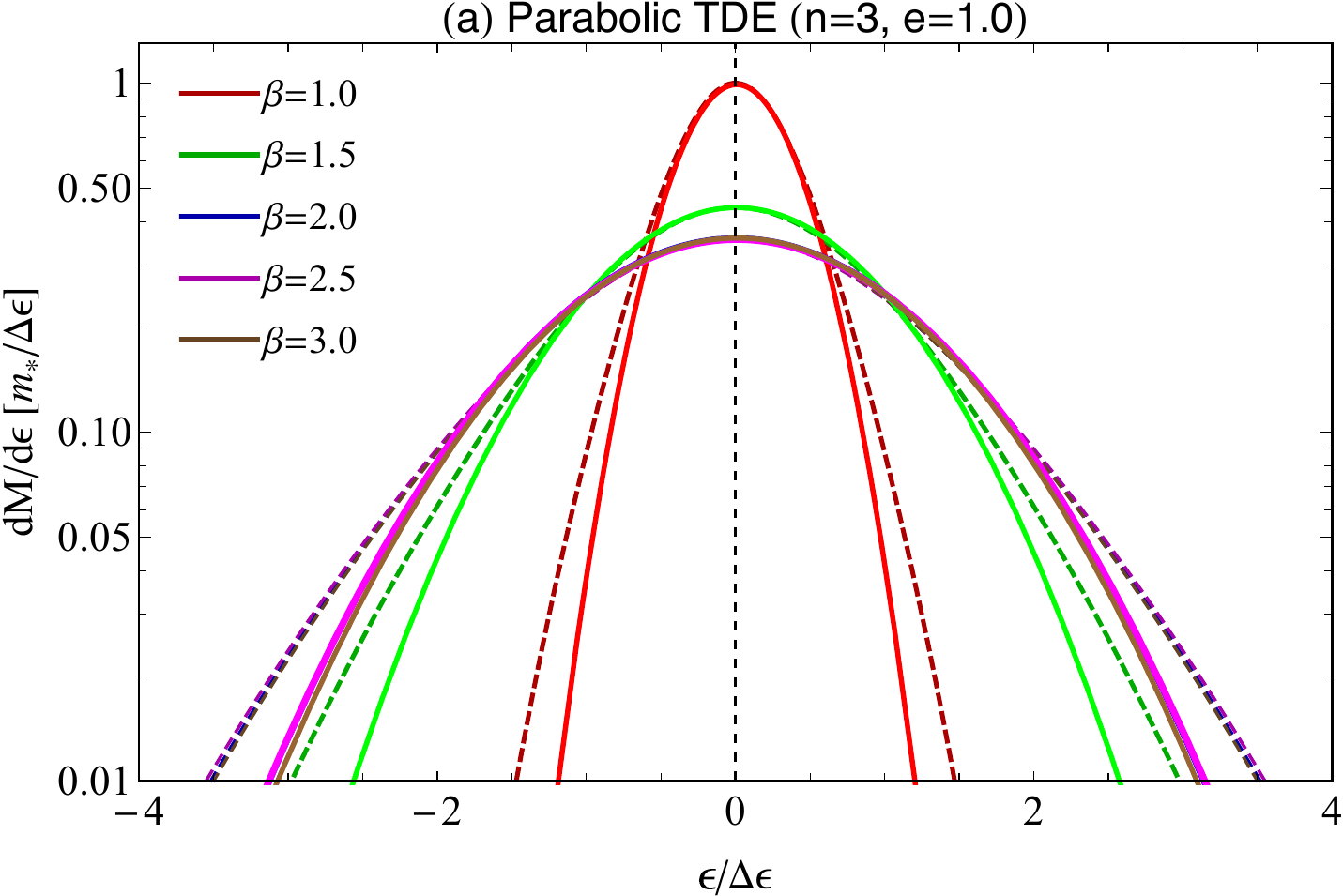}
\includegraphics[width=8cm]{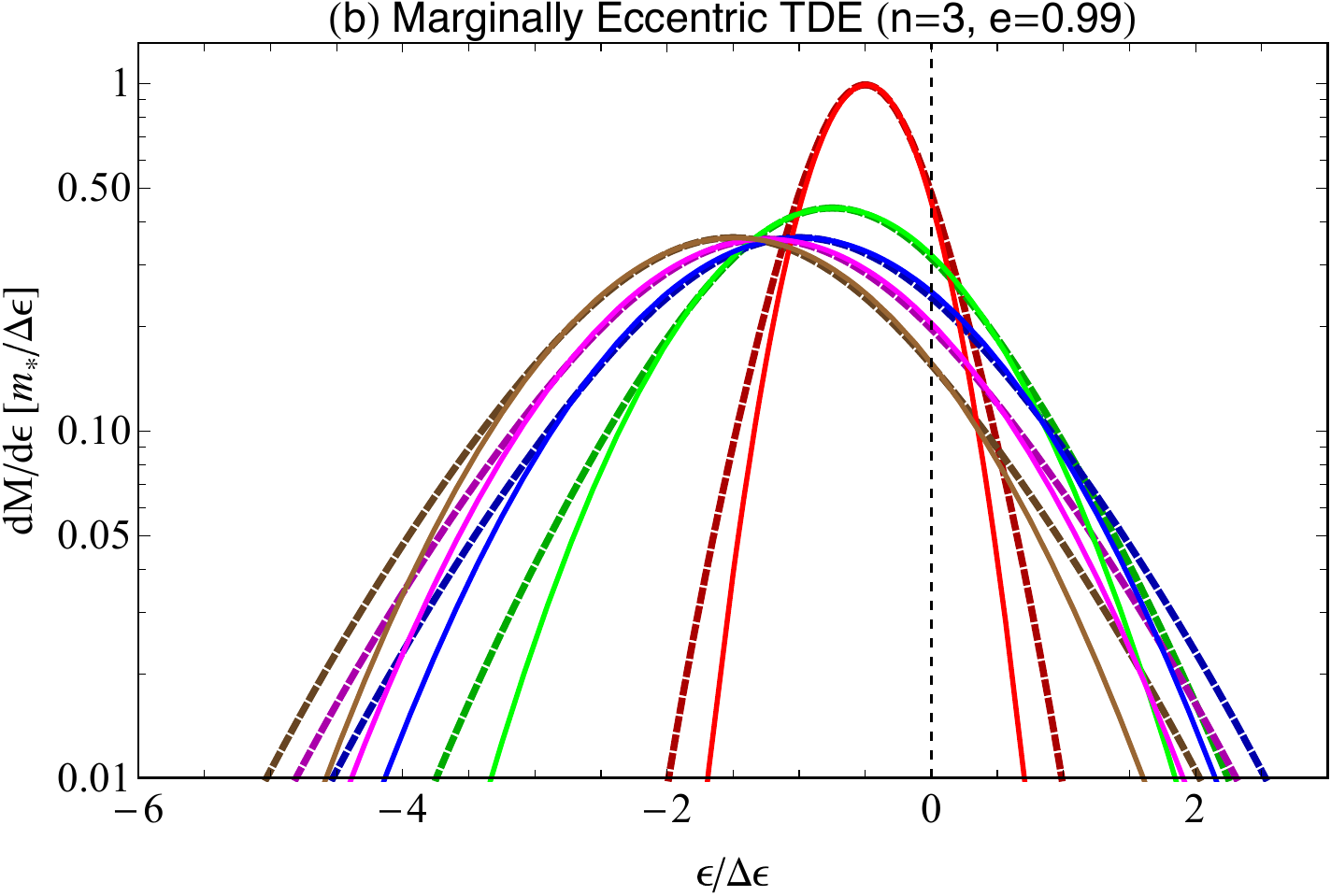}
\\
\includegraphics[width=8cm]{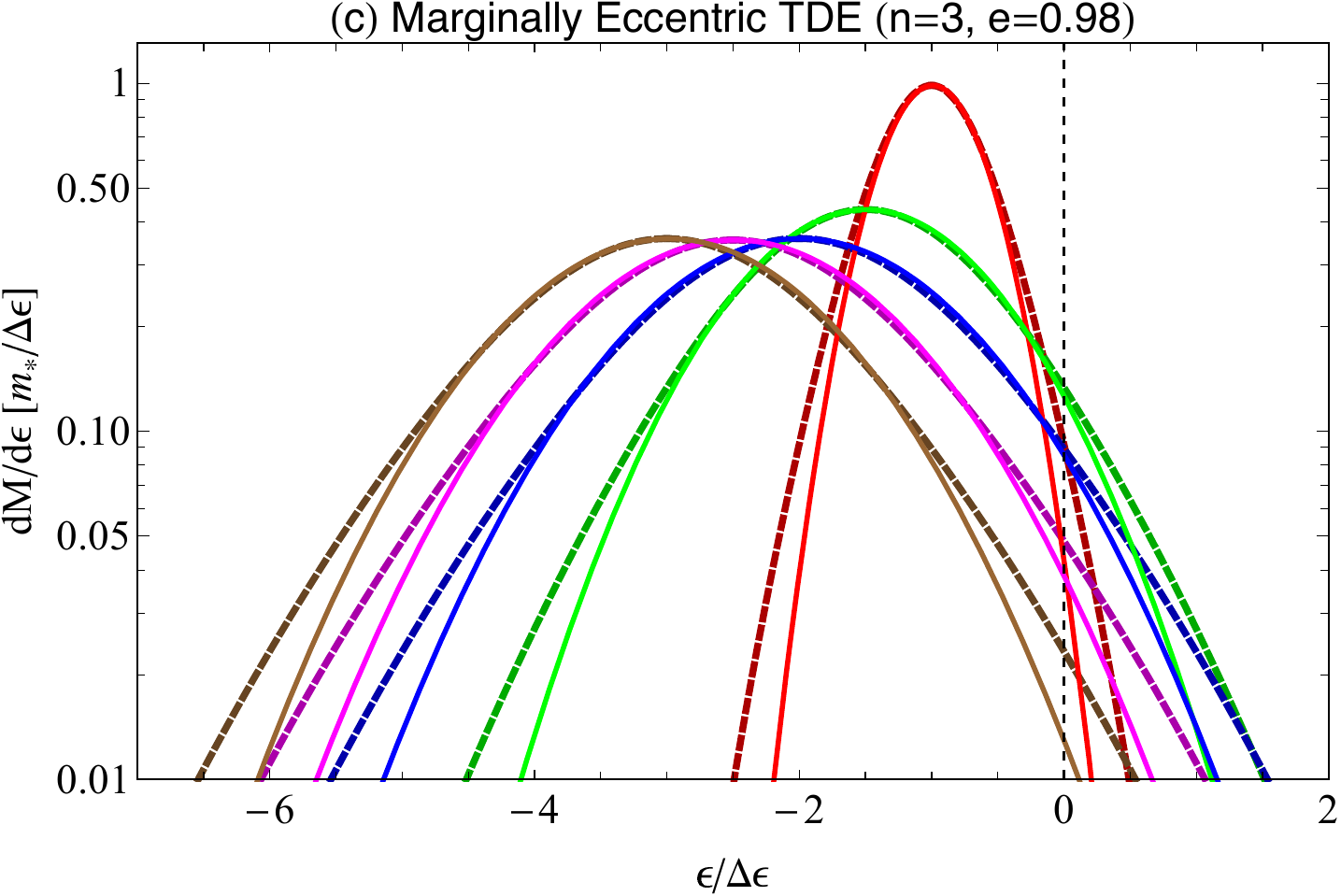}
\includegraphics[width=8cm]{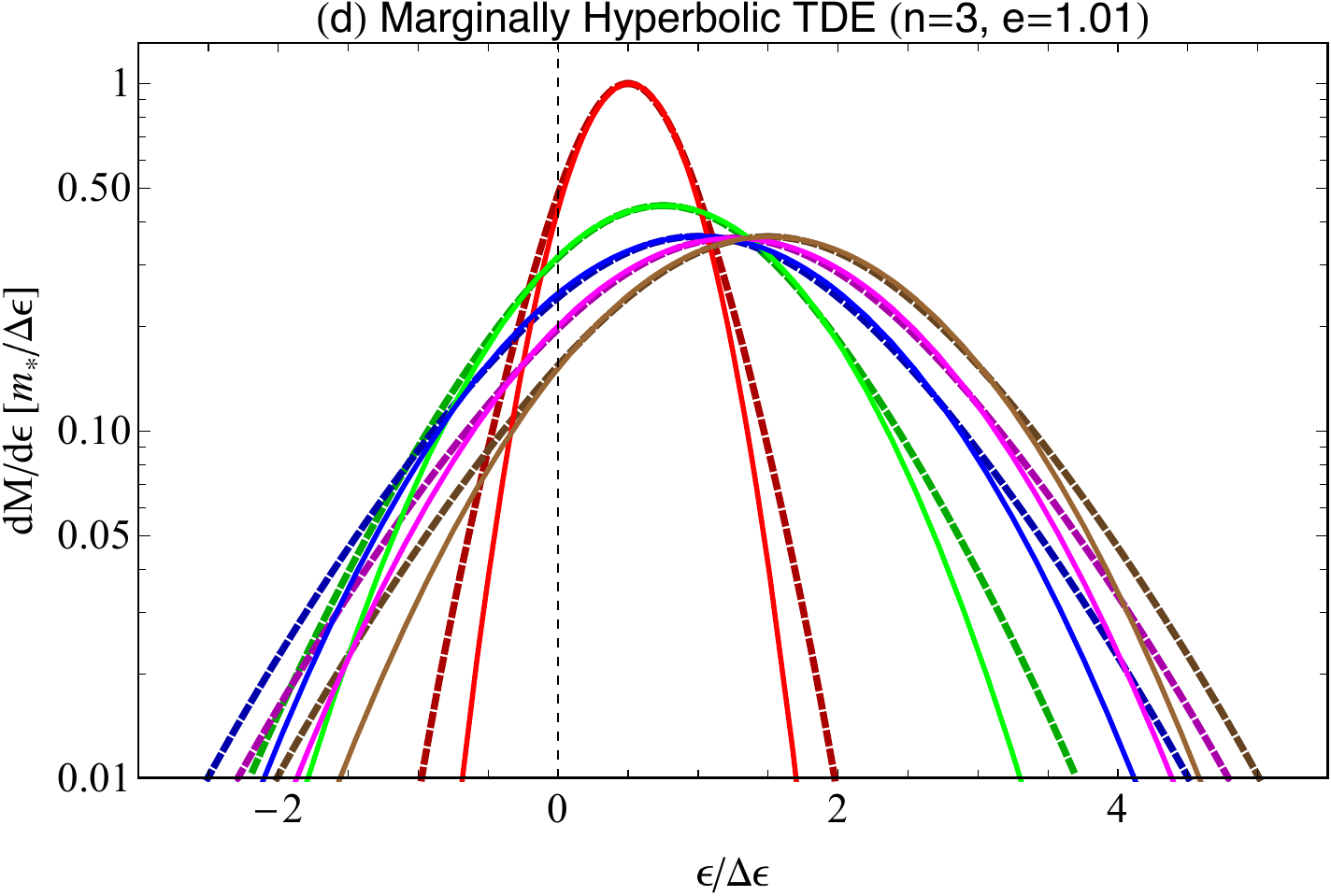}
\\
\includegraphics[width=8cm]{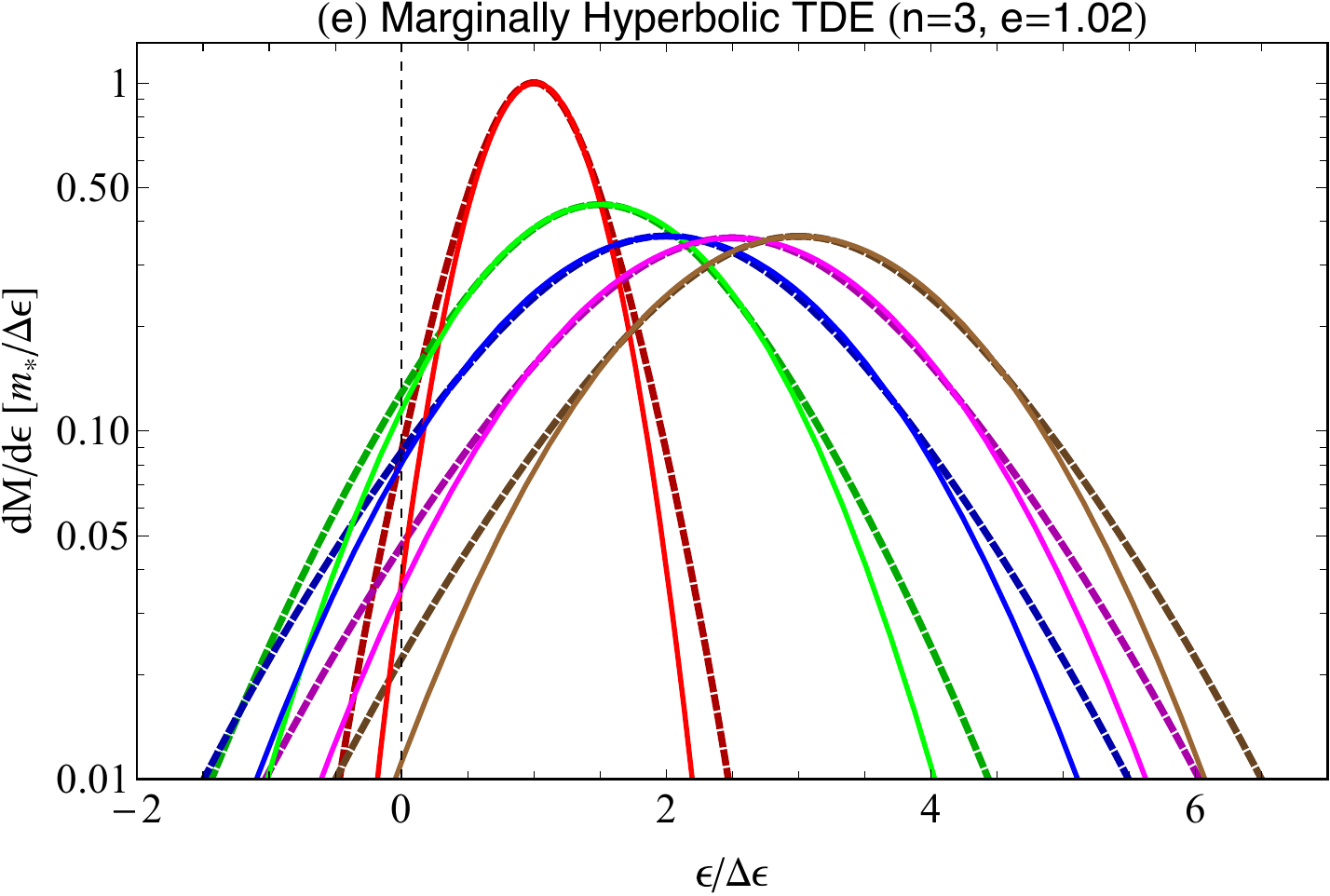}

\caption{
The same format as Figure~\ref{fig:dmde2}, but for the case of $n=3$
}
\label{fig:dmde4}
\end{figure}

Figure~\ref{fig:rms1} depicts the penetration factor dependence of the RMS between 
the simulated data points and the Gaussian-fitted curves. 
Panels (a) and (b) represent the n=1.5 and the n=3 cases, respectively. 
We see from panel (a) that the maximum RMS value is $\sim18\%$ at $\beta=1.5$, 
whereas the maximum RMS value is less than $5\%$ at $\beta=2.5$. 
As shown in panel (b), the maximum RMS value is $\sim24\%$ at $\beta=2.0$, 
whereas the maximum RMS value is less than $7\%$ at $\beta=1.0$. 
We also find that the RMS does not depend on the orbital eccentricity for the $n=1.5$ and $n=3$ cases. 
Figures~\ref{fig:dmde3} and \ref{fig:dmde4} show the comparison between 
the Gaussian-fitted curves with the semi-analytic solutions with $n=1.5$ and 
$n=3$, which is given by equation~(\ref{eq:dmdt2}), respectively. We find from 
the figures that the simulated curves are good agreement with the semi-analytical 
solutions.


As another evaluation of the Gaussian fitting model, 
we compare the mass of the bound part of the stellar debris for both the simulated data and the corresponding Gaussian fitted model.
Each bound mass is calculated by
\begin{eqnarray}
m_{\rm b,Fit}&=&\int^{0}_{\epsilon^{}_{\rm d,b}} \left(\frac{dM}{d\epsilon}\right)_{\rm Fit}\,d\epsilon \nonumber \\
m_{\rm b,SPH}&=&\int^{0}_{\epsilon^{}_{\rm d,b}} \left(\frac{dM}{d\epsilon}\right)_{\rm SPH}\,d\epsilon, \nonumber
\end{eqnarray}
where $\epsilon_{\rm d,b}=-(GM_{\rm bh}/a_{\rm d}+{GM_{\rm bh}}/a)/2$ 
and note that $\epsilon_{\rm d,b}\rightarrow-GM_{\rm bh}/a_{\rm d}$ because of $a\rightarrow\infty$ for parabolic TDEs.
We can then evaluate the error rate as $\Delta{m}=|m_{\rm b,SPH}-m_{\rm b,Fit}|/m_{\rm b,SPH}$. 
For all the models of the marginally eccentric and parabolic TDEs, 
the error rate is distributed over $0.01\lesssim\Delta{m}\lesssim0.04$. 
On the other hand, the mass of the bound part of the stellar debris is so tiny for the marginally hyperbolic TDEs 
that we instead evaluate the unbound mass of the stellar debris. For the simulated data and the corresponding fitted data, 
each unbound mass is calculated by
\begin{eqnarray}
m_{\rm ub,Fit}&=&\int^{\epsilon_{\rm d,ub}}_0 \left(\frac{dM}{d\epsilon}\right)_{\rm Fit}\,d\epsilon \nonumber \\
m_{\rm ub,SPH}&=&\int^{\epsilon_{\rm d,ub}}_0 \left(\frac{dM}{d\epsilon}\right)_{\rm SPH}\,d\epsilon, \nonumber
\end{eqnarray}
where $\epsilon_{\rm d,ub}=(GM_{\rm bh}/a_{\rm d}+{GM_{\rm bh}}/a)/2$. For all the models of the marginally hyperbolic TDEs, the error rate is distributed over $0.01\lesssim\Delta{m}\lesssim0.09$, where $\Delta{m}=|m_{\rm ub,SPH}-m_{\rm ub,Fit}|/m_{\rm ub,SPH}$. We note that the 
largest error rate ($\Delta{m}\sim0.09$) is seen in the case of $n=1.5$, $\beta=2.5$, and $e=1.02$, and the error rate ranges $0.01\lesssim{\Delta{m}}\lesssim0.04$ in the other cases. While the deviation between the simulated data and the corresponding fitted data is less than $5\%$ for all the models of marginally eccentric and parabolic TDEs, it is less than $10\%$ for the cases of marginally hyperbolic TDEs.


%
%

%
%

%
%

%
%

%
%


%
\clearpage
\subsection{Evaluation of tidal spread energy index: k}
\label{subsec:k}
%

Assuming that the standard deviation of the Gaussian-fitted curve, $\Delta\mathcal{E}_{\rm sim}$, 
corresponds to the analytical spread energy $\Delta\mathcal{E}$, we can evaluate the power-law index 
of spread in tidal energy  from equation~(\ref{eq:delebeta}) by
\begin{eqnarray}
k=\frac{\log(\Delta\mathcal{E}_{\rm sim}/\Delta\epsilon)}{\log\beta}.
\label{eq:k-eva}
\end{eqnarray}

\begin{figure}[!ht]
\center
\includegraphics[width=8cm]{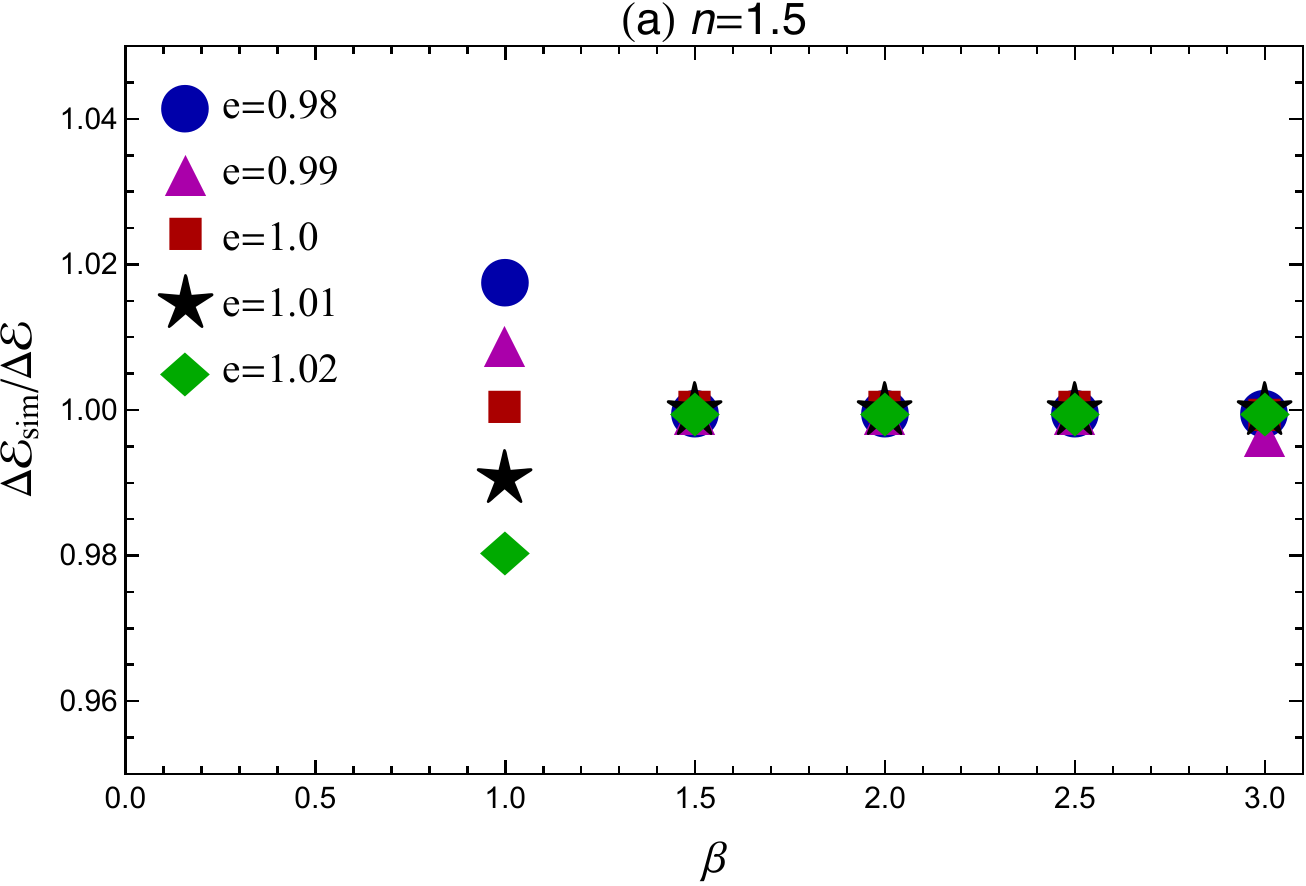}
\includegraphics[width=8cm]{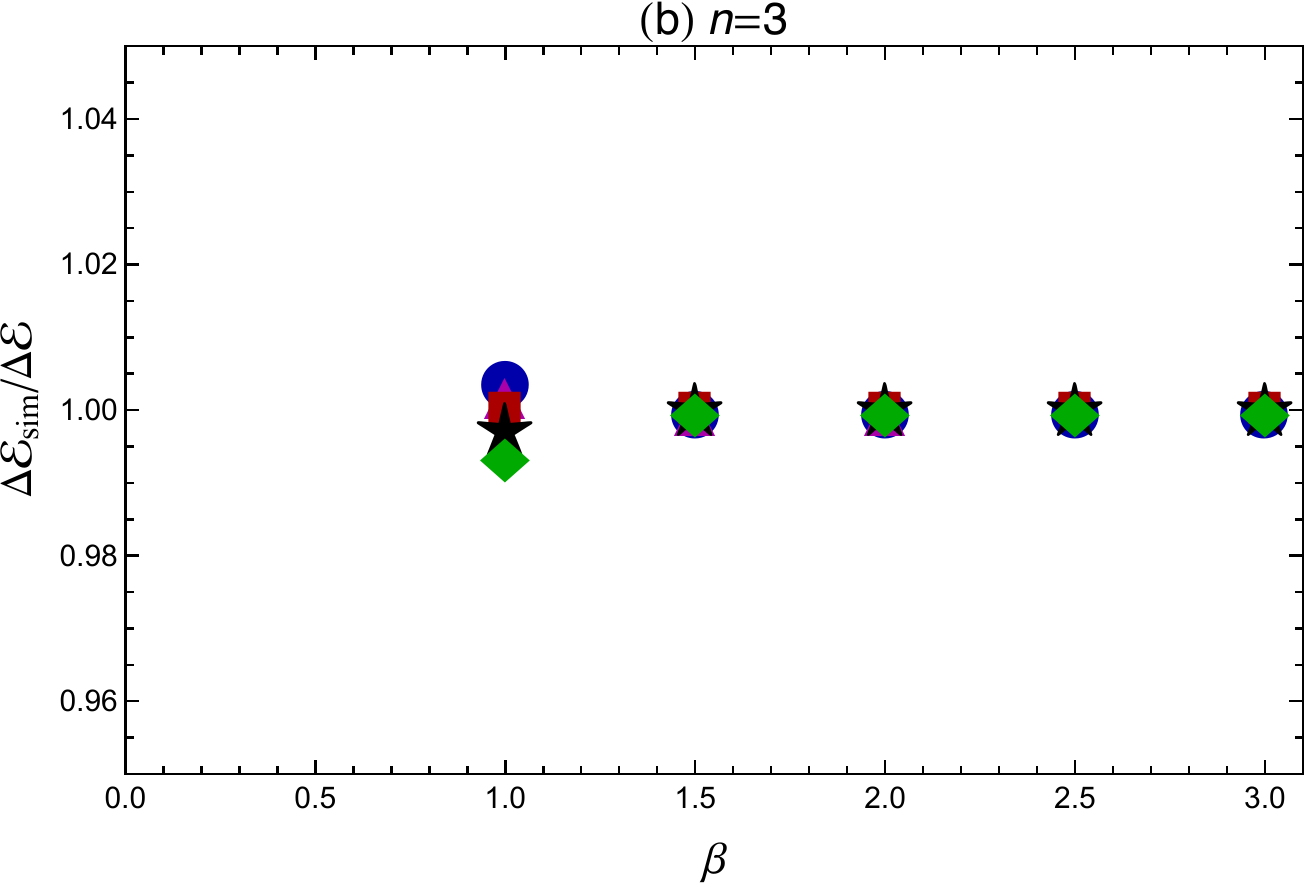}\\
\includegraphics[width=8cm]{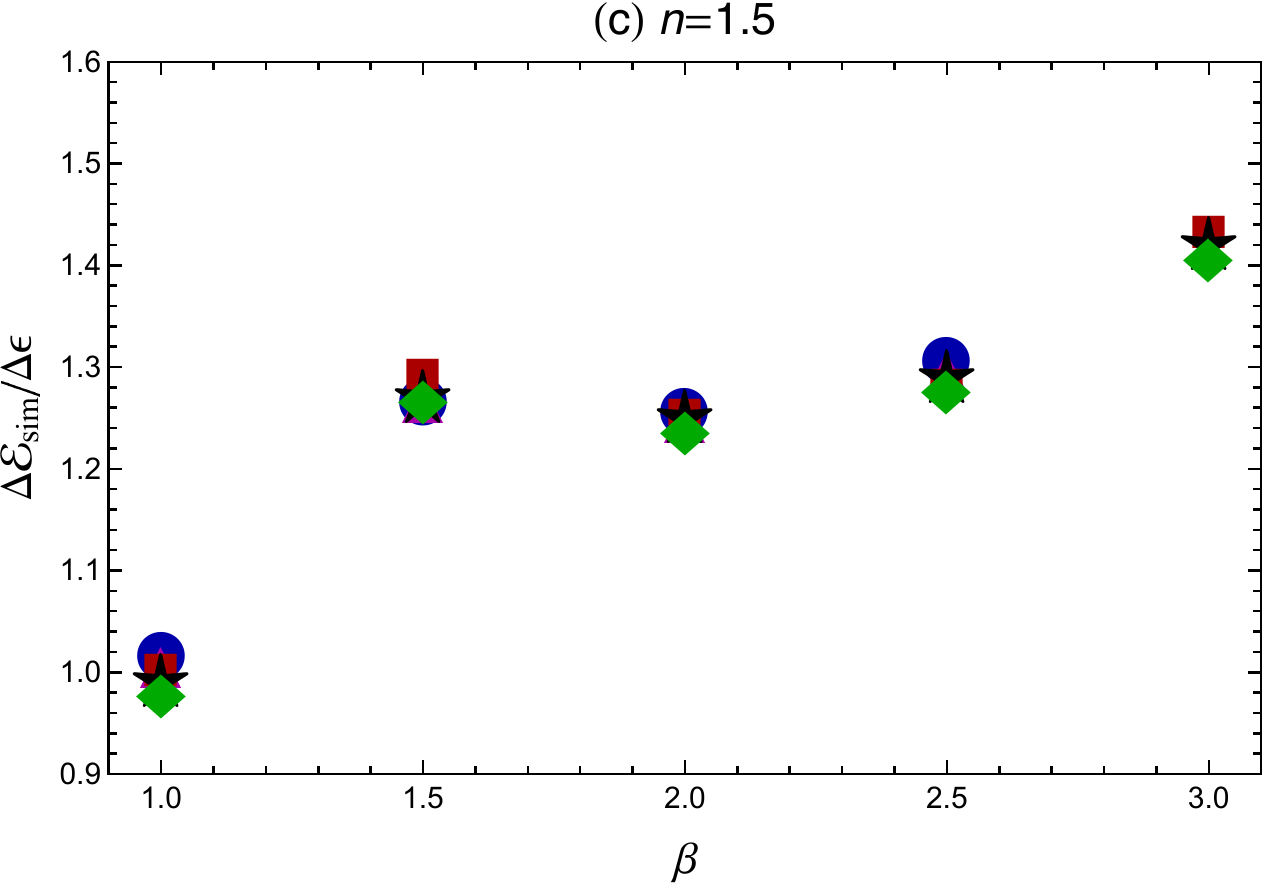}
\includegraphics[width=8cm]{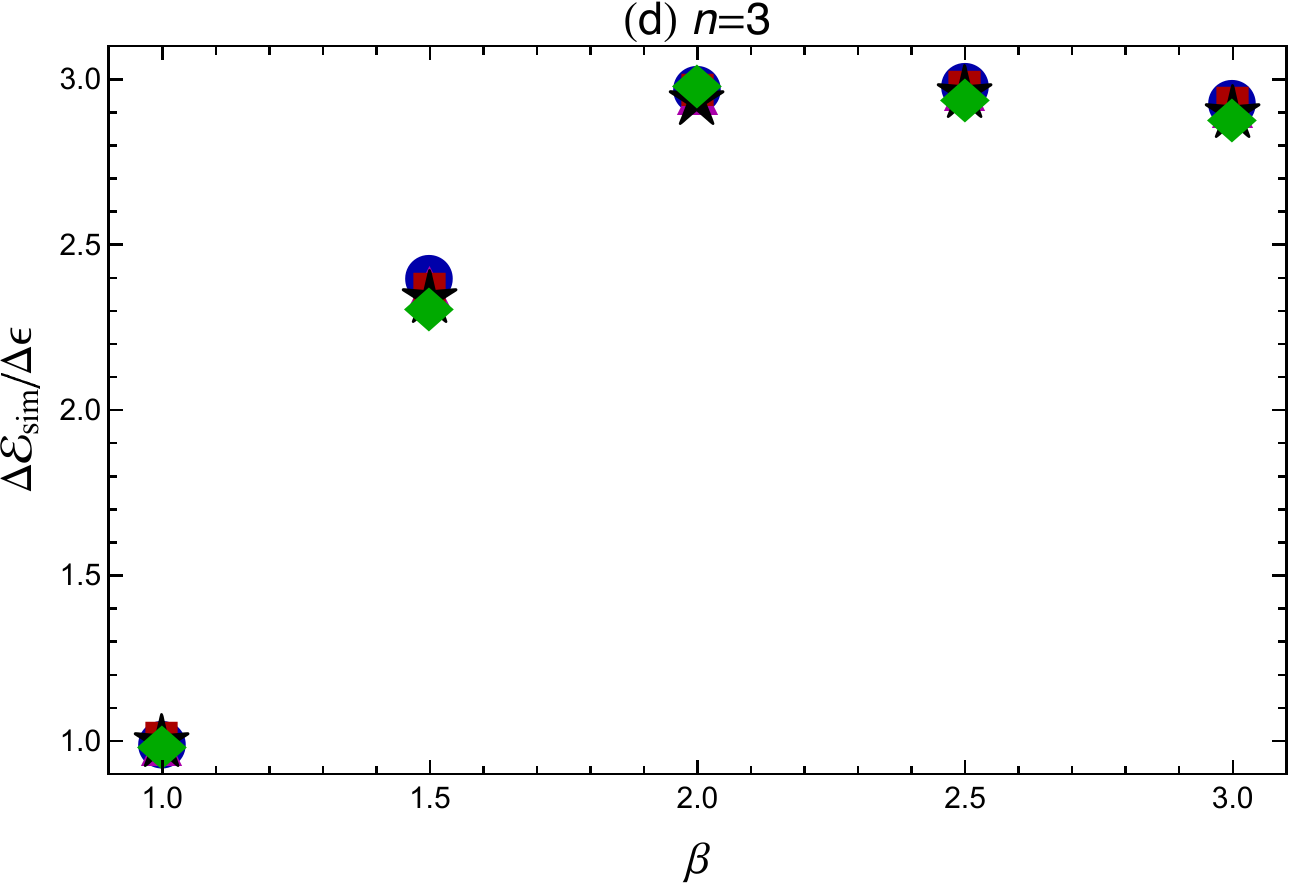}
\caption{
Dependence of the simulated spread energy on the penetration factor. 
Panels (a) and (b) show $\beta$-dependence of $\Delta\mathcal{E}_{\rm sim}/\Delta\mathcal{E}$ 
for the $n=1.5$ and $n=3$ cases, respectively. Panels (c) and (d) depict $\beta$-dependence 
of $\Delta\mathcal{E}_{\rm sim}/\Delta\epsilon$ for the $n=1.5$ and $n=3$ cases, respectively.
Note that the relation between respective normalization is given by 
$\Delta\mathcal{E}=\Delta\epsilon\,\beta^k$ (see also equation~\ref{eq:delebeta}).  
In each panel, the blue circles, magenta triangles, red squares, black stars, and green 
rhombuses represent results for $e=0.98$, $0.99$, $1.0$, $1.01$ and $1.02$, respectively.
}
\label{fig:debeta1}
\end{figure}

Figure~\ref{fig:debeta1} shows the dependence of the simulated spread energy on the penetration factor. 
Panels (a) and (b) show $\beta$-dependence of $\Delta\mathcal{E}_{\rm sim}/\Delta\mathcal{E}$ 
for the $n=1.5$ and $n=3$ cases, respectively. Panels (c) and (d) depict $\beta$-dependence 
of $\Delta\mathcal{E}_{\rm sim}/\Delta\epsilon$ for the $n=1.5$ and $n=3$ cases, respectively.
In each panel, the blue circles, magenta triangles, red squares, black stars and green rhombuses 
represent results of $e=0.98$, $0.99$, $1.0$, $1.01$ and $1.02$, respectively. 
We find from panels (a) and (b) that the simulated spread energies are in good agreement with 
the analytical values for the $n=1.5$ and $n=3$ cases, although the error of $2\%$ is obtained 
for $\beta=1$. In addition, $\Delta\mathcal{E}_{\rm sim}/\Delta\epsilon$ slightly increases for $\beta=1$ 
as the orbital eccentricity decreases. We also find from panels (c) and (d) that the simulated 
spread energy, overall, increases beyond $\Delta\epsilon$ as the penetration factor increases.

 $k$ with different penetration factors for the $n=1.5$ and $n=3$ cases, respectively. Panels (a) to (d) show results for $\beta=1.5, \beta=2, \beta=2.5$, and $\beta=3$, respectively. These two figures show that the value of $k$ is distributed between $0.2$ and $0.9$ for $n=1.5$, while the value of $k$ takes between $0.95$ and $2.2$ for the $n=3$ case. The detailed values of $k$ can be seen at the sixth column of Tables~\ref{tbl:1} and \ref{tbl:2}. We also find that the value of $k$ is larger than 2 only for the $\beta=1.5$ and $n=3$ case.

\begin{figure}[!ht]
\center
\includegraphics[width=8cm]{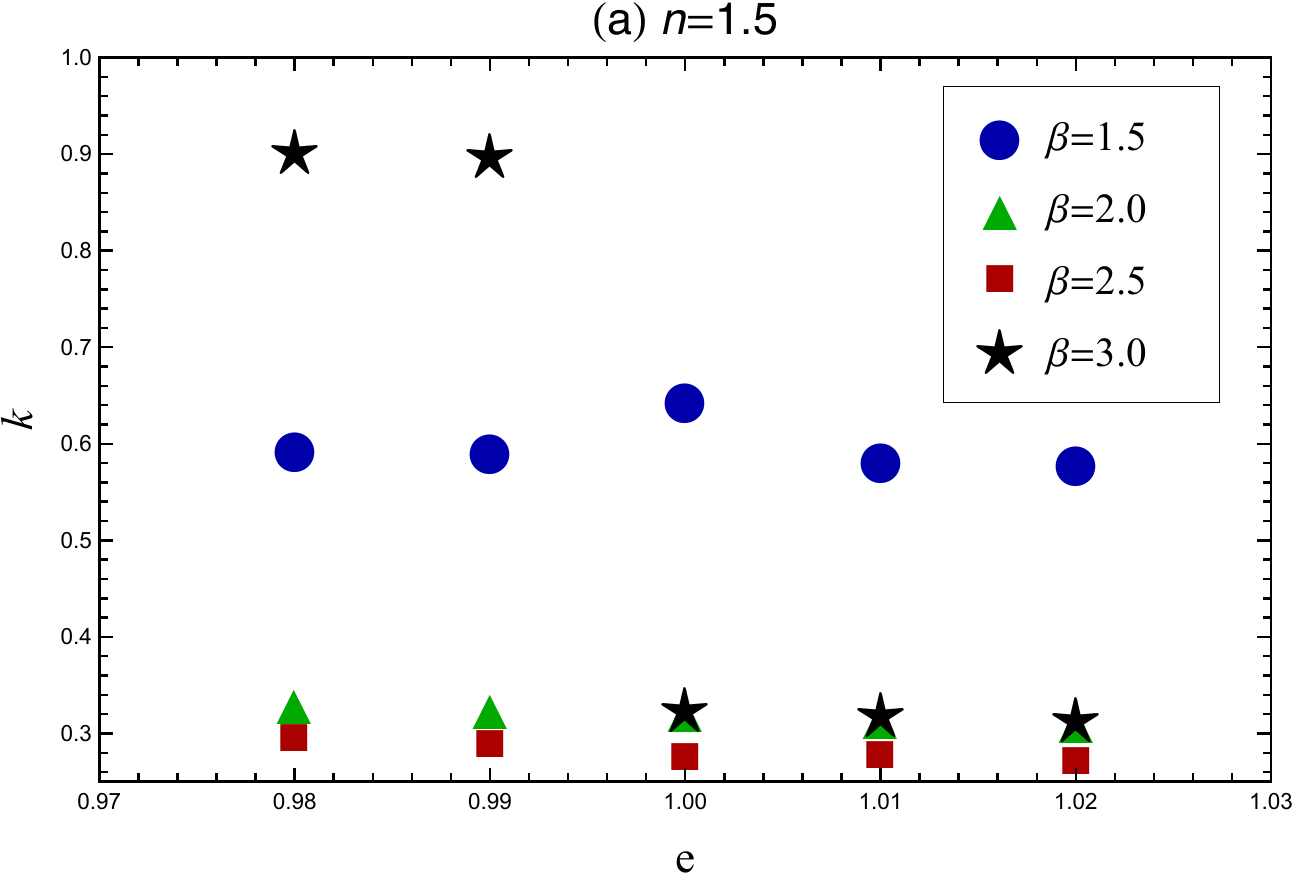}
\includegraphics[width=8cm]{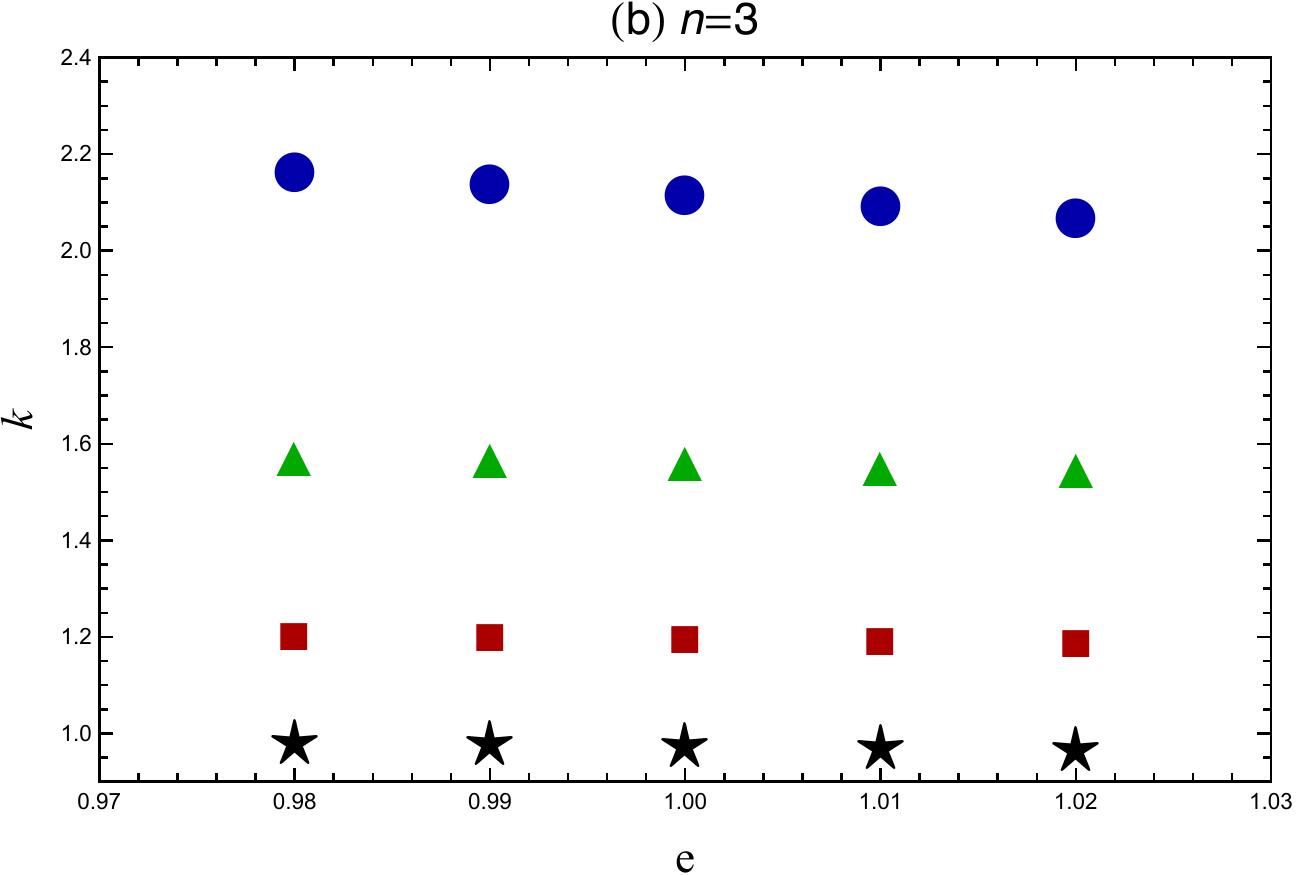}\\
\caption{
Orbital eccentricity dependence of the tidal spread energy index. 
Panels (a) and (b) represent it for the $n=1.5$ and $n=3$ cases, respectively.
The blue circles, green triangles, red squares, and black stars denote 
the value of $k$ for $\beta=1.5$, $\beta=2$, $\beta=2.5$, and $\beta=3$, respectively.
}
\label{fig:k-e}
\end{figure}

Figure~\ref{fig:k-e} depicts the dependence of the tidal spread energy index on the orbital eccentricity with different penetration factors: $\beta=1.5, \beta=2, \beta=2.5$, and $\beta=3$. Each panel shows a different polytrope. These two figures show that the value of $k$ is distributed between $0.2$ and $0.9$ for $n=1.5$, while the value of $k$ takes between $0.95$ and $2.2$ for the $n=3$ case. The detailed values of $k$ can be seen at the sixth column of Tables~\ref{tbl:1} and \ref{tbl:2}. It is found from panel (b) that the value of $k$ is higher as $\beta$ is smaller for a $n=3$ polytrope, and 
this tendency is independent of the orbital eccentricity. We also find that the value of $k$ can be larger than 2 only for the $\beta=1.5$ and $n=3$ case.

%
%

%
%

%
\clearpage
\subsection{Mass fallback rates}
\label{sec:dmdt}
%

In this section, we evaluate the mass fallback rate of 
each model by using equation (\ref{eq:dmdt}), where 
$dM/d\epsilon$ is given by the Gaussian fitted curves.

\begin{figure}[!ht]
\center
\includegraphics[width=8cm]{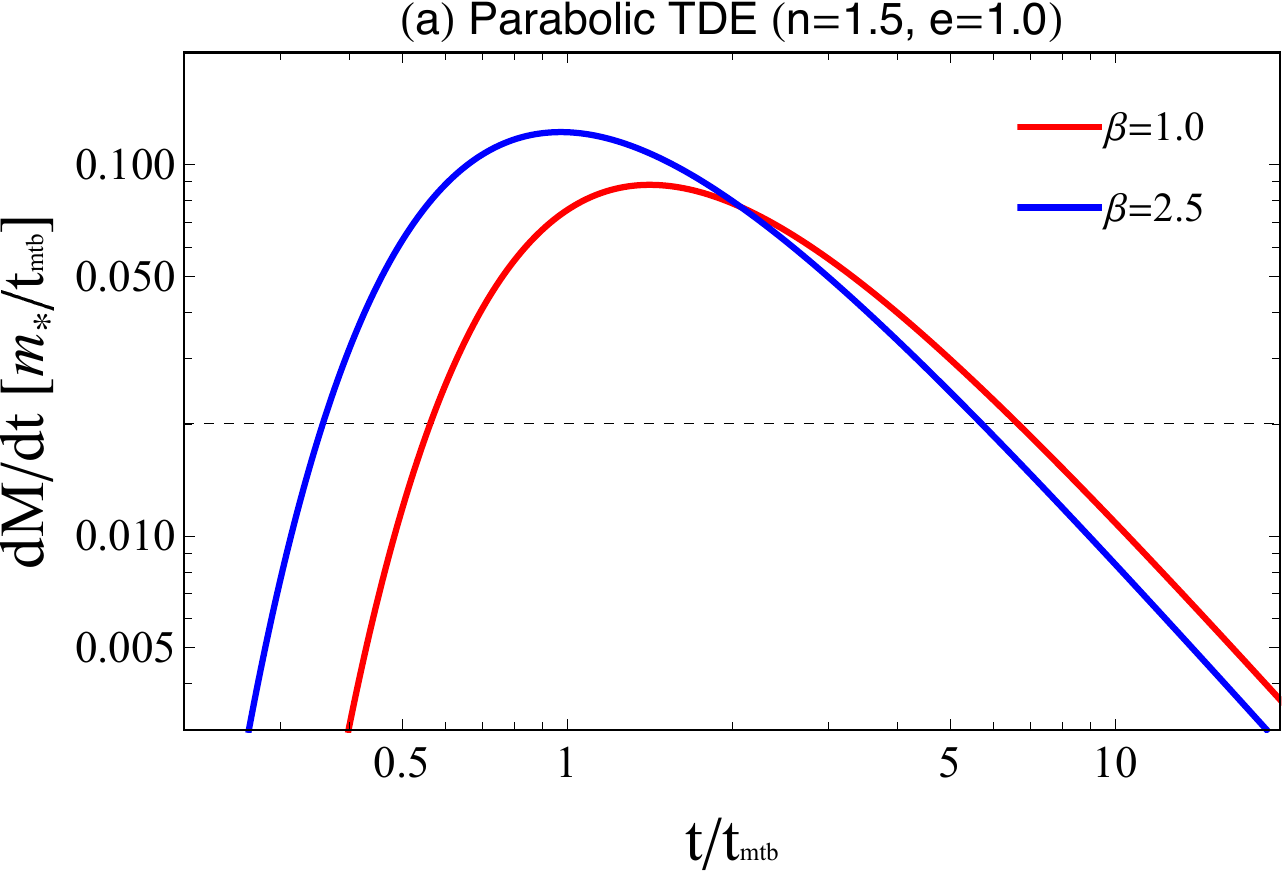}
\includegraphics[width=8cm]{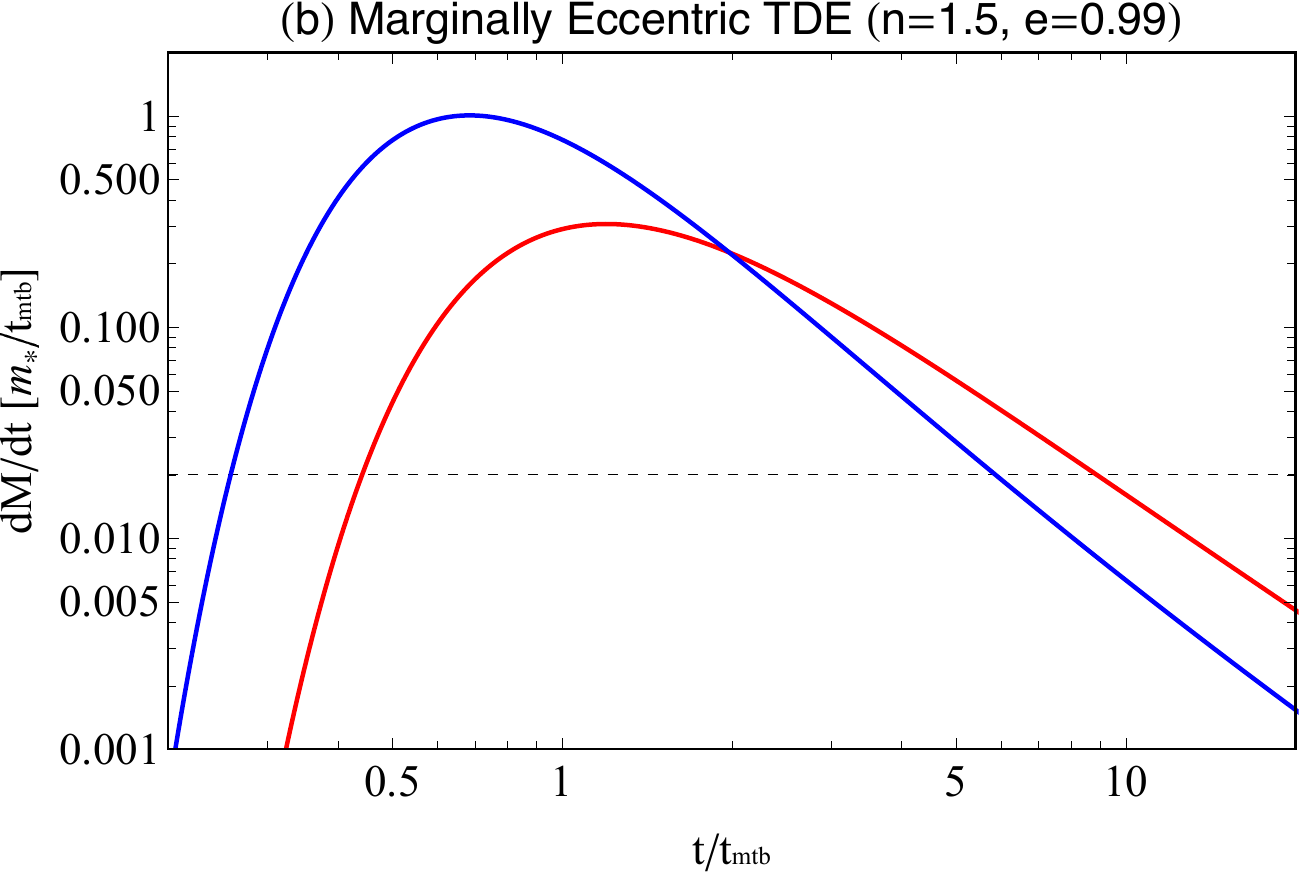}\\
\includegraphics[width=8cm]{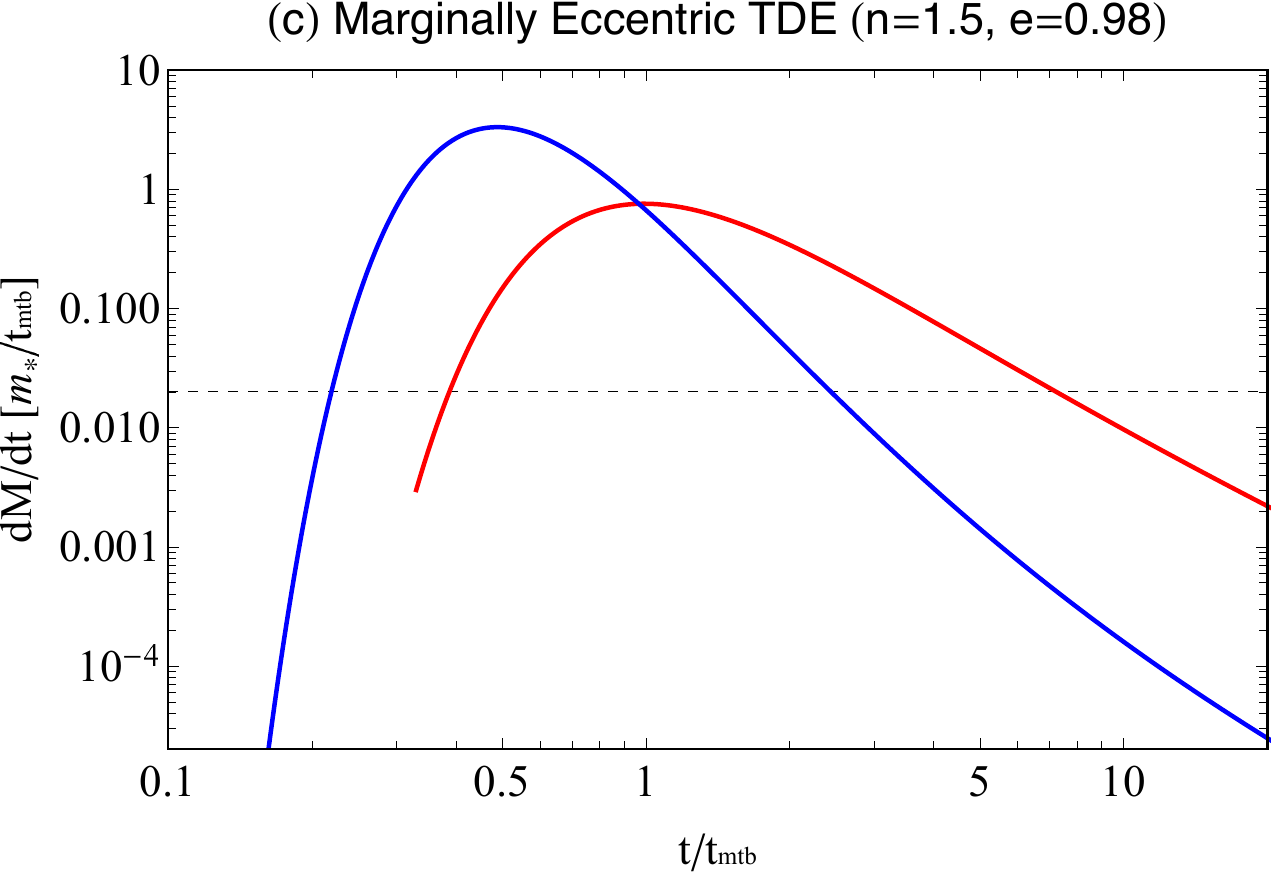}
\includegraphics[width=8cm]{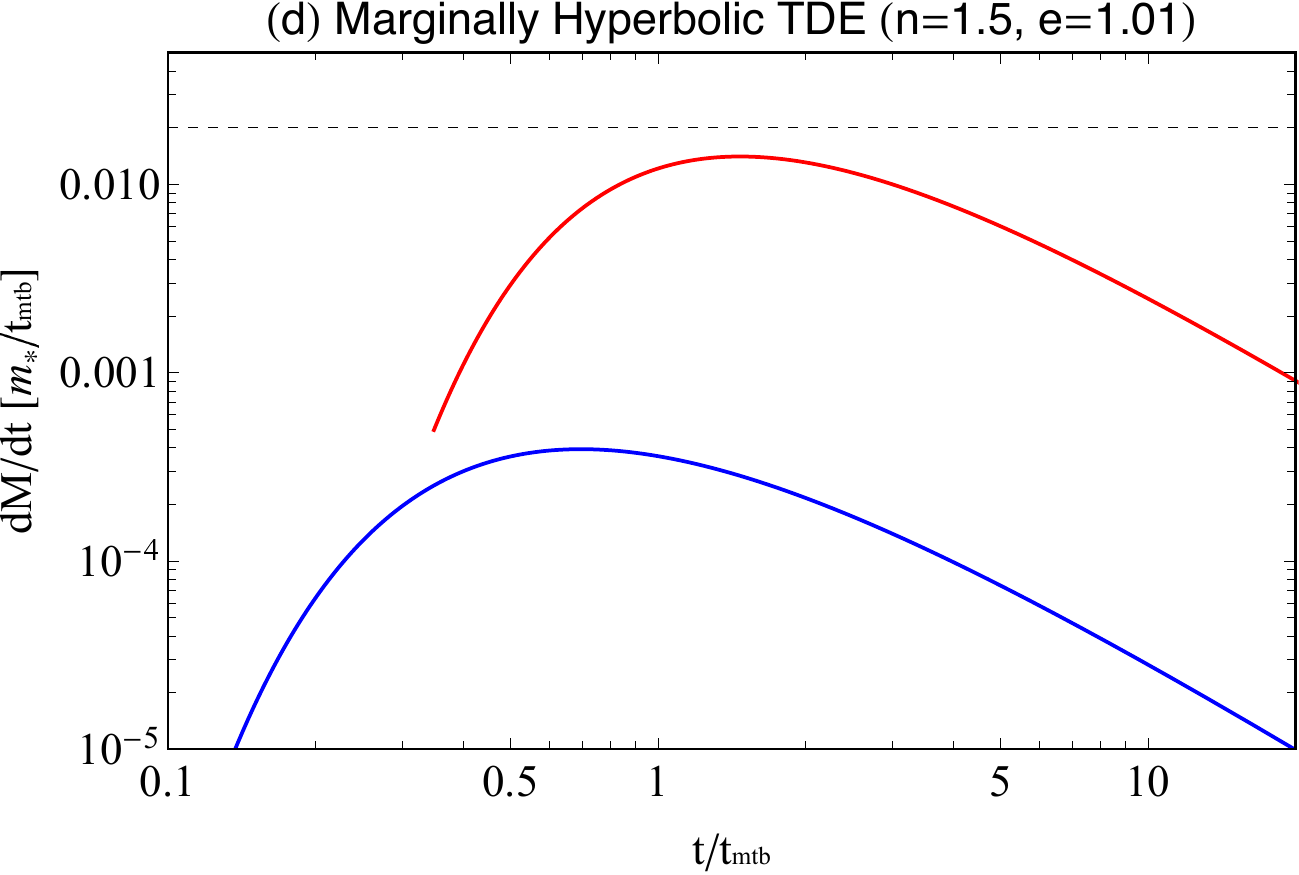}\\
\caption{
Gaussian-fitted mass fallback rates for a n=1.5 polytrope. 
They are normalized by $m_*/t_{\rm tmb}=0.11\,M_{\odot}/{\rm yr}
\left(M_{\rm bh}/10^{6}M_{\odot}\right)^{-1/2}
\left(M_{*}/M_{\odot}\right)
\left(r_{*}/R_{\odot}\right)^{-3/2}
\left(\beta/1\right)^{3k/2}$, where $k$ is obtained 
from Table~\ref{tbl:1}.
The red and blue solid lines show the normalized mass fallback rates of $\beta=1$ and $\beta=2.5$, respectively.
The dashed line shows the normalized Eddington accretion rate.
Each panel shows a different orbital eccentricity.
}
\label{fig:dmdt1}
\end{figure}

\begin{figure}[!ht]
\center
\includegraphics[width=8cm]{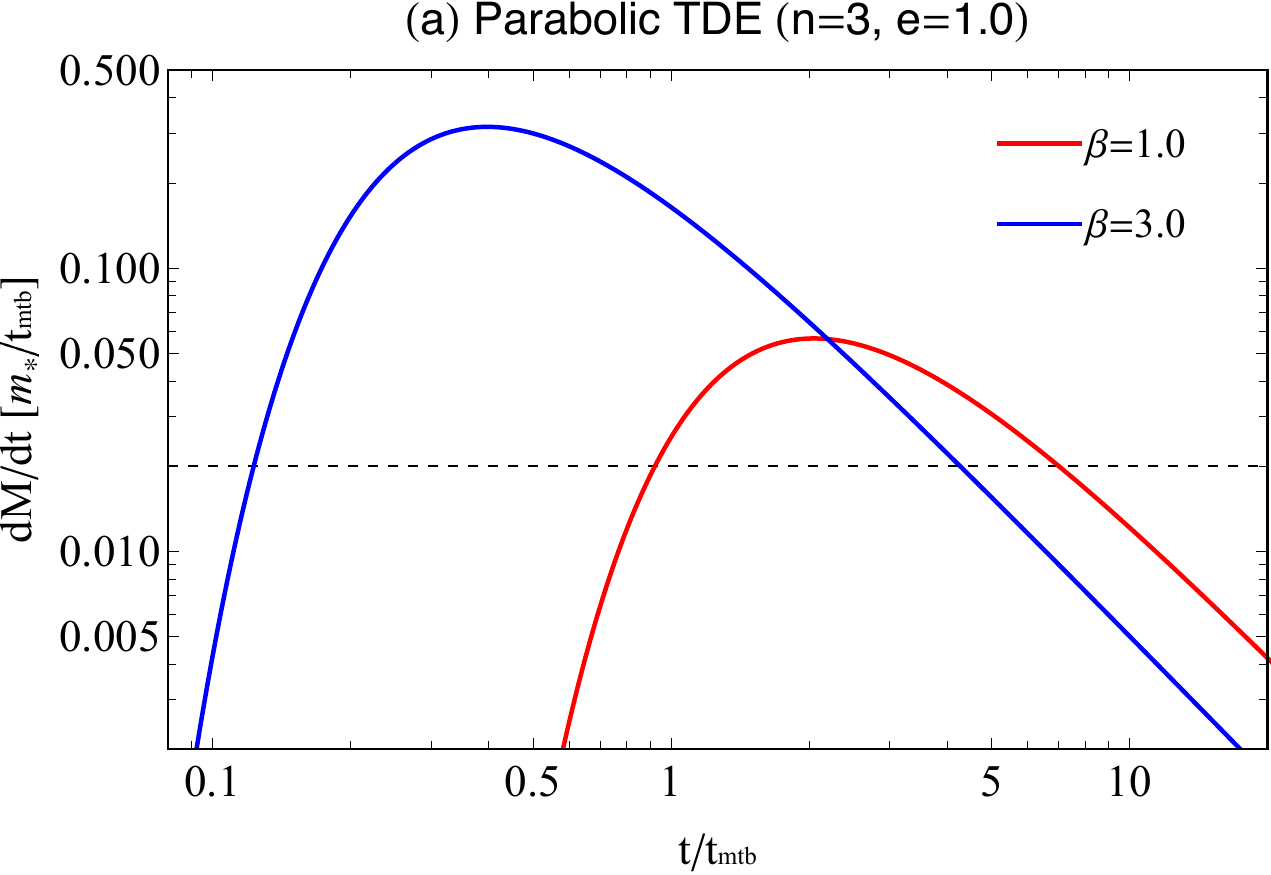}
\includegraphics[width=8cm]{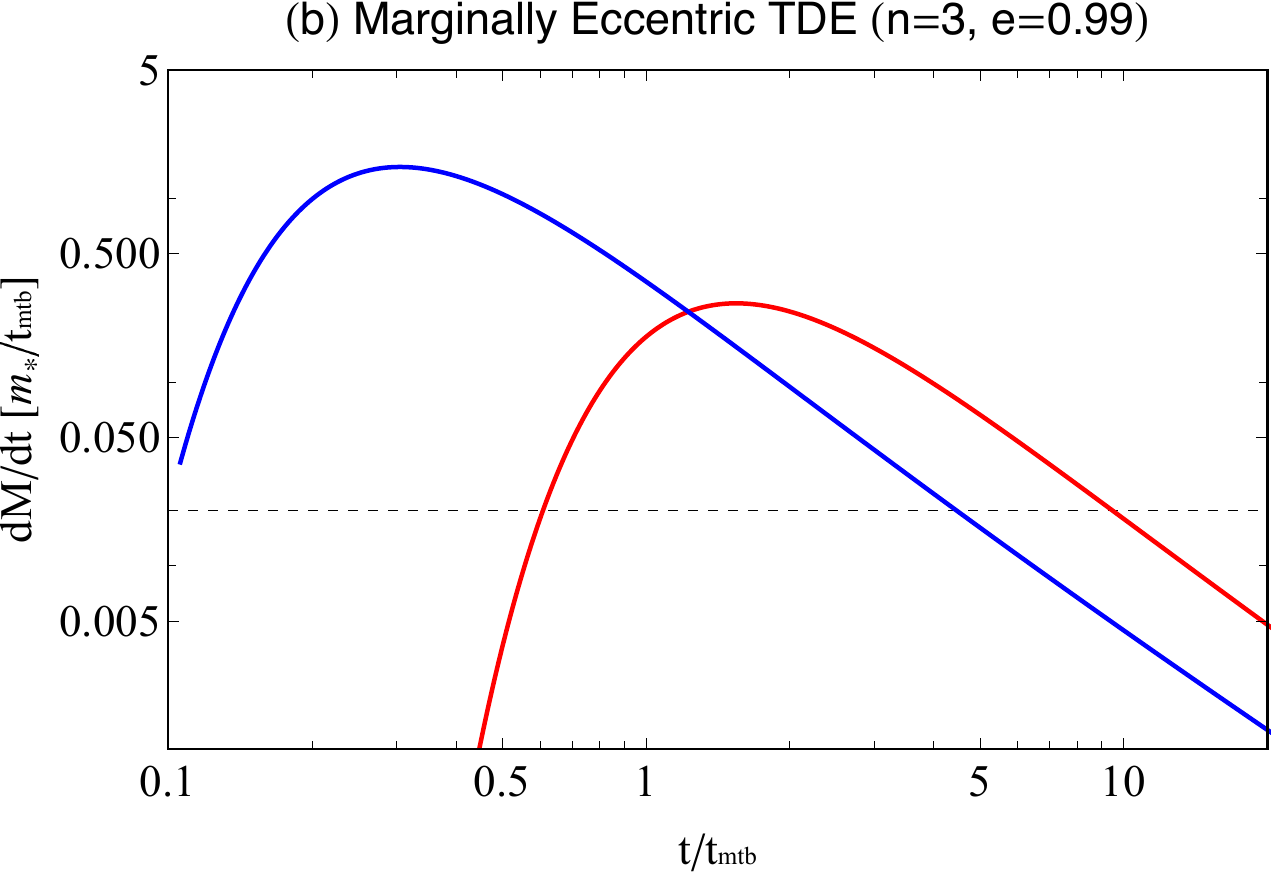}\\
\includegraphics[width=8cm]{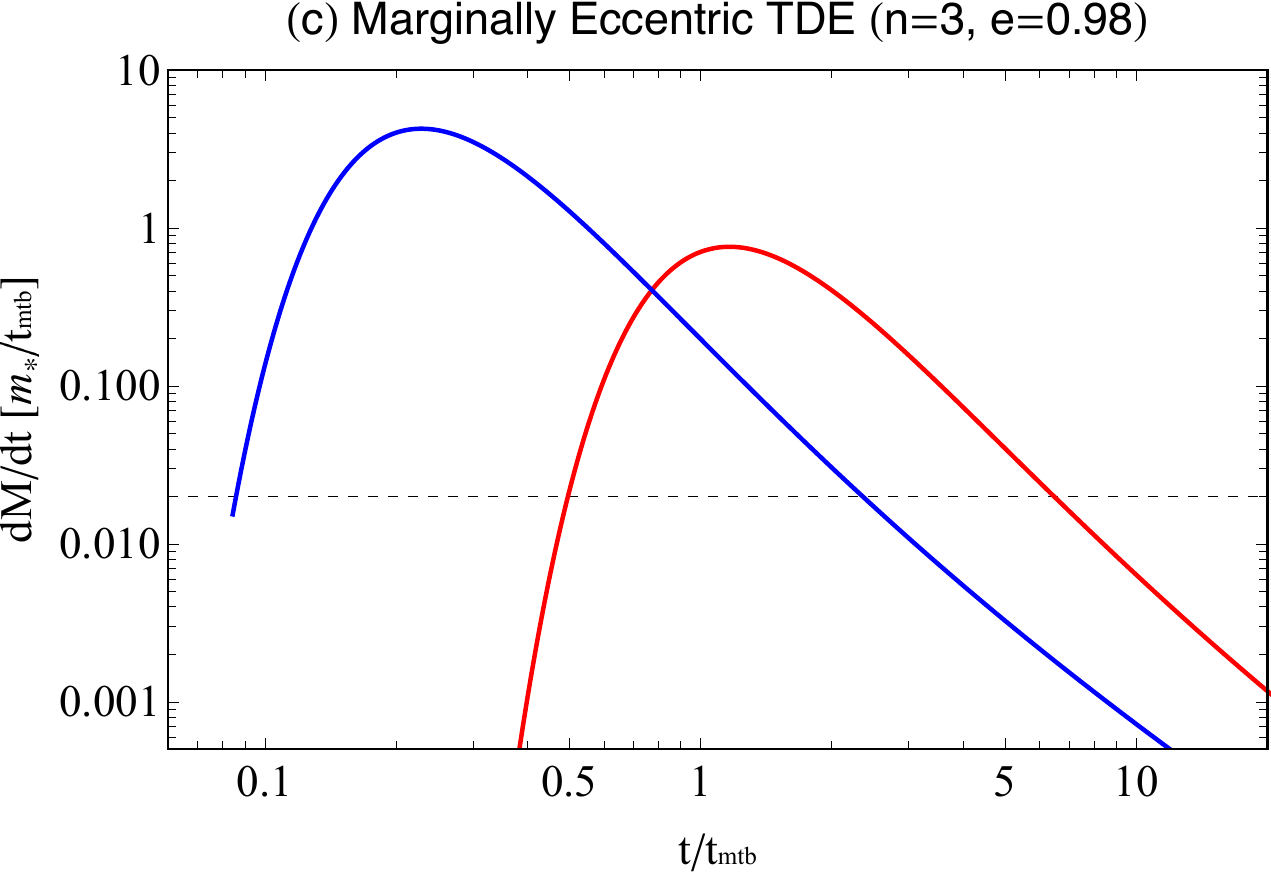}
\includegraphics[width=8cm]{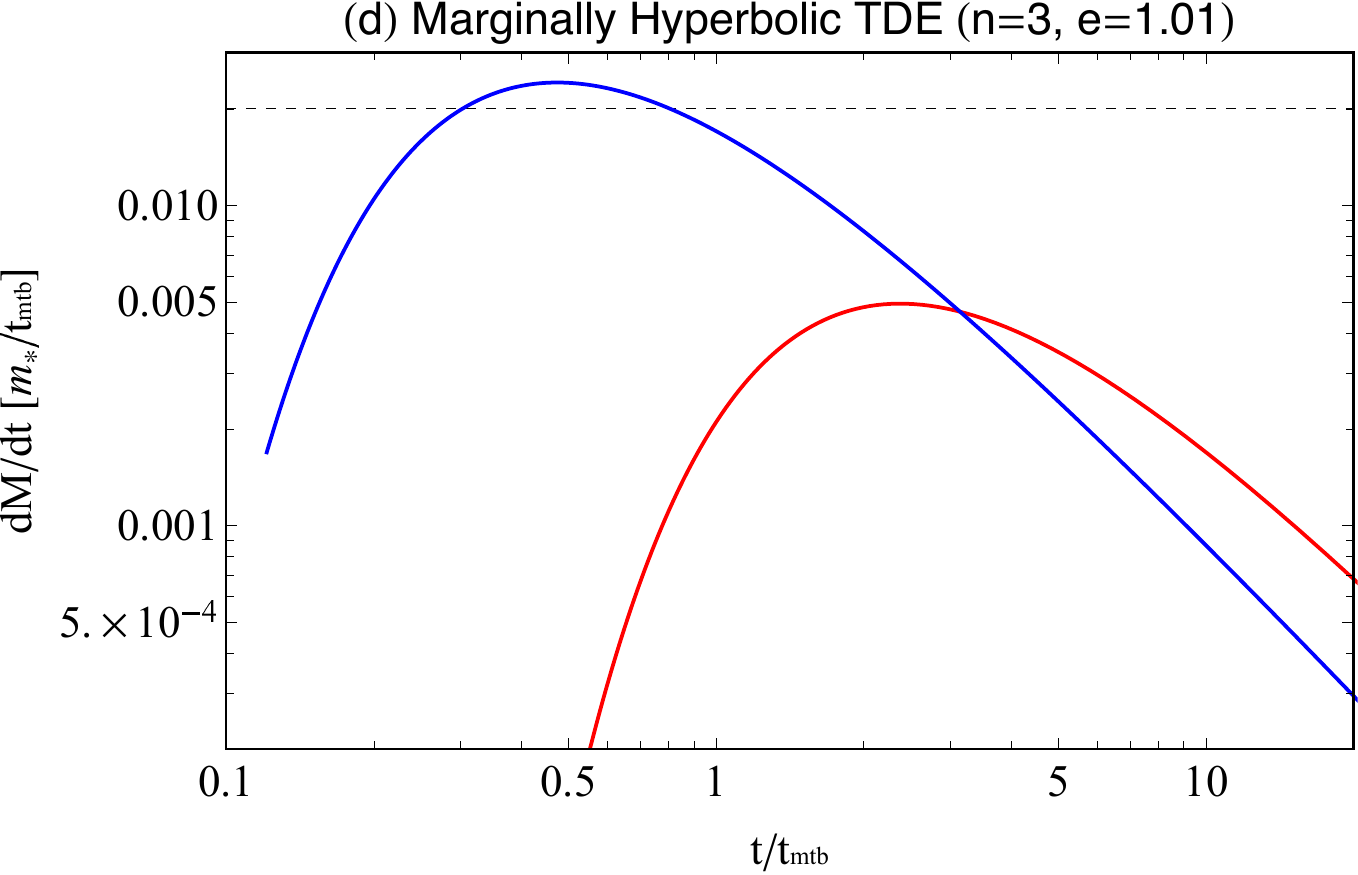}\\
\caption{
The same format as Figure~\ref{fig:dmdt1}, 
but for the $n=3$ case. The red and blue solid lines show 
the normalized mass fallback rates of $\beta=1$ and $\beta=3$, respectively.
}
\label{fig:dmdt2}
\end{figure}

Figures~\ref{fig:dmdt1} and \ref{fig:dmdt2} depict the Gaussian-fitted mass fallback rates for $n=1.5$ and $n=3$ polytropes, respectively.
The red and blue curves represent the mass fallback rates of $\beta=1$ and $\beta=2.5$ in Figure~\ref{fig:dmdt1}, while they correspond to the $\beta=1$ and $\beta=3$ cases in Figure~\ref{fig:dmdt2}. Panels (a)-(d) depict the mass fallback rate of the parabolic TDE ($e=1.0$), marginally eccentric TDEs ($e=0.99$ and $e=0.98$), and  the marginally hyperbolic TDE ($e=1.01$), respectively. The figures indicate that the peak of the mass fallback rate is higher as $\beta$ is higher, except for the marginally hyperbolic TDE of the $n=1.5$ case. This implies that the integral part of equation~\ref{eq:dmdt2} (i.e., stellar density profile) more tightly correlates with the penetration factor compared with the $1/\beta^k$ term. The reason why the $\beta=1$ rate is overall higher than the $\beta=2.5$ rate for the marginally hyperbolic TDE is that the amount of the bound debris of $\beta=1$ is larger than that of $\beta=2.5$ (see Panel (d) of Figure~\ref{fig:dmde1}). Overall, the mass fallback rates of marginally eccentric TDEs 
are an order of magnitude or more larger than those of the parabolic TDE, while the mass fallback rates of the marginally hyperbolic TDEs are less than or comparable to the Eddington rate and about one or a few orders of magnitude smaller than those of the parabolic TDEs.

\begin{figure}[!ht]

\includegraphics[width=8cm]{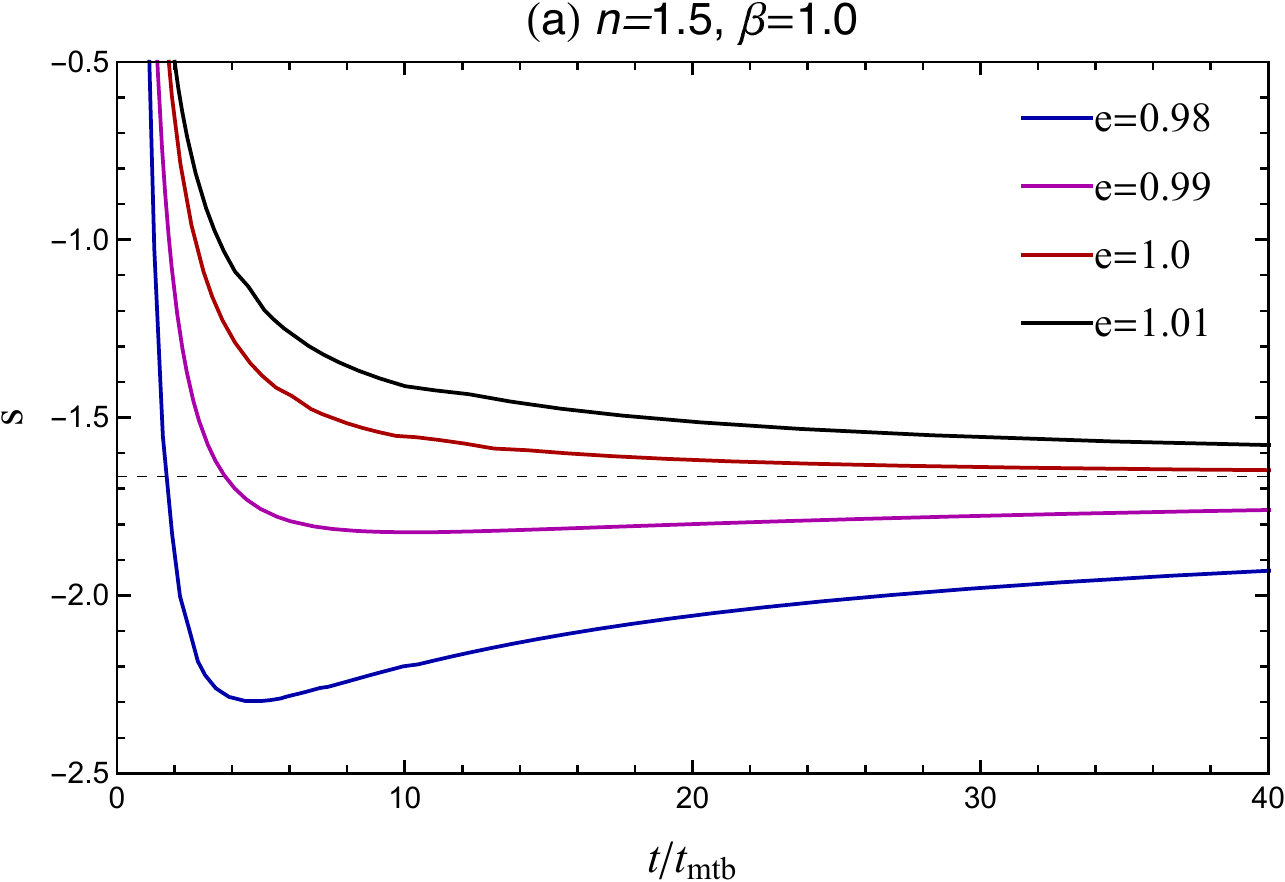}
\includegraphics[width=8cm]{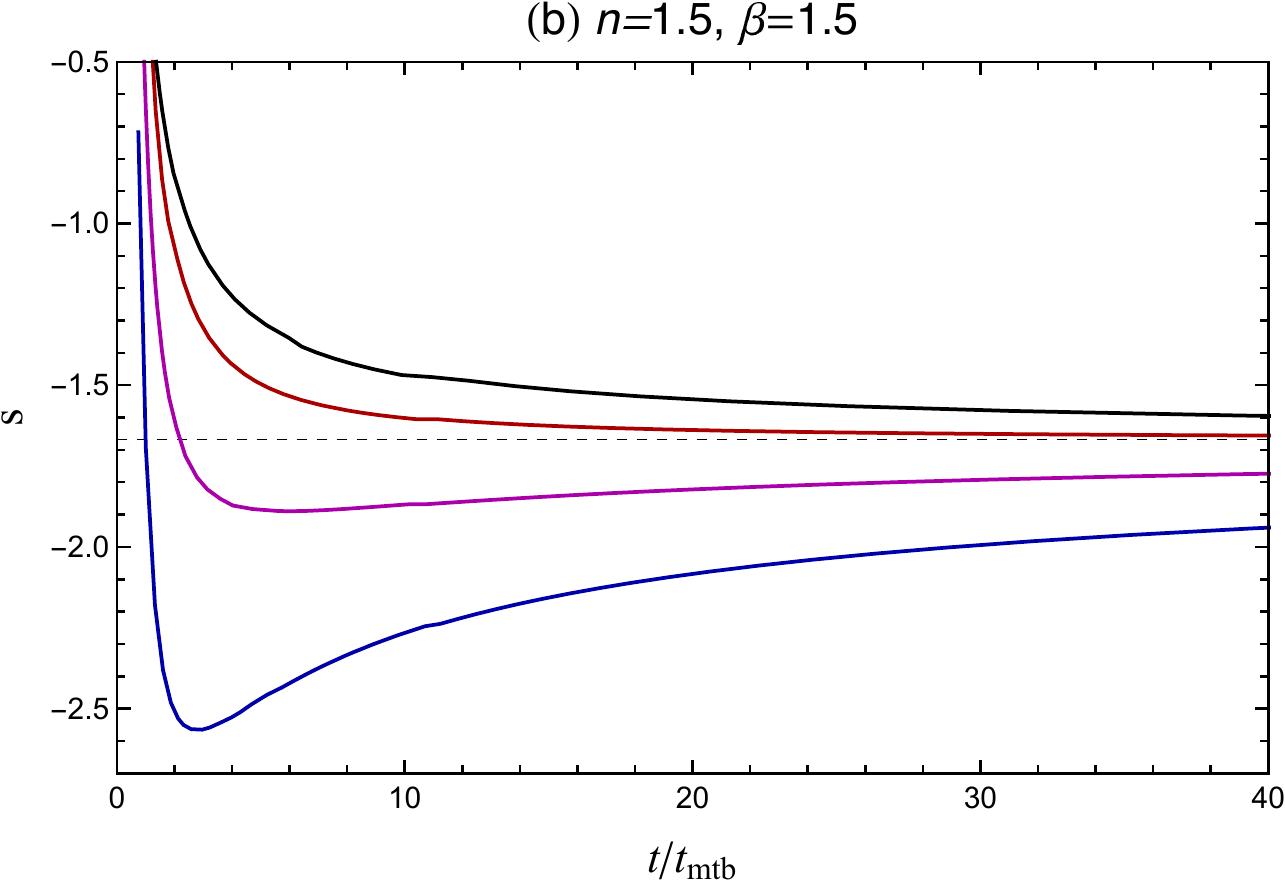}
\\
\includegraphics[width=8cm]{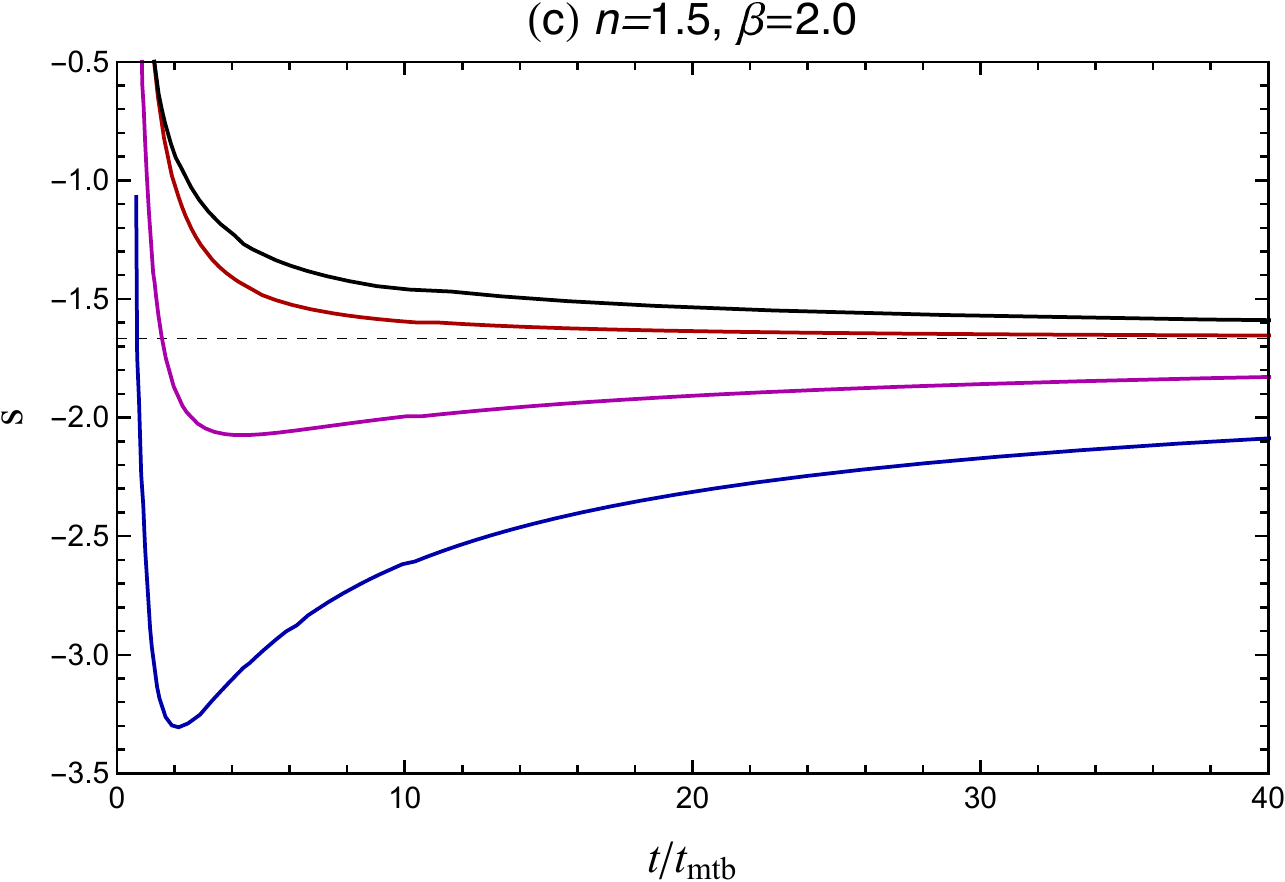}
\includegraphics[width=8cm]{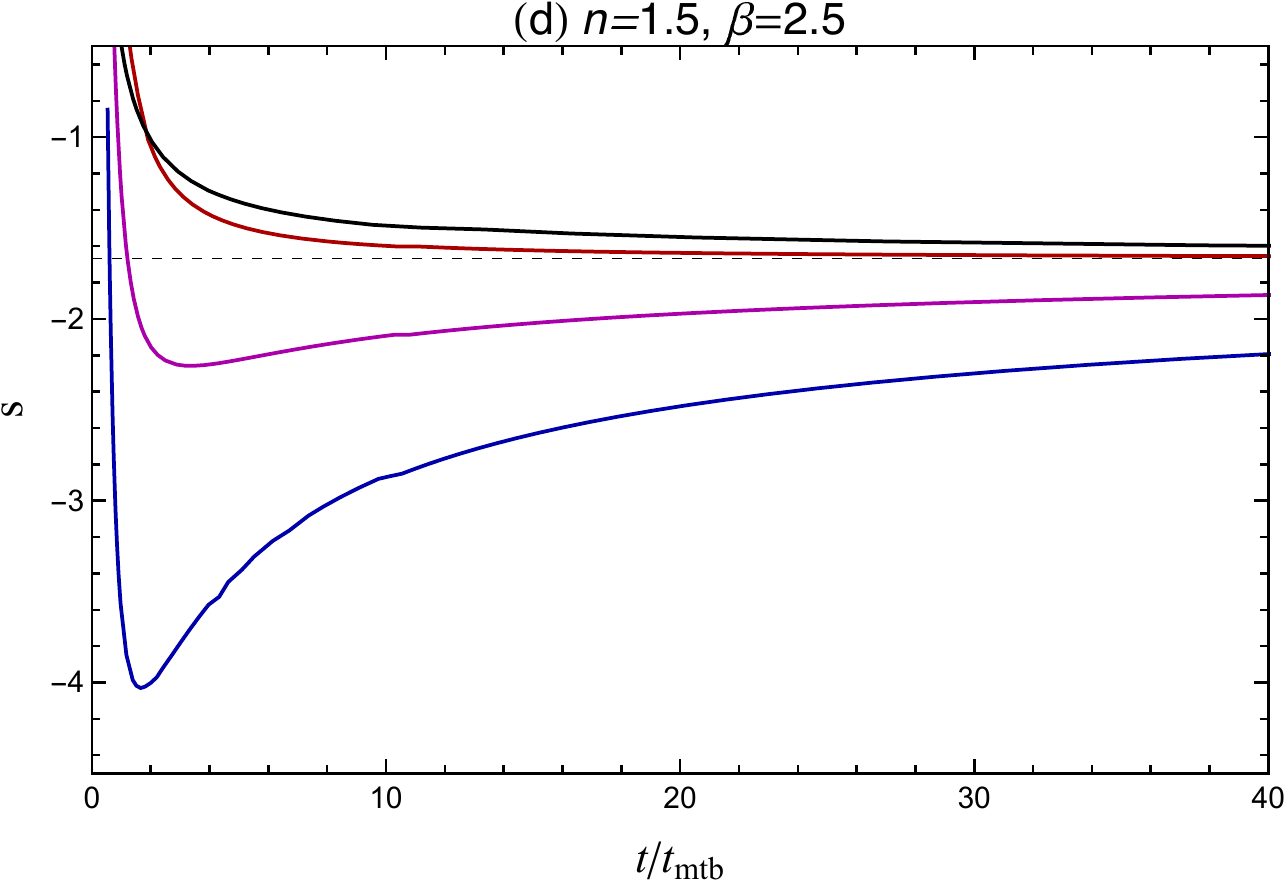}
\\
\includegraphics[width=8cm]{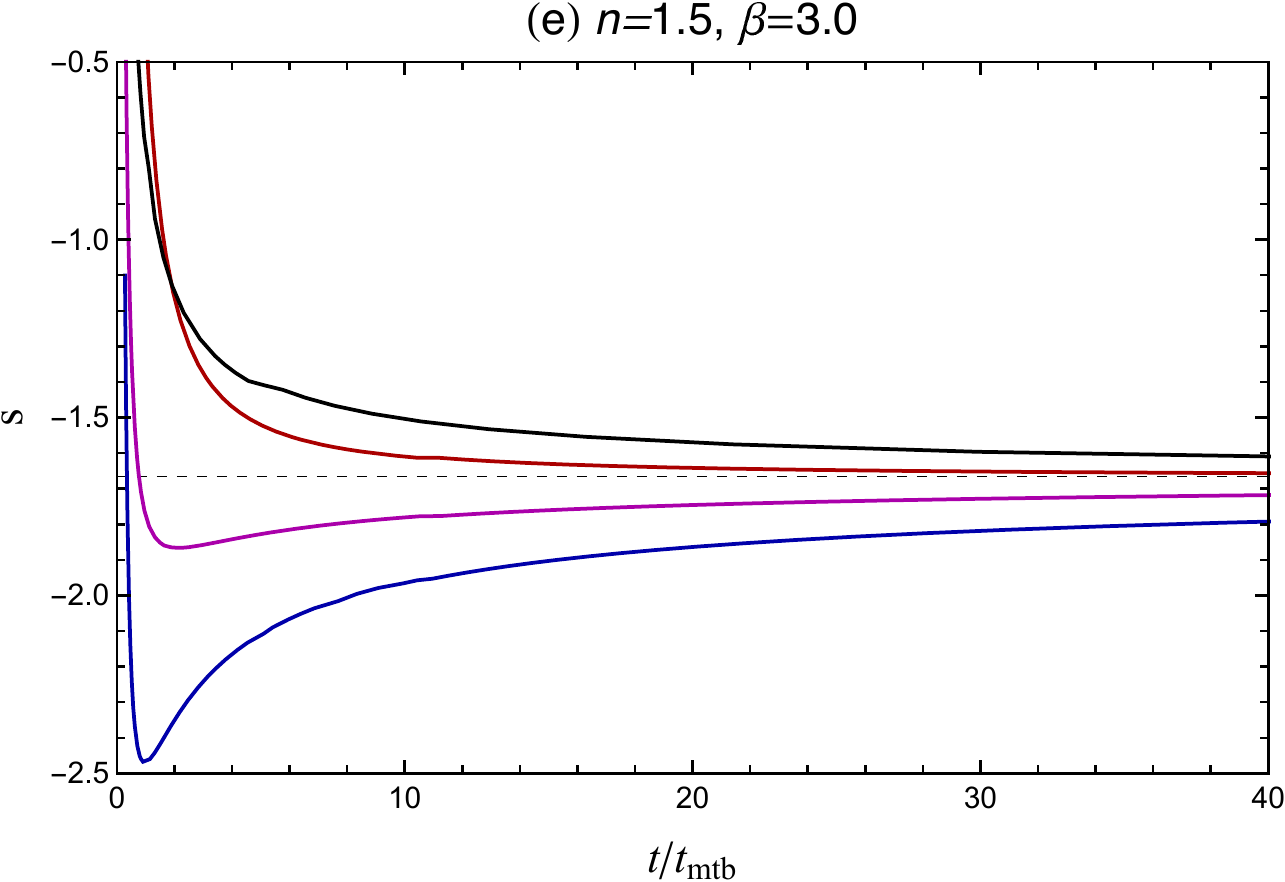}

\caption{
The slope of the Gaussian-fitted mass fallback rate as a function of time normalized 
by $t_{\rm mtb}$ for a $n=1.5$ polytrope. 
Each panel shows a different penetration factor.
The blue, magenta, red, and black solid lines denote 
the slopes of $e=0.98$, $e=0.99$, $e=1.0$, and $e=1.01$, respectively.
The dashed line represents $-5/3$, which is the slope of a standard TDE.
}
\label{fig:s-t1}
\end{figure}

\begin{figure}[!ht]

\includegraphics[width=8cm]{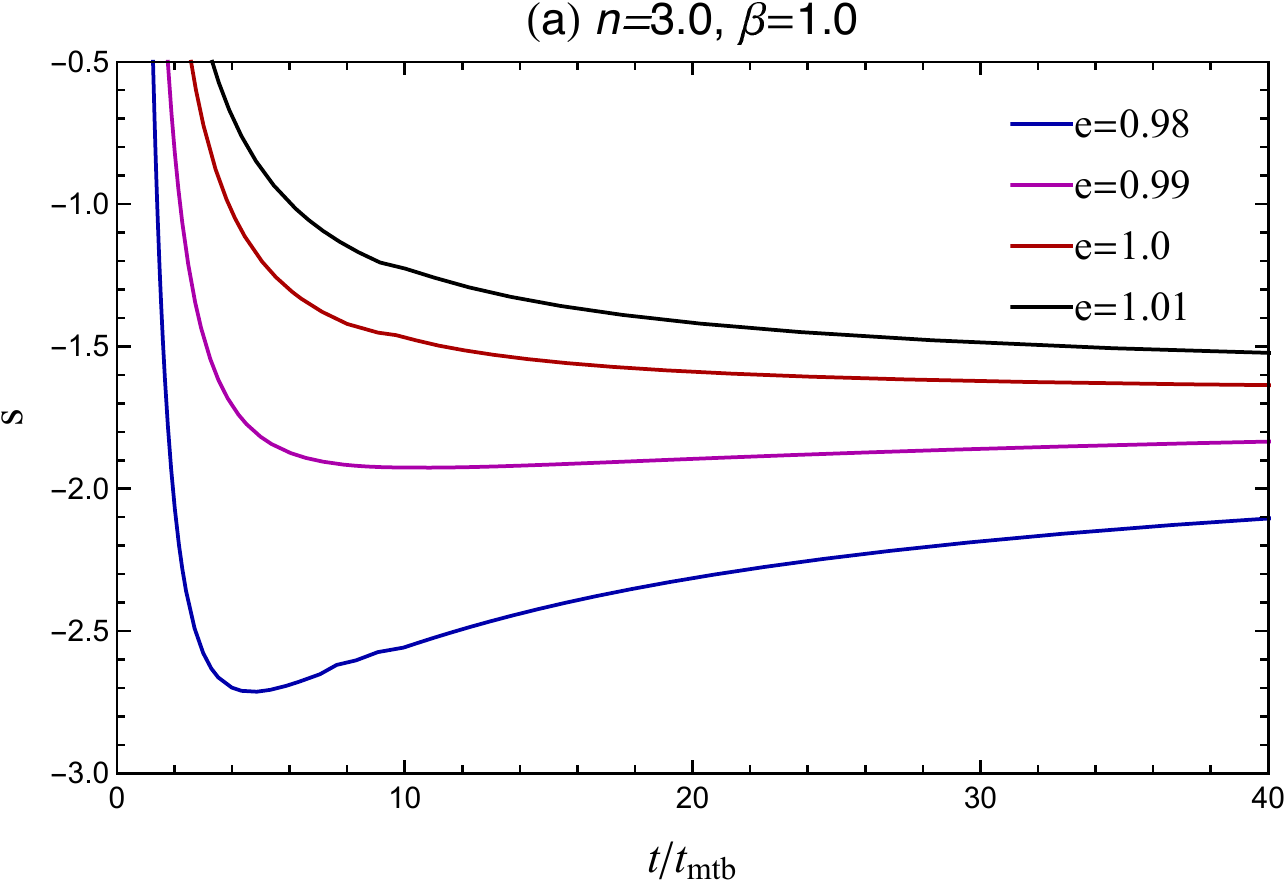}
\includegraphics[width=8cm]{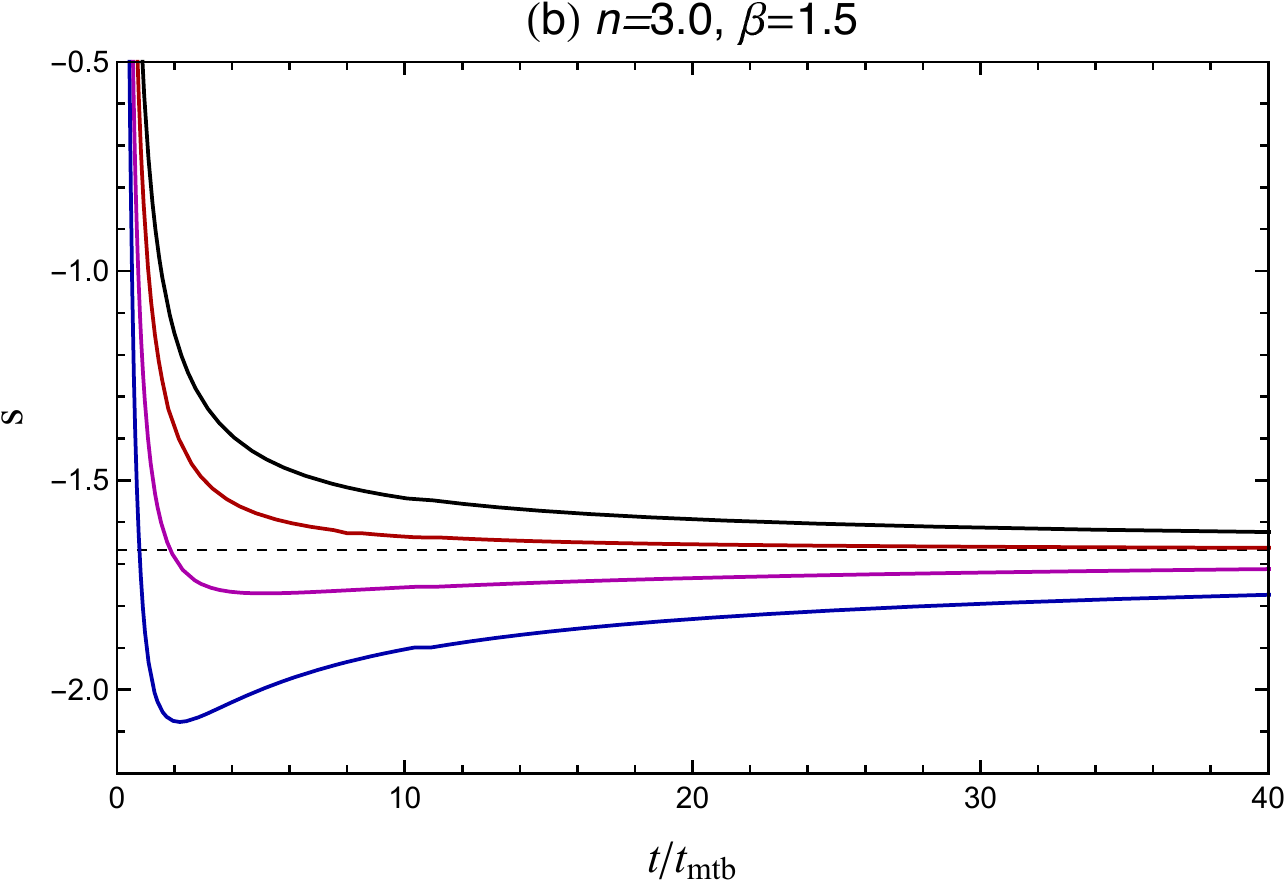}
\\
\includegraphics[width=8cm]{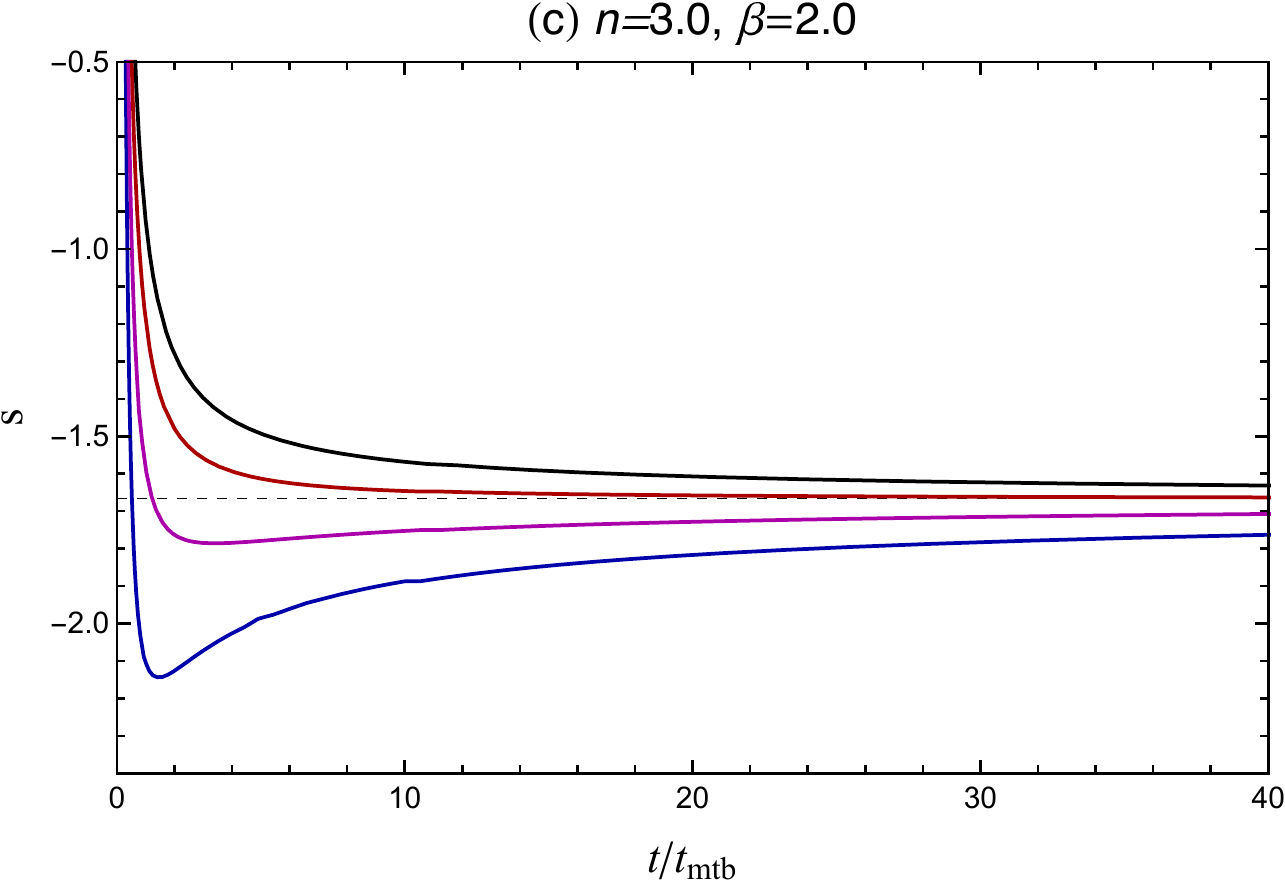}
\includegraphics[width=8cm]{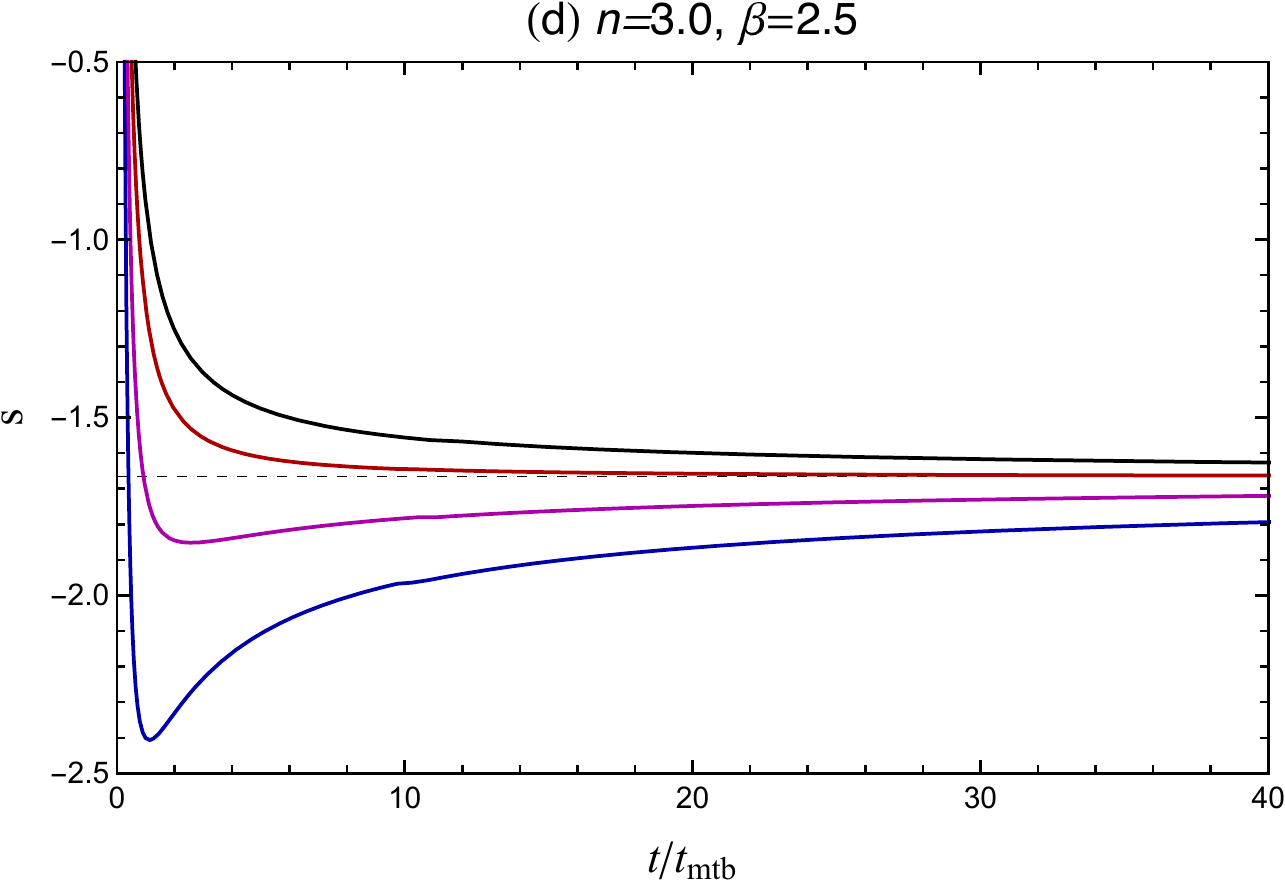}
\\
\includegraphics[width=8cm]{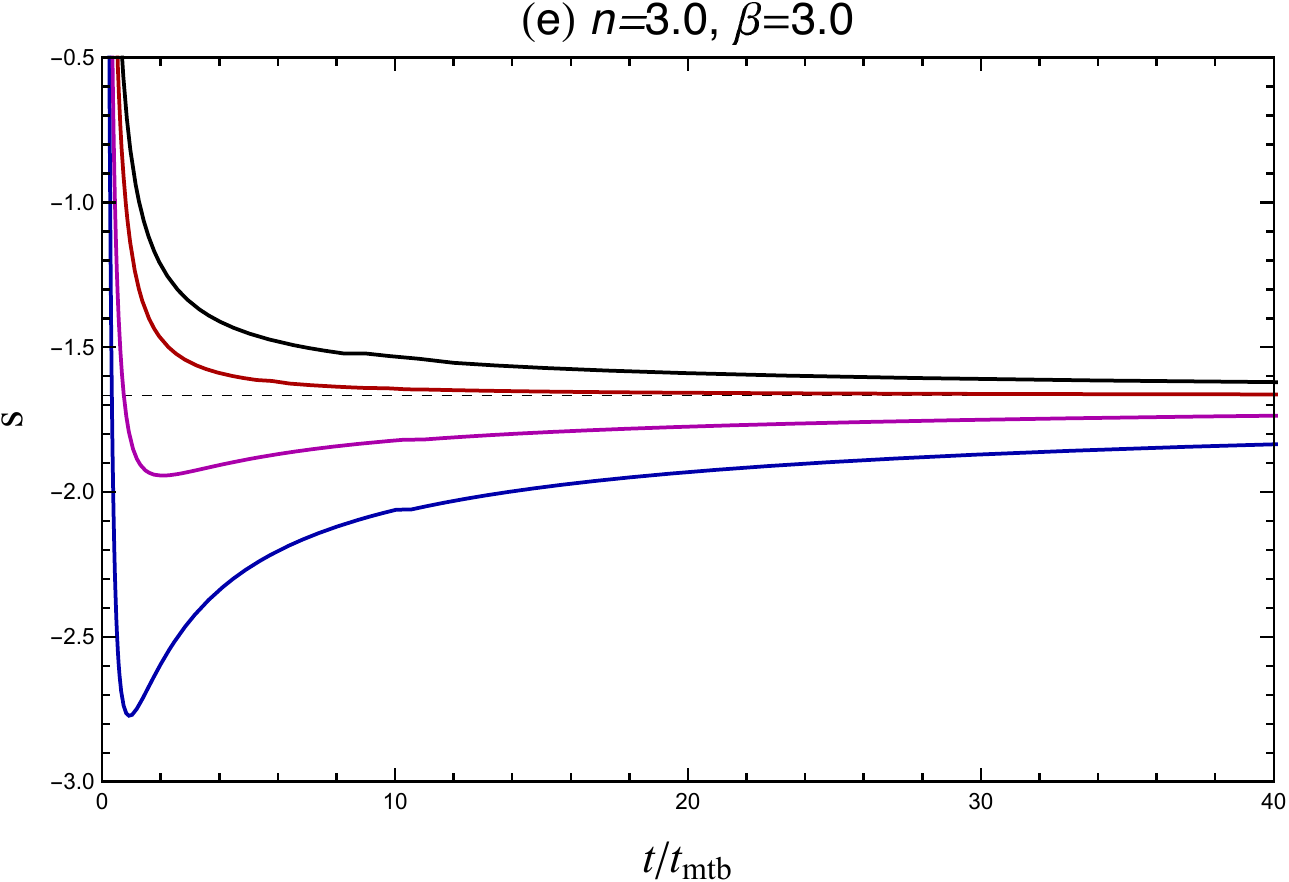}

\caption{The same format as Figure~\ref{fig:s-t1}, but for the $n=3$ case.}
\label{fig:s-t2}
\end{figure}

Figure~\ref{fig:s-t1} depicts the slope of the Gaussian-fitted mass fallback rate as
a function of time normalized by $t_{\rm mtb}$ for the case of a $n=1.5$ polytrope.
Assuming that $dM/dt=At^{s}$, the slope of the mass fallback rate is given by 
\begin{eqnarray} 
s=t\left(\frac{d^2M/dt^2}{dM/dt}\right),
\end{eqnarray}
where the proportionality coefficient, $A$, is a constant value. 
The solid blue, magenta, red, and black lines represents the slope of 
$e=0.98$, $e=0.99$, $e=1.0$, and $e=1.01$ cases, respectively, whereas 
the dashed line denotes the asymptotic slope of the mass fallback rate 
for the standard case, $-5/3$. Each panel depicts a different penetration factor.
We find that the mass fallback rates of all types of TDEs are flatter than $t^{-5/3}$ at early times, while they are different at very late times for respective TDEs. The mass fallback rate asymptotically approaches to $t^{-5/3}$ for parabolic TDEs, is steeper than $t^{-5/3}$ for marginally eccentric TDEs, and is flatter for marginally hyperbolic TDEs. 
The time evolution of the slope in the $n=3$ case is qualitatively same as the evolution of the slope 
in the $n=1.5$ case as shown in Figure~\ref{fig:s-t2}.

These outcomes are theoretically explained in the following way. 
Assuming $dM/d\epsilon=B{t}^{\alpha}$ ($B>1$), the slope of the 
mass fallback rate can be written by
\begin{equation}
s=\alpha-\frac{5}{3},
\label{eq:slope}
\end{equation}
where we adopt the Keplerian third law to equation~(\ref{eq:dmdt}).
Equation (\ref{eq:slope}) indicates that the mass fallback rate is flatter (steeper) than $t^{-5/3}$ 
if $\alpha$ is positive (negative). Because $d^2M/d\epsilon^2=3\left(2\pi GM_{\rm bh}\right)^{-2/3}B \alpha t^{\alpha +2/3}$ 
where we used equation~(\ref{eq:dedt}) for the derivation, $\alpha$ should be positive (negative) 
if the slope of $dM/d\epsilon$ about $\epsilon$ is positive (negative). It is clearly seen from 
Figure~{\ref{fig:dmde1}} that since the inclination of $dM/d\epsilon$ at the far negative side 
of $\epsilon$ is positive for all types of TDEs and all the given range of the penetration factor, 
$\alpha$ should be positive at early times. That is why the mass fallback rate is, independently of $\beta$, flatter than $t^{-5/3}$ at early times for all types of TDEs. On the other hand, it is clear from Figure~\ref{fig:dmde1} that the inclination of $dM/d\epsilon$ of parabolic TDEs is nearly flat at zero energy so that $\alpha\approx0$ for all the given range of the penetration factor. Thus, the asymptotic slope of the mass fallback rate should be, independently of $\beta$, $-5/3$ at late times. In marginally eccentric TDEs, the inclination of $dM/d\epsilon$ is negative at zero energy (i.e., $\alpha<0$) for all the given range of the penetration factor so that the asymptotic slope is steeper than $-5/3$ at late times. In marginally hyperbolic TDEs, the inclination of $dM/d\epsilon$ is positive at zero energy (i.e., $\alpha>0$) for all the given range of the penetration factor so that the asymptotic slope is flatter than $-5/3$ at late times.


%
%

%
%

\clearpage

%
%
\clearpage

%
%

%
%

%
\section{Discussion and Conclusions}
\label{sec:con}
%
We have revisited the mass fallback rates of marginally 
bound to unbound TDEs by taking account of the penetration factor ($\beta$), 
tidal spread energy index ($k$), orbital eccentricity ($e$), 
and stellar density profile with a polytropic index ($n$). 
We have compared the semi-analytical solutions with 3D 
SPH simulation results. Our primary conclusions are 
summarized as follows:
\begin{enumerate}
\item 
We have analytically derived the formulations of both the differential mass distribution and 
corresponding mass fallback rate and obtained the semi-analytical solutions for them. 
Both the differential mass distribution and corresponding mass fallback rate depend on the 
penetration factor, tidal spread energy index, orbital eccentricity (or semi-major axis), 
and stellar density profile (see equations~\ref{eq:semidmde} and \ref{eq:kimidmde} for differential 
mass distributions and equations~\ref{eq:dmdt1} and \ref{eq:dmdt2} for mass fallback rates).
\item
The differential mass distributions obtained by the SPH simulations show good agreement with 
the Gaussian-fitted curves, with errors of $\sim5\%$ to $18\%$ for $n=1.5$, and 
$\sim7\%$ to $24\%$ for $n=3$. We find that the Gaussian-fitted curves are in good agreement with 
the semi-analytical solutions, indicating that the analytically derived mass fallback 
rates can match the simulated rates within the range of the fitting accuracy.
\item The simulated spread in debris energy is larger than $\Delta\epsilon=GM_{\rm bh}/r_{\rm t}(r_*/r_{\rm t})$ 
as the penetration factor increases for all the cases, which is consistent with our assumption that the spread 
in debris energy is proportional to $\beta^k$ (see also equation (4) of \citealt{2013MNRAS.435.1809S}).
While the tidal spread energy index is distributed over $0.27\lesssim{k}\lesssim0.9$ 
for $n=1.5$ case for any value of $\beta$ and $e$, it is distributed over 
$0.97\lesssim{k}\lesssim1.58$ for $n=3$ and any value of $\beta$ and $e$ 
except for the $\beta=1.5$ case. In this case, $k$ ranges from $2.08$ to $2.17$ depending 
on the orbital eccentricity.

\item
We have updated the two critical eccentricities to classify five types of TDEs (see also \citealt{2018ApJ...855..129H}), based on the spread in energy being proportional to $\beta^k$, as follows: $e_{\rm crit,1}=1-2(M_{\rm bn}/m_*)^{-1/3}\beta^{k-1}$ and $e_{\rm crit,2}=1+2(M_{\rm bh}/m_*)^{-1/3}\beta^{k-1}$ (see equation~\ref{eq:ec1}). Again, TDEs can be classified by five different types: eccentric ($e<e_{\rm crit,1}$), marginally eccentric ($e_{\rm crit,1}\lesssim{e}<1$), purely parabolic ($e=1$), marginally hyperbolic ($1<e<e_{\rm crit,2}$), and hyperbolic ($e\gtrsim{e_{\rm crit,2}}$) TDEs, respectively.

\item
The mass fallback rates of the marginally eccentric TDEs are an order of magnitude or more larger than those of the parabolic TDEs, while the mass fallback rates of the marginally hyperbolic TDEs are less than or comparable to the Eddington rate and about one or a few orders of magnitude smaller than those of the parabolic TDEs.

\item
We find that the mass fallback rates of all the types of TDEs are flatter than $t^{-5/3}$ at early times, 
while they are different at late times for the respective TDEs. The mass fallback rate asymptotically 
approaches to $t^{-5/3}$ for the parabolic TDEs, is steeper than $t^{-5/3}$ for the marginally eccentric TDEs, 
and is flatter for the marginally hyperbolic TDEs. The flatter nature at early times is because the inclination 
of the differential mass-energy distribution at the far side of negative debris energy is positive, while the 
steeper (flatter) nature at late times is because it is negative (positive) at zero energy. The reason why the 
asymptotic slope is $-5/3$ is that the slope of the differential mass distribution at zero energy is nearly 
flat (see also the detailed argument in the last two paragraphs of Section~\ref{sec:dmdt}).

\item 
When the orbital eccentricity ranges for $1.0<e<e_{\rm crit,2}$ (i.e., marginally hyperbolic TDEs), 
only a little fraction of stellar mass can fall back to the black hole, which leads to the formation of 
advection dominated accretion flow (ADAF) or radiatively inefficient accretion flows (RIAF). The 
marginally hyperbolic TDEs can be an origin of ADAFs (or RIAFs) around dormant SMBHs. 
If the orbital eccentricity is more than $e_{\rm crit,2}$ (i.e., hyperbolic TDEs), no stellar debris 
can fall back to the black hole, which leads to a failed TDE.
\end{enumerate}

Given the black hole mass, stellar mass and mass, and stellar density profile with a polytrope index, we see from Figures~\ref{fig:dmdt1} and \ref{fig:dmdt2} that the mass fallback rate depends on the orbital eccentricity and penetration factor, and the orbital eccentricity enhances the peak of mass fallback rate than the penetration factor. The peak of the mass fallback rate can change by orders of magnitude over the range of $0.98\le{e}\le1.01$ for both $n=1.5$ and $n=3$ cases. 
On the other hand, the mass fallback rate also strongly depends on the stellar types (i.e., stellar mass and radius): a white dwarf disruption shows a much larger fallback rate than the main sequence cases, whereas a red giant disruption represents a significantly lower rate \citep{2017ApJ...841..132L}. TDEs of different stellar types are distinguishable, e.g., through the spectral line diagnosis because these stars have different compositions. The degeneracy between the stellar type and the orbital eccentricity should, then, be solvable by comparison with the TDEs of the same stellar types.

Partial tidal disruptions are the events where the outer layers of a star are tidally stripped by the black hole tidal forces. According to \cite{2013ApJ...767...25G, 2017A&A...600A.124M}, a star can be partially disrupted if $\beta\lesssim0.9$ for a $n=1.5$ polytrope and if $\beta\lesssim2.0$ for a $n=3$ polytrope. No partial tidal disruption is seen for n=1.5 because of $\beta\ge1.0$ in our simulations, while a star should be partially disrupted for the $n=3$ case with $\beta=1.0$ and $\beta=1.5$. In these cases, the mass fallback rate is thought to be asymptotically proportional to $t^{-9/4}$ \citep{2013ApJ...767...25G,2019ApJ...883L..17C} because of the influence of the survived core on the stellar debris. However, the asymptotic slope of the mass fallback rate is $-5/3$ in our parabolic TDE case and is steeper only for the marginally eccentric TDEs. It suggests that there can be a degeneracy between the partial disruption of stars on parabolic orbits and marginally eccentric TDEs. We will study the partial disruption of stars on bound orbits to examine whether this degeneracy is solvable in a forthcoming paper.


Recent observations suggest that the peak luminosities of some candidates are significantly sub-Eddington in optical/UV \citep{2017ApJ...842...29H,2019ApJ...872..151M} and X-ray TDEs \citep{2017ApJ...838..149A}. 
These are inconsistent with the assumption that the bolometric luminosity is proportional to the mass fallback rate because the mass fallback rate significantly exceeds the Eddington accretion rate in the case of main-sequence star disruptions. 
Two main ideas to solve this problem have been proposed. 
One is the partial TDEs, where the mass fallback rate can be lower \citep{2001ApJ...549..467I,2013ApJ...767...25G}. Another idea is that the radiative efficiency is very low for an elliptical disk accretion \citep{2017MNRAS.467.1426S}. 
We propose a new approach that tidal disruption of stars on marginally hyperbolic orbits causes the low luminosity TDEs. As future works, this possibility should be tested by comparing it with the other two existing ideas.


%
If $4\lesssim\beta\lesssim\beta_{\rm max}$, where $\beta_{\rm max}=r_{\rm t}/r_{\rm S}\approx24\,(M_{\rm bh}/10^6\,M_\odot)^{-2/3}(m_*/M_\odot)^{-1/3}(r_*/R_\odot)$ is the maximum possible value of the penetration factor, the general relativistic (GR) effects get significantly important. In this case, the spread in tidal energy would not follow the simple power law of the penetration factor anymore. For example, if $\beta=10$ and $k\sim2$, we are not sure if the spread in debris energy can be 100 times larger than the standard case, as predicted by the simple scaling (see equation~\ref{eq:delebeta}). It is also unclear how high and steep the early time 
mass fallback rate is in this case. We will examine how much GR can affect $\beta$-dependence of spread in debris energy 
in tidal disruption of such a deep-plunging star, although some existing studies show deviation from 
the scaling law \citep{2015ApJ...805L..19E,2019MNRAS.488.5267D,2019MNRAS.485L.146S}.

\cite{2019ApJ...886..114H} proposed that high-energy neutrinos with $\sim7.5\,{\rm TeV}\,(M_{\rm bh}/10^6\,M_\odot)^{5/3}$ can be emitted from an ADAF and/or RIAF formed after tidal 
disruption of a star by the decay of charged pions originated in ultra-relativistic protons.
In the standard TDE theory, the RIAF phase would start at $t_{\rm RIAF}\sim10^{10}\,{\rm s}$ for $10^6\,M_\odot$ black hole after a solar-type star disruption. For marginally unbound TDEs, the RIAF phase would start at about four orders of magnitude earlier than the standard case, i.e., $t_{\rm RIAF}\sim10^{6}\,{\rm s}$ for $10^6\,M_\odot$ black hole and a solar-type star. Because the neutrino energy generation rate is estimated to be $L_\nu{t}_{\rm RIAF}\mathcal{R}$, where $L_\nu$ and $\mathcal{R}$ are the neutrino luminosity and the TDE rate respectively, such a short timescale can significantly enhance the energy generation rate even if the event rate of marginally hyperbolic TDEs would be subdominant.

%
%
\section*{Acknowledgments}
%
The authors thank Matthew Bate, Atsuo Okazaki, Takahiro Tanaka, and Nicholas C. Stone 
for their helpful comments and suggestions. The authors thank the anonymous referee 
for constructive comments and suggestions that helped to improve the manuscript.
The authors also acknowledge the Yukawa Institute for Theoretical Physics (YITP) at Kyoto University. 
Discussions during the YITP workshop YITP-T-19-07 on International Molecule-type Workshop 
"Tidal Disruption Events: General Relativistic Transients” were useful to complete this work.
This research has been supported by Basic Science Research 
Program through the National Research Foundation of Korea (NRF) funded by the 
Ministry of Education (2016R1A5A1013277 and 2017R1D1A1B03028580 to K.H.), 
and also supported by the National Supercomputing Center with supercomputing 
resources including technical support (KSC-2019-CRE-0082 to G.P. and K.H.).
%

\end{document}